\documentclass{ieeeaccess}
\usepackage[T1]{fontenc}
\usepackage{cite}
\usepackage{amsmath,amssymb,amsfonts}
\usepackage{algorithmic}
\usepackage{graphicx}
\usepackage{textcomp}
\usepackage{makecell}
\usepackage{multirow}
\usepackage{rotating}
\usepackage{booktabs}

\def\BibTeX{{\rm B\kern-.05em{\sc i\kern-.025em b}\kern-.08em
    T\kern-.1667em\lower.7ex\hbox{E}\kern-.125emX}}
    
\usepackage{subfigure}
\usepackage[usenames,dvipsnames,svgnames,table]{xcolor}
\usepackage{comment}
\usepackage{enumerate}
\usepackage{enumitem}

\usepackage{framed}

\usepackage{pifont}
\newcommand{\cmark}{\ding{51}}%
\usepackage{url}
\usepackage{breakurl}
\usepackage[breaklinks]{hyperref}

\usepackage{multicol}
\usepackage{caption}
\usepackage{longtable}

\graphicspath{
  {./}
  {./figures/}
}

\begin{document}
\history{Received 11 May 2023, accepted 10 June 2023, date of publication 21 June 2023, date of current version 28 June 2023.}
\doi{10.1109/ACCESS.2023.3288334}

\title{5G Multi-access Edge Computing: a Survey on Security, Dependability, and Performance}

\author{
\uppercase{Gianfranco Nencioni}\authorrefmark{1},
\uppercase{Rosario G. Garroppo}\authorrefmark{2}, and
\uppercase{Ruxandra F. Olimid}\authorrefmark{3}}
\address[1]{Department of Electrical Engineering and Computer Science, University of Stavanger, 4021 Stavanger, Norway (e-mail: gianfranco.nencioni@uis.no)}
\address[2]{Department of Information Engineering, University of Pisa, 56126 Pisa, Italy (e-mail: rosario.garroppo@unipi.it)}
\address[3]{Department of Computer Science, University of Bucharest, 030018 Bucharest, Romania (e-mail: ruxandra.olimid@fmi.unibuc.ro)}

\tfootnote{This work was supported by the Norwegian Research Council through the 5G-MODaNeI project (no. 308909) and the Italian Ministry of Education and Research (MIUR) in the framework of the FoReLab project (Departments of Excellence).}

\markboth
{G. Nencioni, R. G. Garroppo, and R. F. Olimid: 5G Multi-access Edge Computing: Security, Dependability, and Performance}
{G. Nencioni, R. G. Garroppo, and R. F. Olimid: 5G Multi-access Edge Computing: Security, Dependability, and Performance}

\corresp{Corresponding author: Gianfranco Nencioni (email: gianfranco.nencioni@uis.no).}

\begin{abstract}
The Fifth Generation (5G) of mobile networks offers new and advanced services with stricter requirements. Multi-access Edge Computing (MEC) is a key technology that enables these new services by deploying multiple devices with computing and storage capabilities at the edge of the network, close to end-users. MEC enhances network efficiency by reducing latency, enabling real-time awareness of the local environment, allowing cloud offloading, and reducing traffic congestion.
New mission-critical applications require high security and dependability, which are rarely addressed alongside performance. This survey paper fills this gap by presenting 5G MEC's three aspects: security, dependability, and performance. The paper provides an overview of MEC, introduces taxonomy, state-of-the-art, and challenges related to each aspect. Finally, the paper presents the challenges of jointly addressing these three aspects.




\end{abstract}


\titlepgskip=-15pt

\maketitle

\section{Introduction} \label{sec:introduction}

The Fifth Generation (5G) of mobile networks is currently under deployment.
The main innovation is the provision of wireless connectivity for various usage scenarios \cite{itu-r}:
\emph{enhanced Mobile Broadband (eMBB)}, for services with very high data rate requirements (up to 20Gb/s);
\emph{massive Machine-Type Communication (mMTC)}, developed for connecting a huge number of Internet-of-Things (IoT) devices (up to one million devices/km$^2$);
\emph{Ultra-Reliable Low-Latency Communication (URLLC)}, for services requiring high reliability and very low latency (up to 1ms).
eMBB allows improving the services provided by the Fourth Generation (4G) of mobile networks.
mMTC enhances the services that are now provided by Low-Power Wide Area Networks, such as Long-Term Evolution MTC (LTE-M) and Narrowband IoT (NB-IoT).
URLLC enables innovative advanced services, such as mission-critical applications, industrial automation, and enhanced Vehicular to Everything (V2X) such as platooning or remote driving.
eMBB services are under deployment, and 5G smartphones have already been produced and sold by many manufacturers.
Currently, mMTC is not under deployment since many network operators have deployed LTE-M and NB-IoT in relatively recent times.
URRLC is a usage scenario that is more immature and challenging, and it is attracting a lot of attention from the research community.

One of the technologies that enable 5G to provide URLLC services is the \emph{Multi-Access Edge Computing (MEC)}. MEC consists in the deployment of storage and computing platforms at the edge of the (radio) access network. In this way, MEC is enabling the delivery of services with low latency but can also enable context awareness and task offloading. Moreover, MEC is also the enabler of the edge intelligence, which is anticipated to be one of the main innovations of the Sixth Generation (6G) of mobile networks~\cite{6g-edge-intelligence}.

MEC is the name given by the European Telecommunications Standards Institute (ETSI) which has an Industry Specification Group (ISG) \cite{etsi-mec-isg} that is standardizing MEC since 2014. Before 2017, MEC was standing for "Mobile Edge Computing", but ETSI decided to generalize the standard to other access technologies, not only 4G and 5G but also fixed-access networks and Wireless Local Area Networks (WLAN).

ETSI MEC is not the only standardization effort on edge computing. \emph{Fog computing} and \emph{cloudlet} are the two main alternatives.
The cloudlet was proposed in 2009 \cite{cloudlet-origin} and can be considered the first effort on edge computing. The cloudlet consists of a micro-cloud close to the mobile device.
Fog computing has been first proposed by Cisco in 2011. Since 2015, fog computing is promoted and standardized by the OpenFog consortium \cite{openfog}. Fog computing has been introduced as an extension of the cloud computing paradigm from the core to the edge of the network. Fog Computing consists of a three-layer architecture where clouds, fog nodes, and IoT devices interact.
There is also another technology called Mobile Cloud Computing (MCC), but it is not edge computing since it consists of offloading tasks from mobile users to the cloud.
This paper uses as a reference the ETSI MEC, but many considerations can also be generalized for the other edge computing solutions, and, in our study, works on alternative edge computing are also included.

As already mentioned, URLLC is the most innovative and challenging usage scenario for 5G and the one that requires MEC.
To support URLLC, MEC has to cope with high requirements of ultra-reliability, which means security and dependability, and low latency, which is a performance indicator.
This is one of the reasons for which security, dependability, and performance are three critical aspects of MEC.

While MEC security has been investigated to some extent in recent years and several surveys are available, there are fewer surveys available on performance and even less on dependability.
To the best knowledge of the authors, this is the first work that jointly investigates the security, dependability, and performance of 5G MEC.

This paper has the following contributions:
\begin{itemize}
    \item State of the art and challenges on the security, dependability, and performance of 5G MEC. Each aspect is addressed \emph{individually} by using a similar structure and content organization. This organization helps to better jointly investigate and compare the three aspects. 
    \begin{itemize}
    \item First, the \emph{taxonomy} of the investigated aspect is introduced. In this way, experts on the other aspects can better understand the investigated aspect. 
    \item Second, the \emph{state of the art} is presented. The state of the art is divided into standardization efforts and academic publications.
    \item Finally, the \emph{challenges} are presented and organized according to the ETSI MEC architecture.
    \end{itemize}
    \item Challenges in \emph{jointly} addressing security, dependability, and performance in 5G MEC. The joint provision of these three aspects and the related trade-offs are analyzed and discussed by including also the future perspective of 6G.
\end{itemize}

\begin{figure}[!t]
  \centering
  \includegraphics[width=\columnwidth]{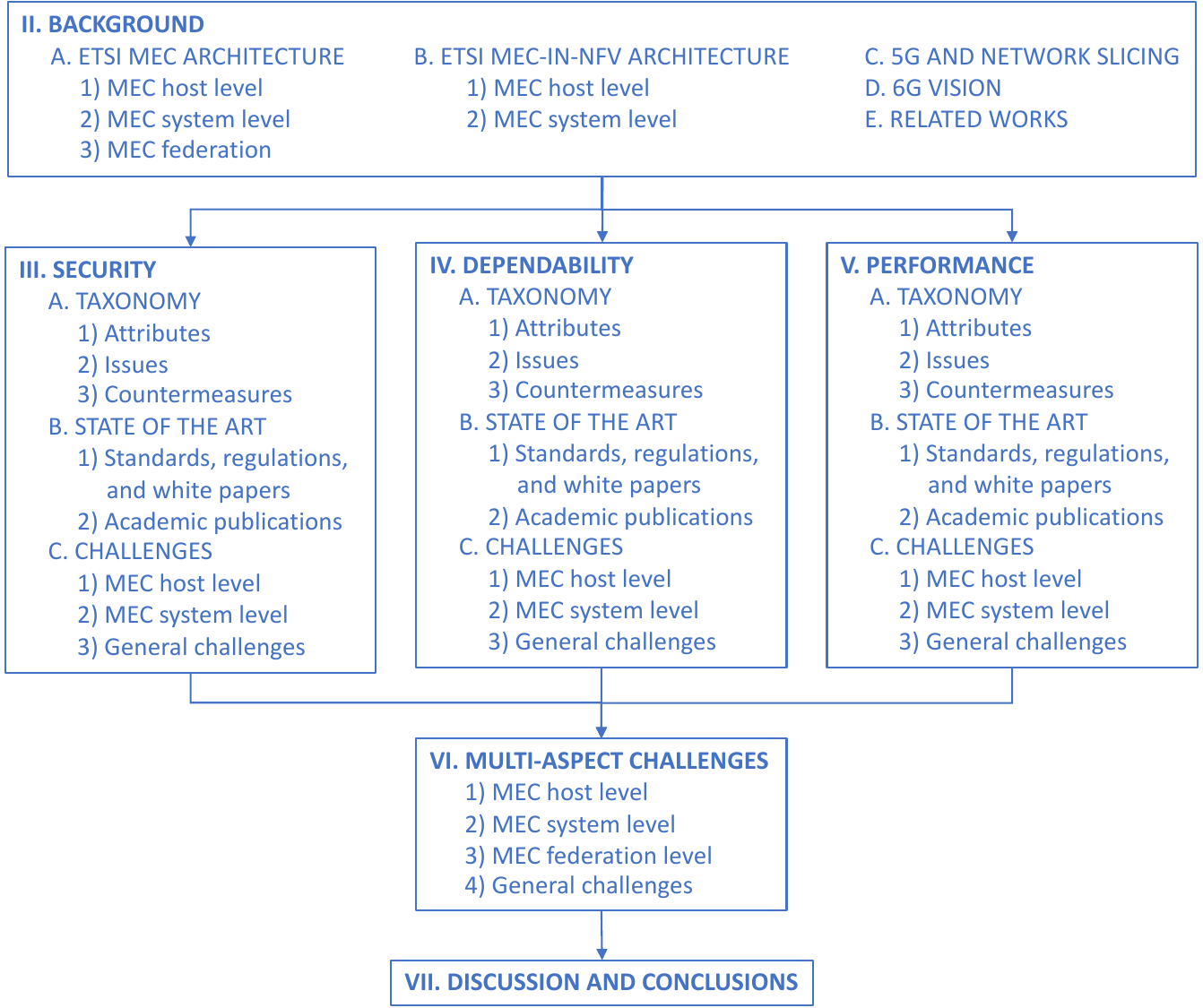}
  \caption{Paper Flow}
  \label{fig:paper-flow}
\end{figure}

The paper flow is depicted in Figure~\ref{fig:paper-flow}.
The next section introduces the necessary background concepts and definitions of 5G MEC.
Sections \ref{sec:security},  \ref{sec:dependability}, and  \ref{sec:performance} present the state of the art and the challenges of 5G MEC related to security, dependability, and performance, respectively.
Section \ref{sec:discussion} discusses the challenges and trade-offs of jointly addressing security, dependability, and performance aspects.
Finally, Section~\ref{sec:conclusions} presents the conclusions.

\section{Background} \label{sec:background}

In the ETSI specifications \cite{ETSI:MEC001}, MEC is defined as \emph{"system which provides an IT service environment and cloud-computing capabilities at the edge of the access network which contains one or more type of access technology, and in close proximity to its users"}. 

Before presenting the state of the art of MEC focusing on the three different aspects, the fundamental concepts of MEC and the related enabling technologies are presented. 

Table~\ref{tab:acronyms} lists the main acronyms that will be used in the rest of the paper. Most of the MEC acronyms are defined in \cite{ETSI:MEC001, ETSI:MEC003}. Note that what is currently defined as MEC was previously defined as Mobile Edge. For example, MEO was the acronym for Mobile Edge Orchestrator. 

\begin{table}[!t]
\caption{List of Acronyms}
\begin{center}
\begin{tabular}{p{1.5cm}p{6cm}}
\toprule
5G &Fifth Generation of mobile networks\\
BSS &Business Support System\\
CN &Core Network\\
CFS &Customer-Facing Service\\
DNS &Domain-Name System\\
EM &Element Management\\
ETSI &European Telecommunications Standards Institute\\
GR & Group Report (ETSI) \\
GS & Group Specification (ETSI) \\
LCM &Life-Cycle Management\\
MANO &Management and Orchestration\\
MEAO &MEC Application Orchestrator\\
MEC &Multi-access Edge Computing\\
MEF &MEC Federator\\
MEFB &MEC Federation Broker\\
MEFM &MEC Federation Manager\\
MEH &MEC Host\\
MEO &MEC Orchestrator\\
MEP &MEC Platform\\
MEPM &MEC Platform Manager\\
MEPM-V &MEC Platform Manager - NFV\\
NFV &Network Function Virtualization\\
NFVI &NFV Infrastructure\\
NFVO &NFV Orchestrator\\
OSS &Operation Support System\\
RAN &Radio Access Network\\
SDN &Software-Defined Networking\\
SST &Slice/Service Type\\
VIM &Virtualization/Virtualized Infrastructure Manager\\
VM &Virtual Machine\\
VNF &Virtualized Network Function\\
VNFM &VNF Manager\\
\bottomrule
\label{tab:acronyms}
\end{tabular}
\end{center}
\end{table}

\subsection{ETSI MEC Architecture}

\begin{figure}[!t]
  \centering
  \includegraphics[width=0.5\textwidth]{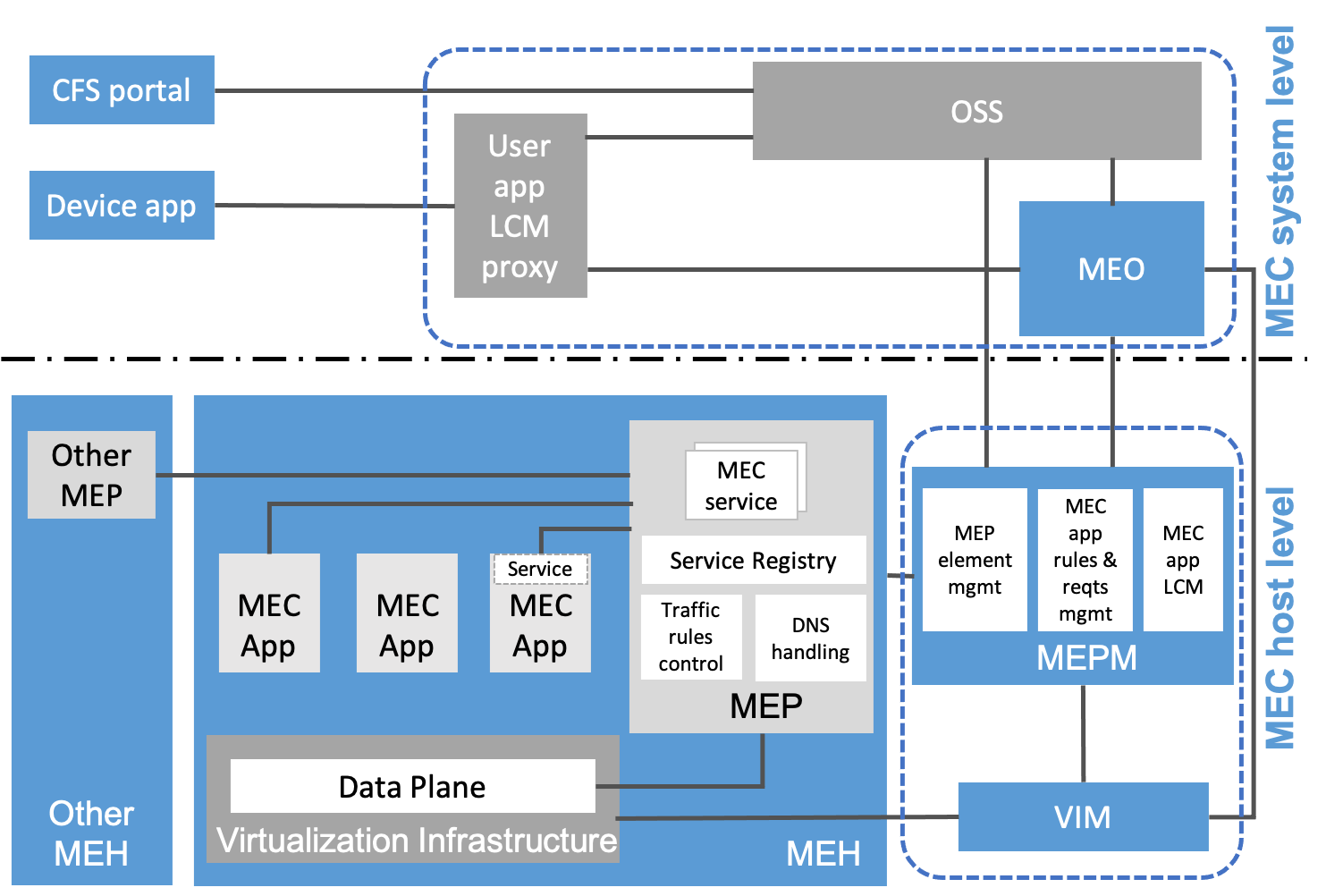}
  \caption{ETSI MEC Reference Architecture \cite{ETSI:MEC003}}
  \label{fig:MEC-arch}
\end{figure}

A MEC system is defined as a collection of \emph{MEC Hosts} (MEHs) and \emph{MEC management} necessary to run MEC applications~\cite{ETSI:MEC001}.
Figure \ref{fig:MEC-arch} illustrates the ETSI MEC general reference architecture, divided into two levels: the \emph{MEC host level} and the \emph{MEC system level} \cite{ETSI:MEC003}.

\subsubsection{MEC host level}
A MEH contains a \emph{virtualization infrastructure} that provides computation, storage, and networking resources to run \emph{MEC applications} (MEC Apps) and a \emph{MEC Platform} (MEP). MEC applications run as Virtual Machines (VMs) and can offer and consume \emph{MEC services}~\cite{ETSI:MEC003}.
The MEP is a collection of functionalities required to run the MEC applications. The MEP can also host MEC services. More information on MEC applications can be found in \cite{ETSI:5G-MEC_Software} (development) and in \cite{ETSI:MEC011} (enablement).

The \emph{MEC host-level management} is composed of  the \emph{MEC Platform Manager} (MEPM) and the \emph{Virtualization Infrastructure Manager} (VIM).
The MEPM manages MEC applications' life cycle, rules, and requirements (e.g., required resources, latency), and provides element management functions to the MEP. 
The VIM manages the allocation and monitoring of the virtual resources and transmits faults and performance reports to the MEPM.
OpenStack is commonly used to implement VIM \cite{OpenStack}.

\subsubsection{MEC system level}
The \emph{MEC system-level management} is composed of the \emph{MEC Orchestrator} (MEO), the operator's \emph{Operations Support System} (OSS), and the \emph{user application Life-Cycle Management (LCM) proxy}~\cite{ETSI:MEC003}.

The MEO is the core component that has an overview of the complete MEC system. MEO on-boards the application packages (performs integrity and authenticity checks, as well as compliance with the operator policies), selects the appropriate MEH(s) for MEC application instantiation based on the related rules and requirements, and triggers the MEC application instantiation, termination, and relocation (if needed).

The operator's OSS receives requests from the \emph{Customer-Facing Service (CFS) portal} or from the \emph{device applications}, via the \emph{user application LCM proxy}.
The CFS portal allows operators' third-party customers to select and order MEC applications or receive service level information concerning provisioned applications.
A device application is an application in a device that can interact with the MEC system. In response to a user request via a device application, an \emph{user application} is instantiated in the MEC system.
The user application LCM proxy allows the device applications to request the on-boarding, instantiation, and termination of user applications in the MEC system.
If the OSS decides to grant a request it transmits it to the MEO for further processing.

\subsubsection{MEC federation}

The MEC architecture can be extended to allow inter-MEC system communication. To this purpose, the \emph{MEC federation} is defined as \emph{"a federated model of MEC systems enabling shared usage of MEC services and applications"}~\cite{ETSI:MEC035}.

The MEC architecture in Figure~\ref{fig:MEC-arch} can the extended by adding a \emph{MEC Federator} (MEF), which may be composed of the functionalities of the \emph{MEC Federation Broker} (MEFB) and the \emph{MEC Federation Manager} (MEFM). The MEF interfaces with other MEFs enabling the information exchange. It also interfaces with at least one MEO (the one the MEF is belonging to). A MEF may serve as a single point of contact for multiple MEFs acting as a broker between MEFs~\cite{ETSI:MEC003}. 

\medskip
Table \ref{tab:MEC-elements} summarizes the functionalities of the main architectural elements, as explained in~\cite{ETSI:MEC003}.


\renewcommand{\labelitemi}{-}
\begin{table}[!t]
\caption{Key Elements in the MEC Architecture}
\begin{center}
\begin{tabular}{p{1.6cm}p{6.1cm}}
\toprule
\textbf{Element} & \textbf{Functionalities}\\
\midrule
\multicolumn{2}{l}{\textbf{MEH}}\\
MEP &
\vspace{-0.3cm}
\begin{itemize}[leftmargin=*]
	\item Offers the necessary environment for the MEC applications.
	\item Hosts MEC services.
	\item Instructs the data plane based on received traffic rules.  
	\item Configures the DNS proxy/server. 
	\item Provides access to persistent storage.
	\item Provides timing information.
\end{itemize} 
\\
Virtualization Infrastructure &
\vspace{-0.3cm}
\begin{itemize}[leftmargin=*]
	\item Provides compute, storage, and network resources for the MEC applications.
	\item Data plane routes traffic among MEC applications, services, DNS proxy/server, and various access networks.
\end{itemize} 
\\
MEC App &
\vspace{-0.3cm}
\begin{itemize}[leftmargin=*]
	\item Discovers, advertises, consumes, and offers MEC services.
	\item Performs support procedures related to the life cycle of the application.
\end{itemize} 
\\
\midrule
\multicolumn{2}{l}{\textbf{MEC Host-level Management}}\\
MEPM &
\vspace{-0.3cm}
\begin{itemize}[leftmargin=*]
	\item Manages the life cycle, rules, and requirements of the applications.
	\item Prepares the virtualization infrastructure to run a software image.
	\item Provides element management functions to the MEP.
\end{itemize} 
\\
VIM & 
\vspace{-0.3cm}
\begin{itemize}[leftmargin=*]
	\item Manages the allocation and monitoring of the virtual resources.
	\item Prepares the virtualization infrastructure to run a software image.
	\item Transmits faults and performance reports to the MEPM.
\end{itemize} 
\\
\midrule
\multicolumn{2}{l}{\textbf{MEC System-level Management}}\\
MEO &
\vspace{-0.3cm}
\begin{itemize}[leftmargin=*]
	\item Maintains an overview of the complete MEC system.
	\item On-boards  the  application  packages.
	\item Selects  the appropriate  MEH(s)  for  MEC  application  instantiation.
	\item Triggers MEC application instantiation, termination, and relocation.
\end{itemize} 
\\
OSS &
\vspace{-0.3cm}
\begin{itemize}[leftmargin=*]
	\item Decides on granting requests for instantiation and terminating of applications.
\end{itemize} 
\\
User app LCM proxy &
\vspace{-0.3cm}
\begin{itemize}[leftmargin=*]
	\item Permits the device applications to request the on-boarding, instantiation, and termination of user applications.
	\item Informs the device applications about the status of the user applications.
\end{itemize} 
\\
\midrule
\multicolumn{2}{l}{\textbf{MEC Federation}}\\
MEF &
\vspace{-0.3cm}
\begin{itemize}[leftmargin=*]
	\item Registers the MEC system information sent by a MEO.
        \item Discovers MEC systems.
        \item May act as a one to many intermediary between MEFs (broker capability).
        \item Exchanges the  MEC system information.
        \item Manages the MEC application lifecycle across different MEC systems.
        \item Monitors the MEC application across different MEC systems.
\end{itemize} 
\\
\bottomrule
\end{tabular}
\label{tab:MEC-elements}
\end{center}
\end{table}



\subsection{ETSI MEC-in-NFV Architecture}

MEC uses a virtualization platform to run the MEC application in the MEH.
NFV is a virtualization platform where the network functions are decoupled from the hardware by using virtual hardware abstraction.
It is, therefore, beneficial to reuse the infrastructure and the infrastructure management of NFV~\cite{ETSI:MEC002}.

\begin{figure}[!b]
  \centering
  \includegraphics[width=0.49\textwidth]{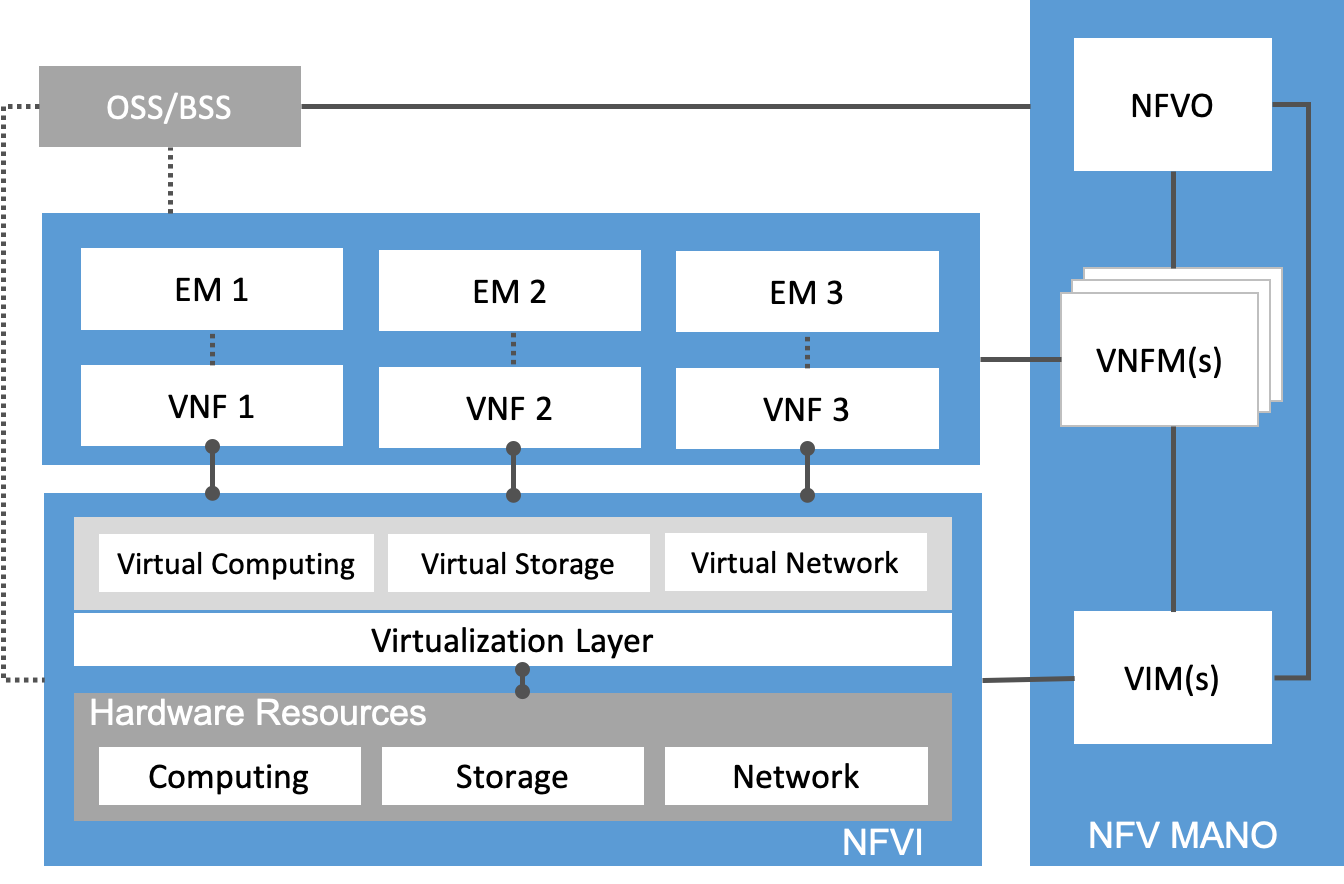}
  \caption{ETSI NFV Reference Architectural Framework~\cite{ETSI:NFV002}}
  \label{fig:NFV-arch}
\end{figure}

Figure \ref{fig:NFV-arch} illustrates the ETSI NFV reference architecture~\cite{ETSI:NFV002}. 
To provide network services, the NFV  is composed of \emph{Virtualized Network Functions} (VNFs), an underlying \emph{NFV Infrastructure (NFVI)}, and a \emph{NFV Management and Orchestration} (MANO).

A VNF is a software implementation of a network function, which is decoupled from the hardware resources it uses. The VNFs rely on the NFVI, where the needed virtualized resources (computing, storage, and network) are obtained from the hardware resources through the virtualization layer. A VNF can be deployed, by case, over one or several VMs \cite{ETSI:NFV002}, where VMs are partitioned on the resources of a hardware host by software programs called \emph{hypervisors}.

The NFV MANO is composed of three main components: the \emph{NFV Orchestrator} (NFVO), the \emph{VNF Manager} (VNFM), and the \emph{Virtualized Infrastructure Manager} (VIM). The NFVO is the highest hierarchical level of the NFV MANO and is responsible for the creation and LCM of network services.
The VNFMs are instead responsible for the LCM of the VNFs, which are locally managed by the \emph{Element Management} (EM) systems. A VNFM can serve one or multiple VNFs.
Finally, the VIM controls and manages the NFVI resources (e.g., it is in charge of the inventory of software, computing, storage, and network resources, increasing resources for VMs, improving energy efficiency, collection of infrastructure fault operations, capacity planning, and optimization) \cite{ETSI:NFV002}.

An NFV-based network service is composed of an ordered set of VNFs between two end-points, where the traffic is steered through them.
This composition to provide an NFV-based network service is similar to the one specified by the Service Function Chaining (SFC)~\cite{SFC}. 

\begin{figure}[!t]
  \centering
  \includegraphics[width=0.49\textwidth]{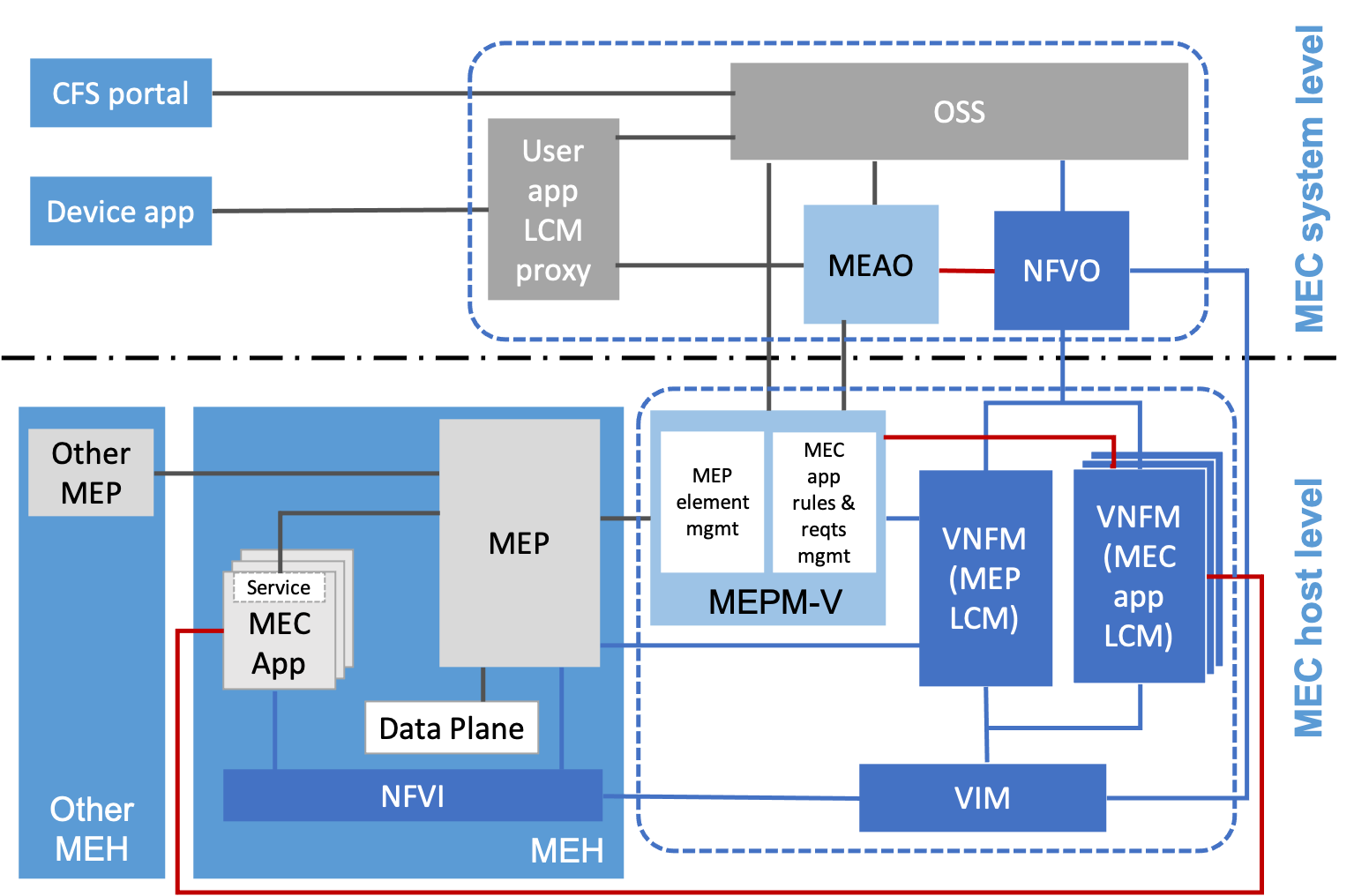}
  \caption{MEC-in-NFV Reference Architecture \cite{ETSI:MEC017}}
  \label{fig:NFV-in-MEC}
\end{figure}

\medskip
Figure \ref{fig:NFV-in-MEC} illustrates the MEC architecture deployed by using NFV \cite{ETSI:MEC017}.
For continuity and clarity, the architectural changes are explained separately for each of the two layers.

\subsubsection{MEC host level}
On the host side, both the MEC applications and the MEP are deployed as VNFs, while the virtual infrastructure is deployed as NFVI. The virtualization infrastructure, as the NFVI, can be implemented with various virtualization technologies, such as hypervisor-based or container-based solutions, but also mixing or/and nesting virtualization technologies \cite{ETSI:MEC027}. On the host management side, the MEPM is substituted by the \emph{MEC Platform Manager - NFV} (MEPM-V) and a VNFM. The MEPM-V has the same responsibilities as the MEPM. The VNFM is delegated the management of the VNF life cycle~\cite{ETSI:MEC017}.
The VIM maintains similar functionalities.

\subsubsection{MEC system level}
In the MEC-in-NFV architecture, the MEO is replaced by the \emph{MEC Application Orchestrator} (MEAO) and an NFVO. The MEAO has the same responsibilities as the MEO. However, the MEAO delegates an NFVO to perform the resource orchestration and the orchestration of the MEC applications (as VNFs). The other elements remain unaffected~\cite{ETSI:MEC017}. 

More details about the MEC deployment in an NFV environment can be found in \cite{ETSI:MEC017}.


\subsection{5G and Network Slicing} \label{sec:5g}

MEC is one of the key technologies of 5G, together with NFV and Software-Defined Networking (SDN) \cite{5g-orch}. In particular, MEC provides 5G of contextual information and real-time awareness of the local environment. In \cite{ETSI:5G-MEC,ETSI:MEC031}, ETSI explains how to deploy and integrate MEC in the 5G system.

\begin{figure}[!t]
  \centering
  \includegraphics[width=0.48\textwidth]{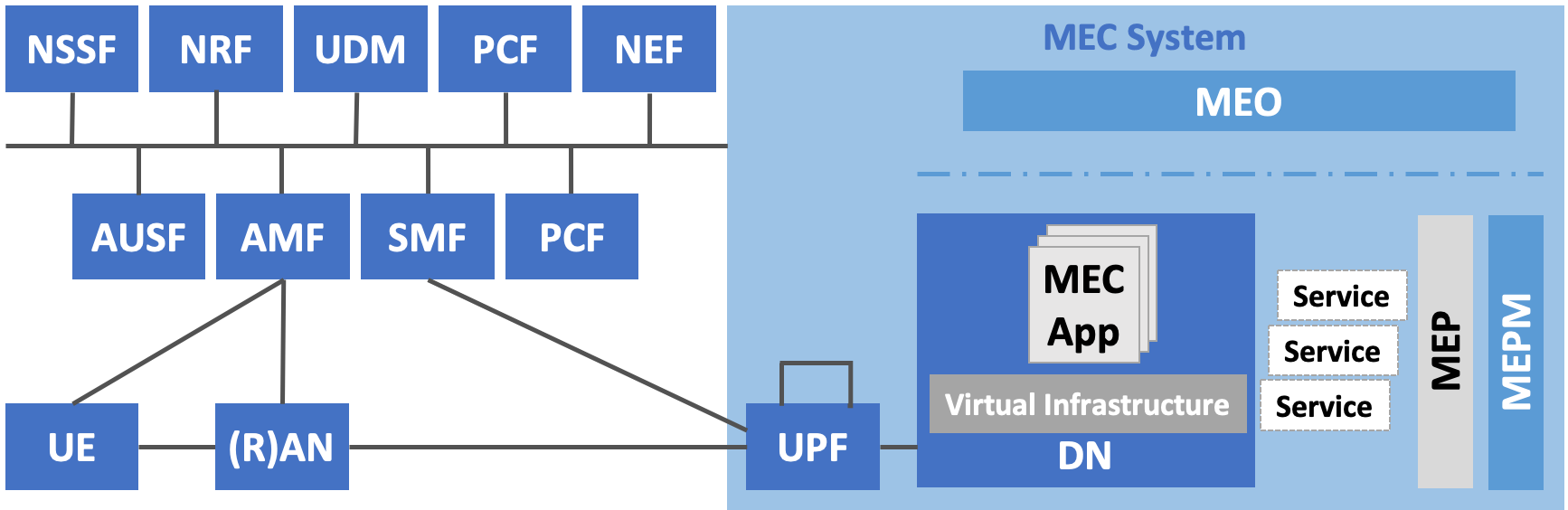}
  \caption{Integrated MEC Deployment in 5G Network~\cite{ETSI:5G-MEC}}
  \label{fig:5G-MEC}
\end{figure}

Figure \ref{fig:5G-MEC} depicts the MEC architecture deployed in a 5G network~\cite{ETSI:5G-MEC}.
In the left part of the figure (in dark blue), there is the 5G Service-Based Architecture (SBA), as described by 3GPP in \cite{3GPP:23.501}, which is composed of the following network function of the 5G Core Network (CN): Network Slice Selection Function (NSSF), Network Resource Function (NRF), Unified Data Management (UDM), Policy Control Function (PCF), Network Exposure Function (NEF), Application Function (AF), Authentication Server Function (AUSF), Access and Mobility Management Function (AMF), and Session Management Function (SMF). Moreover, the 5G SBA is also composed of User Equipment (UE), Radio Access Network (RAN), User Plane Function (UPF), and Data Network (DN). Each function can consume or produce services.

In 5G SBA, MEC is seen as a set of AFs. The MEO and the MEP are acting as AFs. The MEH is instead often deployed as DN.

The NRF contains the registered network functions and their provided services. The services provided by the MEC applications are instead in the service register in the MEP.

Some of the services are available only through the NEF, which acts as a centralized point for service exposure.
The AUSF manages the authentication.

The PCF handles the policies and the rules. The MEP uses the PCF services to impact the traffic steering rules.

The UDM is instead responsible for services related to users and subscriptions. It generates authentication credentials, handles information related to user identification, manages access authorization, and registers users on AMF and SMF.

Finally, connected to the network slicing (which will be presented more in detail later on), the NSSF manages the selection of network slice instances and the allocation of the necessary AMFs.

A key role in the integration of 5G and MEC is performed by the UPF, which can be seen as a distributed and configurable data plane from the MEC.

\begin{figure}[t!]
  \centering
  \includegraphics[width=0.48\textwidth]{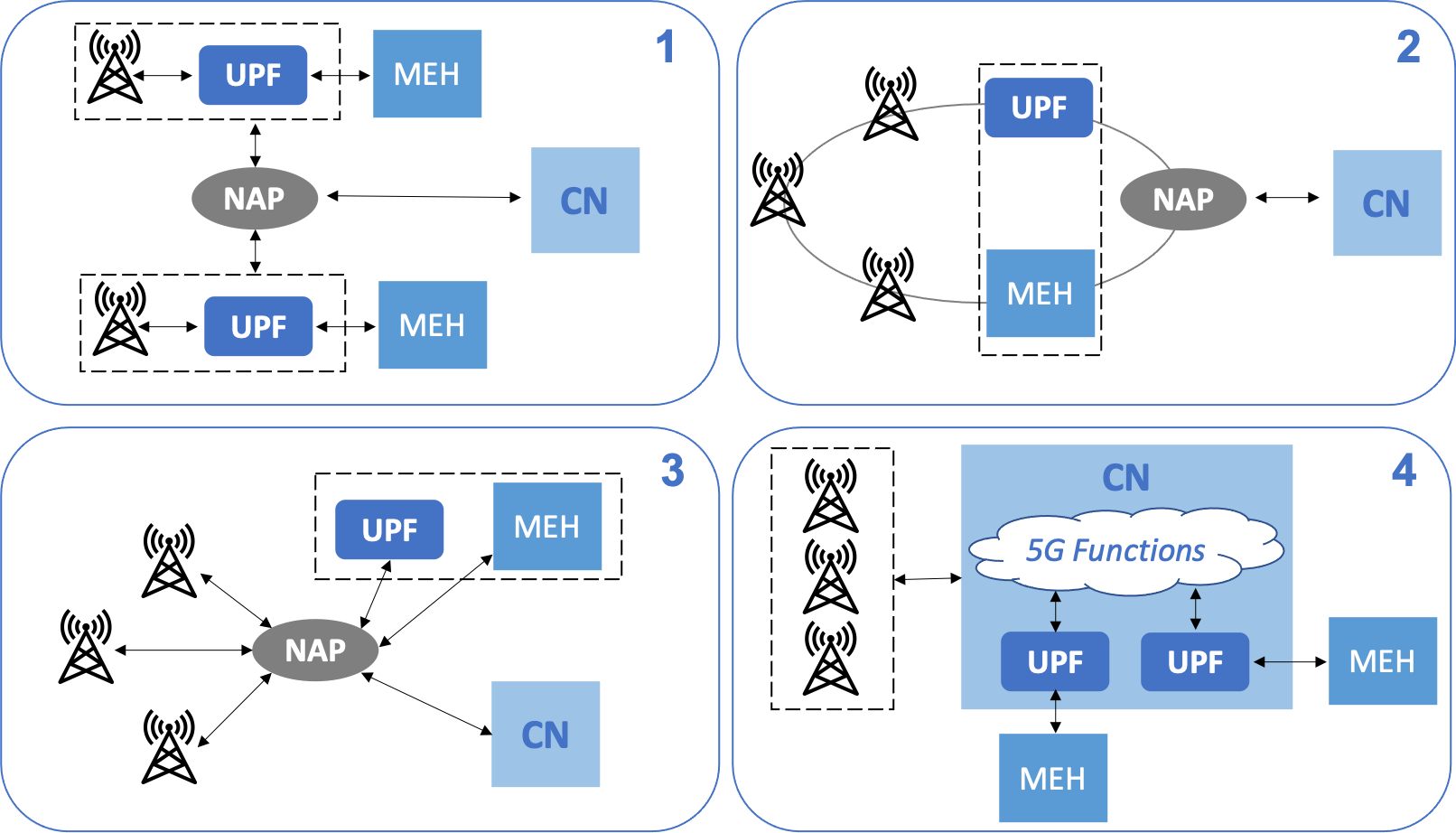}
  \caption{5G-MEC Deployment Scenarios \cite{ETSI:5G-MEC}}
  \label{fig:5G-MEC-depl}
\end{figure}

In \cite{ETSI:5G-MEC}, four scenarios for the deployment of MEHs in a 5G network are presented (see also Figure~\ref{fig:5G-MEC-depl}):
\begin{enumerate}
    \item MEH and the local UPF collocated with the Base Station;
    \item MEH collocated with a transmission node, possibly with a local UPF;
    \item MEH and the local UPF collocated with a Network Aggregation Point (NAP);
    \item MEH collocated with the CN functions (i.e. in the same data centre).
\end{enumerate}

ETSI has also investigated the deployment of MEC in other access technologies, such as 4G \cite{ETSI:5G-MEC_Evol} and WLAN \cite{ETSI:MEC028}. Moreover, ETSI has investigated the deployment in cloud RAN \cite{ETSI:5G-MEC_Cloud}.

Initially, the Radiocommunication sector of the International Telecommunication Union (ITU-R) defined three usage scenarios for 5G and beyond \cite{itu-r}: eMMB, mMTC, and URLLC. The use scenarios have been already described in the introduction
In 3rd Generation Partnership Project (3GPP), the use scenarios are called Slice/Service Types (SSTs). As ITU-R, 3GPP SSTs include eMBB and URLLC, but they do not include mMTC and there are instead Massive Internet of Things (MIoT), Vehicle-to-everything (V2X) service, and High-performance Machine-Type Communication (HMTC)~\cite{3GPP:23.501}.

The ability of 5G to provide services in very different use scenarios is enabled by network slicing. Network slicing allows the flexible and efficient creation of specialized end-to-end logical networks (network slices) on top of shared network infrastructure.
In order to properly operate, the network slices have to be \textit{isolated}. The network slice isolation needs to be valid with respect to security, dependability, and performance. More details on network slice isolation can be found in \cite{netslice-concept-2020}.
In this paper, network slicing is not further described. A more detailed description (together with a security overview) can be found in \cite{5g-netslice-sec}.
In \cite{ETSI:MEC024}, ETSI has defined how MEC supports network slicing.


\begin{table*}[!t]
\caption{ETSI Specifications, Reports, and White Papers}
\begin{center}
\begin{tabular}{p{3cm}p{8cm}p{1cm}p{1.5cm}p{1.5cm}}
\toprule
\textbf{No. \& Ref.} & \textbf{Name} & \textbf{Security} & \textbf{Dependability} & \textbf{Performance}\\
\midrule
GS MEC 002~\cite{ETSI:MEC002} & Phase 2: Use Cases and Requirements & \cmark &\cmark & \cmark  \\
GS MEC 003~\cite{ETSI:MEC003} & Framework and Reference Architecture &  &\cmark &\\
GS MEC-IEG 006~\cite{ETSI:MEC006} & Market Acceleration; MEC Metrics Best Practice and Guidelines &  &\cmark & \cmark \\
GS MEC 009~\cite{ETSI:MEC009} & General principles, patterns and common aspects of MEC Service APIs & \cmark &\\
GS MEC 010-1~\cite{ETSI:MEC010_1} & Mobile Edge Management; Part 1: System, host and platform management & & \cmark  \\
GS MEC 010-2~\cite{ETSI:MEC010_2} & MEC Management; Part 2: Application lifecycle, rules and requirements management & \cmark & \cmark &\\
GS~MEC 011~\cite{ETSI:MEC011} & Edge Platform Application Enablement & \cmark & \cmark &\\
GS MEC 012~\cite{ETSI:MEC012} & Radio Network Information API & \cmark & &  \\
GS MEC 013~\cite{ETSI:MEC013} & Location API & \cmark & &\\
GS MEC 014~\cite{ETSI:MEC014} & UE Identity API & \cmark & &\\
GS MEC 015~\cite{ETSI:MEC015} & Traffic Management APIs & \cmark & \cmark & \cmark \\
GS MEC 016~\cite{ETSI:MEC016} & Device application interface & \cmark & &\\
GS MEC 021~\cite{ETSI:MEC021} & Application Mobility Service API & & \cmark & \cmark \\
GS MEC 024~\cite{ETSI:MEC024} & Support for network slicing & \cmark & \cmark & \cmark \\
GS~MEC 026~\cite{ETSI:MEC026} & Support for regulatory requirements & \cmark & &\\
GS MEC 029~\cite{ETSI:MEC029} & Fixed Access Information API & \cmark & &  \\
GS MEC 030~\cite{ETSI:MEC030} & V2X Information Service API & \cmark & \cmark & \\
GS MEC 031~\cite{ETSI:MEC031} & MEC 5G Integration & \cmark & \cmark & \cmark  \\
GS MEC-DEC 032-1~\cite{ETSI:MEC032_1} & API Conformance Test Specification; Part 1: Test Requirements and Implementation Conformance Statement (ICS) &  & \cmark &\\
GS MEC-DEC 032-2~\cite{ETSI:MEC032_2} & API Conformance Test Specification; Part 2: Test Purposes (TP) & & \cmark & \\
GS MEC 033~\cite{ETSI:MEC033} & IoT API & \cmark & & \\
GS MEC 037~\cite{ETSI:MEC037} & Application Package Format and Descriptor Specification & \cmark & & \cmark \\
GS MEC 040~\cite{ETSI:MEC040} & Federation enablement APIs &\cmark & & \\
\midrule
GR MEC 017~\cite{ETSI:MEC017} & Deployment of Mobile Edge Computing in an NFV environment & & \cmark & \cmark \\
GR MEC 018~\cite{ETSI:MEC018} & End to End Mobility Aspects & \cmark & \cmark & \cmark \\
GR MEC 022~\cite{ETSI:MEC022} & Study on MEC Support for V2X Use Cases & & \cmark & \cmark \\
GR MEC-DEC 025~\cite{ETSI:MEC025} & MEC Testing Framework & & \cmark &\\
GR MEC 027~\cite{ETSI:MEC027} & Study on MEC support for alternative virtualization technologies & \cmark & & \cmark \\
GR MEC 035~\cite{ETSI:MEC035} & Study on Inter-MEC systems and MEC-Cloud systems coordination & \cmark & \cmark & \cmark \\
GR MEC 038~\cite{ETSI:MEC038} & MEC in Park Enterprises deployment scenario &  &  & \cmark \\
GR MEC 042~\cite{ETSI:MEC042} & Guidelines on Interoperability testing &  &  & \cmark \\
\midrule
White paper No. 20~\cite{ETSI:5G-MEC_Software} & Developing Software for Multi-Access Edge Computing  & \cmark & \cmark & \cmark \\ 
White paper No. 23~\cite{ETSI:5G-MEC_Cloud} & Cloud RAN and MEC: A Perfect Pairing & \cmark & & \cmark \\
White paper No. 24~\cite{ETSI:5G-MEC_Evol} & MEC Deployments in 4G and Evolution Towards 5G & \cmark & \cmark & \cmark \\
White paper No. 28~\cite{ETSI:5G-MEC} & MEC in 5G networks & \cmark & \cmark & \cmark \\ 
White paper No. 30~\cite{ETSI:5G-MEC_Enterprise} & MEC in an Enterprise Setting: A Solution Outline & & \cmark & \cmark \\
White paper No. 32~\cite{ETSI:5G-MEC_NetTransf} & Network Transformation; (Orchestration, Network and Service Management Framework) & & \cmark &\\ 
White paper No. 34~\cite{ETSI:5G-MEC_AI} & Artificial Intelligence and future directions for ETSI & & \cmark & \cmark \\
White paper No. 36~\cite{ETSI:5G-MEC_harmonizing} & Harmonizing standards for edge computing - A synergized architecture leveraging ETSI ISG MEC and 3GPP specifications & & \cmark &\\
White paper No. 39~\cite{ETSI:5G-MEC_dns} & Enhanced DNS Support towards Distributed MEC Environment & & \cmark & \cmark \\
White paper No. 46~\cite{ETSI:MEC-security} & MEC security: Status of standards support and future evolutions & \cmark &  & \\
White paper No. 49~\cite{ETSI:MEC-federation} & MEC federation: deployment considerations & \cmark &  & \\
\bottomrule
\end{tabular}
\label{tab:etsi_standards}
\end{center}
\end{table*}

\subsection{6G Vision}
In the next generation of mobile networks, MEC will still be one of the most important technologies.
The key features of 6G will be connected intelligence, programmability, deterministic end-to-end, integrated sensing and communication, sustainability,
trustworthiness, scalability, and affordability \cite{6g-vision}.
As for the previous generations, 6G will introduce improvements of the performance metrics of around one order of magnitude with respect to 5G.

To obtain these enhancements, Artificial Intelligence (AI) and Machine Learning (ML) will be used to enable the system network architecture and control, the edge and ubiquitous computing \cite{Letaief2022}, the radio technology and signal processing \cite{Hoydis2021}, the optical networks, the network and service security, the non-terrestrial network communication, and the special-purpose networks/sub-networks \cite{6g-vision}.

In conclusion, the main innovation of 6G can be summarized as \emph{AI everywhere} to enable the \emph{easy integration of everything}.
In this context, MEC has an important role of bringing AI, enabling distributed (micro)service-based architectures, and helping 6G to reach the "zero delay".


\subsection{Related Works}

In the previous subsections, we have extensively referred to documents produced by the ETSI MEC group. Table~\ref{tab:etsi_standards} shows the ETSI specifications, reports, and white papers that have content in one of the three perspectives that are investigated.


Many surveys and reviews have been published in the recent years.
Many works focus on MEC (first mobile and then multi-access), referring (at least partially) to the ETSI architecture \cite{yu_mobile_2016, hibat_allah_mec_2017, taleb_multi-access_2017, wang_survey_2017, mao_survey_2017, mach_mobile_2017, shahzadi2017multi, Shirazi_2017, bilal2018potentials, peng_survey_2018, roman_mobile_2018, porambage_survey_2018, tanaka_multi-access_2018, abbas_mobile_2018, khan2019edge, yousefpour2019all, RJL2019, li_survey_2019, pham_survey_2020, habibi_fog_2020, filali-MEC-surv-2020, vhora_comprehensive_2020, ranaweera-MEC-security-2021, shah-NetSlice-MEC-2021, mec-implementation, Sarah2023}.
Some works focus exclusively or jointly on fog computing \cite{reliability-madsen-2013, hu_survey_2017, mahmud_fog_2018, wang_survey_2017, Shirazi_2017, bilal2018potentials, roman_mobile_2018, porambage_survey_2018, abbas_mobile_2018, khan2019edge, yousefpour2019all, roadmap-bakshi-2019, habibi_fog_2020, vhora_comprehensive_2020}.
Several works also consider cloudlets \cite{wang_survey_2017, shahzadi2017multi, bilal2018potentials, porambage_survey_2018, abbas_mobile_2018, khan2019edge, yousefpour2019all, habibi_fog_2020, vhora_comprehensive_2020,edge-cps}.
Other works do not focus on any particular architecture but consider generic edge computing \cite{garcia2015edge, huang_exploring_2017, yu_survey_2018, du2018big, ni2019toward, dependability-bagchi-2019, elbamby-wireless-2019, yahuza2020systematic, liu-vehicular-edge-2020, systematic, edge-ml}.

Several works are general surveys \cite{yu_mobile_2016, hu_survey_2017, mahmud_fog_2018, garcia2015edge, shahzadi2017multi, bilal2018potentials, abbas_mobile_2018, khan2019edge, yousefpour2019all, pham_survey_2020, habibi_fog_2020, systematic}.
Some works are focusing on specific perspectives:
security \cite{roman_mobile_2018, du2018big, RJL2019, yahuza2020systematic, ranaweera-MEC-security-2021},
security and resilience \cite{Shirazi_2017},
security and efficiency \cite{ni2019toward},
trustworthiness \cite{edge-cps},
reliability and latency \cite{elbamby-wireless-2019},
dependability \cite{reliability-madsen-2013, dependability-bagchi-2019, roadmap-bakshi-2019}.
Other works focus on specific environments:
vehicular networks \cite{huang_exploring_2017, liu-vehicular-edge-2020},
IoT \cite{yu_survey_2018, porambage_survey_2018, du2018big, ni2019toward},
industrial Internet \cite{li_survey_2019}.
Finally, other works focus on specific tasks or parts:
location trade-off~\cite{hibat_allah_mec_2017},
orchestration \cite{taleb_multi-access_2017},
capabilities on computing, caching, and communication \cite{wang_survey_2017},
communication \cite{mao_survey_2017},
computation offload \cite{mach_mobile_2017},
service adoption and provision \cite{peng_survey_2018},
infrastructure~\cite{tanaka_multi-access_2018},
optimization~\cite{filali-MEC-surv-2020},
ML~\cite{edge-ml}
tools and applications~\cite{vhora_comprehensive_2020},
integration with network slicing \cite{shah-NetSlice-MEC-2021}, 
resource allocation \cite{Sarah2023},
and
implementations~\cite{mec-implementation}.


This work focuses on MEC considering as reference the ETSI architecture, but it also considers research on other architectures. It focuses on three perspectives (security, dependability, and performance) individually and jointly. This work is up to date and does not focus on any particular environment or task.
All the content related to security, dependability, and performance of the above works will be commented in the following sections.
The comparative individual presentation of security, dependability, and performance in 5G MEC and the discussion of jointly addressing these three aspects in 5G MEC are new and will help the researchers to better face the challenges of future 5G MEC systems and beyond.

Several research projects focus directly or indirectly on MEC, a good summary can be found in \cite{porambage_survey_2018,khan2019survey}. This work is part of the 5G-MODaNeI\footnote{https://5g-modanei.ux.uis.no} project, which focuses on dependability and security in 5G MEC.

\section{Security} \label{sec:security}

The research community is active on the security of 5G MEC. Many works highlight the challenges and try to improve the security in 5G MEC. After a brief introduction of the security taxonomy for the readers that are not experts on the topic, we summarize the current research activity, focusing on security-oriented surveys of MEC, and discuss the security challenges.

\subsection{Taxonomy}

\begin{figure}[!t]
  \centering
  \includegraphics[width=0.48\textwidth]{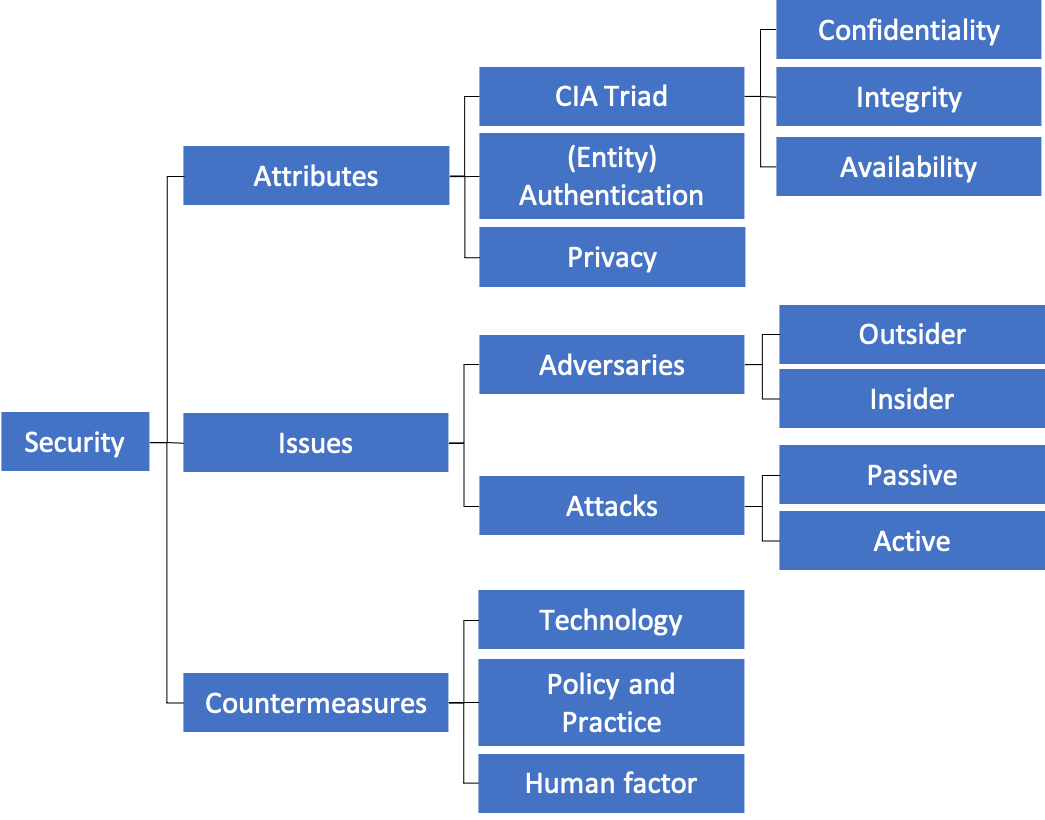}
  \caption{Security Taxonomy}
  \label{fig:sec_overview}
\end{figure}

Security targets some \emph{objectives}, for homogeneity further called \emph{security attributes} (also known as \emph{security requirements} or \emph{security properties}) that are guarded against \emph{adversaries} (or \emph{attackers}). Adversaries are malicious parties that intentionally mount \emph{attacks} with the aim to break one or more security attributes and thus gain illegitimate advantages. Attacks are \emph{security issues} that are possible because of \emph{vulnerabilities} that reside in the system and can be mitigated or defended against by enforcing a variety of \emph{countermeasures}. Figure~\ref{fig:sec_overview} sketches the taxonomy related to security. For more details, the National Institute of Standards and Technology (NIST) glossary containing terms and definitions related to cybersecurity is available at~\cite{NIST:Sec_glossary}. Similarly, for a more in-depth description of related notions, the reader may refer to \cite{crypto_encyclopedia}.

\begin{table}[!t]
\caption{Security Attributes \cite{crypto_encyclopedia,handbook_appl_crypto,NIST:Sec_glossary}
}
\begin{center}
\begin{tabular}{p{2.4cm}p{5.4cm}}
\toprule
\textbf{Attribute} & \textbf{Description}\\
\midrule
Confidentiality (C) & Keep the information secret, except from the authorized parties. \\
Integrity (I) & Ensure that the information has not been altered in an undetected manner by unauthorized  parties. \\
Availability (A) & Assure that the legitimate parties can access a service (or a resource) when they need to.\\
Authentication (Au) & Attest the identity of a party (\emph{entity authentication}) or the source of information  (\emph{data origin authentication}). \\
\midrule
Privacy (P) & 
Control how, when, and to what extent personal information is communicated to other parties. Incorporates different aspects, such as identity (e.g., name, pseudonym), related personal data (that might also become an identity in some context, e.g., phone number), location, user-created data, etc. \\
\bottomrule
\end{tabular}
\label{tab:sec_prop}
\end{center}
\end{table}

\subsubsection{Attributes}

Table \ref{tab:sec_prop} lists the commonly accepted security attributes, as defined in \cite{crypto_encyclopedia,handbook_appl_crypto,NIST:Sec_glossary}. \emph{Confidentiality}, \emph{Integrity}, and \emph{Availability}, known as the \emph{CIA Triad}, are fundamental requirements. Confidentiality is related to privacy and, in this context, guarantees the \emph{privacy of personal data}. Privacy includes other aspects too, such as the \emph{privacy of identity} (in relation to \emph{anonymity} and \emph{unlinkability}) or the \emph{privacy of location}.
Data integrity can be perceived as \emph{data authentication} because it guarantees that the data has not been altered in any way, so it is authentic. For the purpose of this paper, both integrity (or authentication of data) and authentication of entities are valuable requirements. Moreover, \emph{mutual authentication} is a strong form of entity authentication that requests that all the communicating parties authenticate to each other.
Other traditional and generally accepted requirements exist, such as \emph{non-repudiation}, which prevents the denial of previous actions. However, we leave this outside of the very succinct taxonomy presented, as it is not of particular interest to our work. For more discussion on attributes, refer to \cite{crypto_encyclopedia,handbook_appl_crypto}.
These requirements are enforced by security functions, such as e.g., \emph{access control} that protects access to resources against unauthorized parties or \emph{authorization} that officially grants to a party the right to be or to do something.
Based on previous work \cite{liyanage2015software,liyanage2017secure}, Khan et al. accept \emph{visibility} and \emph{centralized policy} as two additional security parameters \cite{khan2019survey}. We omit them here for several reasons, one being that they are not directly related to the security and privacy of the end beneficiaries, but they can be seen more as functions that help to achieve these.


\begin{table*}[!t]
\caption{Main Types of Adversaries and Attacks}
\begin{center}
\begin{tabular}{p{1.5cm}p{1.6cm}p{1cm}p{6.5cm}p{4.5cm}}
\toprule
\textbf{} & \textbf{Type} & \textbf{Main Target} & \textbf{Description} & \textbf{Examples}\\
\midrule
\textbf{Adversary} & Outsider & (all) & The adversary is an outsider, unrelated to the target. & An adversary outside the organization, a malicious competitor \\
& Insider & (all) & The adversary is an insider, somehow related to the target (e.g., the adversary has some insight knowledge, and can access some services/resources). & A malicious employee, a compromised MEC app \\
\midrule
\textbf{Attack} & Passive & (C, P) & The adversary is passive, he/she can listen but not actively interfere with the system. & Eavesdropping, data capture, (offline) (crypt)analysis on collected data. \\
& Active & (all) & The adversary actively interferes by modifying, deleting, injecting messages (or in general, data in any form: storage, computation, or transmission), or altering in any way the input/output or functioning of the system. & \\
& - Deteriorate & (A) & The active adversary damages the functionality of the victim in a black-box fashion, either partially or fully. & Service deprecation, Denial-of-Service (DoS), partial or total physical damage \\
& - Corrupt & (all) & The active adversary compromises the victim, controlling it either partially or fully. & Malware, credential theft, active monitoring \\
& - Impersonate & (all) & The adversary impersonates an entity to gain an advantage. & Man-in-the-Middle (MitM), replay attacks, rogue/fake entities \\
\bottomrule
\end{tabular}
\label{tab:sec_attacks}
\end{center}
\end{table*}

\subsubsection{Issues}

In essence, security issues are caused by \emph{adversaries} that mounts \emph{attacks}.
The attack surface is wide, and many classifications of adversaries and attacks exist in the literature, based on different criteria. Table \ref{tab:sec_attacks} defines the most common types of adversaries and attacks that will be referred to in the paper.

Concerning the position of the adversaries with respect to the target, they can be classified into \emph{outsiders} and \emph{insiders}. An internal adversary has a clear advantage over an external adversary (e.g., has physical access to the network or resources, is authorized to access some data, resources, or services). This makes internal adversaries more powerful than outsiders and insider attacks are usually easier to mount than outsider attacks. 

Further, attacks can be classified into \emph{passive} and \emph{active}~\cite{handbook_appl_crypto,crypto_encyclopedia}. Passive attacks are simpler to mount, cheaper, and more efficient in resource consumption but less powerful than active attacks. They are more difficult to detect because the adversary does not actively interfere and thus changes nothing. 
Passive attacks include \emph{eavesdropping}  and \emph{traffic analysis}. They directly threaten confidentiality but can also precede active attacks by gathering the necessary information to mount these. 
On the other hand, active attacks can directly target any security property by modifying, deleting, and injecting messages (or in general, data in any form: storage, computation, or transmission) or altering in any way the functionality of the target. 
Besides confidentiality, active attacks can also directly damage integrity and authenticity (e.g., \emph{replay attacks}, \emph{Man-in-the Middle Attacks}) or availability (e.g., \emph{Denial of Service - DoS}, \emph{Distributed DoS - DDoS}). 

With respect to the implications for the target, active attacks can be roughly classified into three types: \emph{deteriorate} - partially or fully damage the functionality of a target, \emph{corrupt} - take control over the functionality of a target by either leaking sensitive information or behaving in a specific, desired way, or \emph{impersonate} - fake an entity to gain an advantage of its functionality in the system.
The attacks that deteriorate the good functionality usually aim to damage the availability by consuming resources in excess, so they are strongly related to performance and dependability too. Of course, disruption can also be caused by corrupting a target (by either resetting, stopping, or change its normal functionality). Moreover, an adversary can mount complex attacks that are combinations of several attacks (of different types), and the adversary can be either a single entity or a coalition of entities that collide to mount the attack together. 

Of course, both adversaries and attacks can be referred to in both classifications, so we can very well refer to \textit{active/passive adversaries} and \textit{insider/outsider attacks} too (e.g., an active adversary mounts an active attack, an insider adversary mounts an insider attack). The reader that is unfamiliar with the attacks' exemplification can further refer to \cite{NIST:Sec_glossary,crypto_encyclopedia}.

\subsubsection{Countermeasures}

The countermeasures, also known as \emph{safeguards} \cite{NIST:Sec_glossary}, can be \emph{reactive} or \emph{proactive} and come in a variety of means. As traditionally categorized in the McCumber cube \cite{mccumber1991information}, the safeguards can be enforced in \emph{technology}, \emph{policies and practices}, and the \emph{human factor}. \emph{Technology} can be implemented in either software or hardware, and can consist of cryptographic primitives and protocols (e.g., encryption to protect confidentiality), firewalls, Intrusion Detection Systems (IDS), isolation techniques, and many others. \emph{Policies and practices} consist of rules, regulations, and best practices that specify the expected behavior of involved entities. Examples include authorization policies, incident response procedures, and recovery procedures. Finally, the \emph{human factor} includes education, training, and awareness of users to obey the security policies and make use of technology mechanisms in correspondence to the security goals, make people aware of possible consequences, and become responsible for their acts~\cite{mccumber1991information}.

\subsection{State of the Art}

\subsubsection{Standards, regulations, and white papers}

The ETSI specifications that refer to MEC security aspects (listed in Table~\ref{tab:etsi_standards}) are used as references for presenting the state-of-the-art security requirements and regulations for MEC, presented in Table~\ref{tab:MEC-sec-req}. These include general requirements \cite{ETSI:MEC002},  API-related aspects\cite{ETSI:MEC009,ETSI:MEC011,ETSI:MEC012,ETSI:MEC013,ETSI:MEC014,ETSI:MEC015,ETSI:MEC016,ETSI:MEC029,ETSI:MEC030,ETSI:MEC033,ETSI:MEC040}, network slicing~\cite{ETSI:MEC029}, MEC integration within 5G \cite{ETSI:MEC031}, end-to-end mobility aspects \cite{ETSI:MEC018}. 
Note that N/A here does not mean that security requirements are out of scope or importance for the given element; N/A means that no explicit requirements are listed in the specifications.

White papers by ETSI (also listed in Table~\ref{tab:etsi_standards}) but even other organizations and industry companies \cite{zte:sec,ngmn_sec} refer to the security aspects of MEC within 5G too.

\renewcommand{\labelitemi}{-}
\begin{table*}[!t]
\caption{Security Requirements in the MEC Architecture According to Standardization Documentation}
\begin{center}
\begin{tabular}{p{1.8cm}p{13cm}}
\toprule
\textbf{Element} & \textbf{Security Requirements}\\
\midrule
\multicolumn{2}{l}{\textbf{General}}\\
 &
\vspace{-0.3cm}
\begin{itemize}[leftmargin=*]
	\item The MEC system shall provide a secure environment for running services for the involved actors, in particular the user, the network operator, the operators' third parties, and the platform vendor \cite{ETSI:MEC002}.
	\item The MEC system shall prevent illegal access from dishonest terminals, secured communication is necessary between the radio nodes and the MEC service~\cite{ETSI:MEC002}.
	\item The MEC system should securely collect and store logs, including information about charging \cite{ETSI:MEC002}.
	\item MEC services might require end-to-end security mechanisms \cite{ETSI:MEC002}.
	\item APIs must be secured, including general aspects such as controlling the frequency of the API calls, unexposure of sensitive information via the API, the authentication of a client to RESTful MEC Service API is performed using OAuth 2.0, and APIs shall support HTTP over TLS 1.2 \cite{ETSI:MEC009,ETSI:MEC012,ETSI:MEC014,ETSI:MEC016}.
	\item The MEC system discovery, including security (authentication/authorization, system topology hiding/encryption), charging, identity management, and monitoring aspects, is an essential prerequisite to form a MEC federation \cite{ETSI:MEC035}.
\end{itemize} 
\\
\midrule
\multicolumn{2}{l}{\textbf{MEH}}\\
MEP &
\vspace{-0.3cm}
\begin{itemize}[leftmargin=*]
	\item The MEP shall only provide a MEC app with the information the MEC app is authorized for \cite{ETSI:MEC002}.
	\item The MEP shall provide a secure environment for providing and consuming MEC services when necessary \cite{ETSI:MEC002}.
	\item A MEP should securely communicate to a MEP that might belong to different MEC systems \cite{ETSI:MEC003}.
	\item If the MEP is not dedicated to a single NSI, then the MEP shall support a different set of services and functionalities in distinct NSIs; in particular, the MEP needs to share an infrastructure that allows authentication and authorization at the NSI level and assure isolation of data and services between NSIs \cite{ETSI:MEC024}. 
	\item The MEP discovery is provided by means of the MEC systems exchanging information about their MEPs, i.e., their identities, a list of their shared services, as well as authorization and access policies \cite{ETSI:MEC035}.
\end{itemize} 
\\
Virtualization Infrastructure &
\vspace{-0.3cm}
\begin{itemize}[leftmargin=*]
	\item Virtualization should not introduce any new security threat; in particular, hypervisors should not introduce new vulnerabilities, patch management, state-of-the-art protections and secure boot mechanisms should be in place~\cite{ETSI:NFV002}. 
	\item A proper isolation of VNFs \cite{ETSI:NFV002} and VMs \cite{ETSI:MEC027} should be realized, the interfaces between the NFV components should be secured \cite{ETSI:NFV002}.
\end{itemize} 
\\
MEC App &
\vspace{-0.3cm}
\begin{itemize}[leftmargin=*]
	\item The application instance must satisfy the necessary security constraints \cite{ETSI:MEC002}.
    \item The MEC applications should have access to a persistent storage space \cite{ETSI:MEC002}.
    \item The MEC application should only have access to information for which it is authorized, should manage the access control and integrity of the user content, and should be authorized to consume or provide MEC services~\cite{ETSI:MEC002}.
    \item Upon authentication, the MEC applications should be able to communicate securely regardless if they are placed in the same MEH or different MEHs \cite{ETSI:MEC002}, even located in different MEC systems (in case of multi-operator scenario) \cite{ETSI:MEC002,ETSI:MEC003}.
	\item Operator trusted MEC application (seen as an extension of the MEP functionality) have advanced privileges of secure communication with the MEP \cite{ETSI:MEC002}.
    \item The security should be added to the MEC app package following the requirements defined for NFV \cite{ETSI:NFV004}, where the VNF refers to MEC app \cite{ETSI:MEC037}. 
\end{itemize} 
\\
\midrule
\multicolumn{2}{l}{\textbf{MEC Host-level Management}}\\
MEPM &
\vspace{-0.3cm}
\begin{itemize}[leftmargin=*]
	\item The MEC management should verify the authenticity and integrity of a MEC application~\cite{ETSI:MEC002,ETSI:MEC010_2}.
\end{itemize} 
\\
VIM & 
\vspace{-0.3cm}
\begin{itemize}[leftmargin=*]
	\item N/A (see Virtualization Infrastructure)
\end{itemize} 
\\
\midrule
\multicolumn{2}{l}{\textbf{MEC System-level Management}}\\
MEO &
\vspace{-0.3cm}
\begin{itemize}[leftmargin=*]
	\item The MEO checks the integrity and authenticity of the application packages \cite{ETSI:MEC003,ETSI:MEC010_2}.
	\item The MEO authorizes the requests of the OSS (e.g., application instantiation and termination, fetch on-boarded application package, query application package information) \cite{ETSI:MEC010_2}.
	\item MEAO adapts the orchestration operations based on the available NSIs and their requirements, including security requirements too \cite{ETSI:MEC024}.
\end{itemize} 
\\
OSS &
\vspace{-0.3cm}
\begin{itemize}[leftmargin=*]
	\item N/A
\end{itemize} 
\\
User app LCM proxy &
\vspace{-0.3cm}
\begin{itemize}[leftmargin=*]
	\item N/A
\end{itemize} 
\\
\midrule
\multicolumn{2}{l}{\textbf{MEC Federation}}\\
MEF &
\vspace{-0.3cm}
\begin{itemize}[leftmargin=*]
	\item The MEF should enable to exchange information in a secure manner among MEPs and MEC applications that belong to different MEC systems \cite{ETSI:MEC003,ETSI:5G-MEC_Cloud}.
        \item The MEF should face the security threats given by the heterogeneous scenario and edge resource sharing among operators (together with edge computing service providers and partners) \cite{ETSI:MEC-security,ETSI:MEC-federation}. These threats can be related to the access network, the architecture, the core network, the MEC elements, or other \cite{ETSI:MEC-security}.
        \item For security reasons, the information of MEP should be hidden between federated MEC systems \cite{ETSI:MEC040}.
\end{itemize} 
\\
\bottomrule
\end{tabular}
\label{tab:MEC-sec-req}
\end{center}
\end{table*}
\begin{table*}[!t]
\caption{Academic Surveys Related to MEC Security}
\begin{center}
\begin{scriptsize}
\begin{tabular}{p{0.7cm}p{1.9cm}p{0.5cm}p{6.3cm}p{6.3cm}}
\toprule
\textbf{Ref.} & \textbf{Aspect} & \textbf{MEC only} & \textbf{Main contribution} & \textbf{Relevance to MEC security} \\
\midrule
\cite{AKLOYG18} & 5G security & No & Gives an overview of 5G security challenges and solutions, referring to other technologies (e.g., SDN, NFV, cloud). & Mentions briefly some MEC security challenges, considered together with mobile clouds. \\
\cite{roman_mobile_2018} & Edge general & No & Analyses the security threats, challenges, and mechanisms in all edge paradigms. & Discusses MEC from the perspectives of security, dependability, and performance but keeps the discussion decoupled from the architecture. \\
\cite{du2018big} & MEC/IoT security & Yes & Considers MEC privacy issues in the context of heterogeneous IoT. &  Focuses on big data privacy issues in MEC with respect to data aggregation and data mining, and considers ML privacy-preserving techniques. \\
\cite{porambage_survey_2018} & MEC/IoT security & Yes & Overviews the MEC technology for the realization of IoT applications. & Reviews papers and discusses problems, challenges, and possible solutions for IoT security, privacy, and trust in relation to MEC.\\
\cite{abbas_mobile_2018} & MEC general & Yes & Presents a survey on different aspects related to MEC (e.g., computation, storage, energy efficiency, research infrastructure), also including security and privacy. & Presents some security mechanisms and privacy issues related to MEC keeping the discussion quite general. \\
\cite{Shirazi_2017} & Edge security and resilience & No & Reviews some security and resilience aspects in MEC and fog in relation to cloud technologies. & Generally discusses security requirements and challenges in MEC together with fog and in comparison with the cloud.  \\
\cite{bilal2018potentials} & Edge general & No & Presents an overview of potential, trends, and challenges of edge computing. & Refers only briefly to security and privacy. \\
\cite{zhang2018data} & Edge security & No & Surveys the data security and privacy-preserving in edge computing. & Analysis the data security and privacy challenges and countermeasures.\\
\cite{rapuzzi2018building} & Edge security & No & Discusses network threats in fog and edge computing. & Indicates threats, challenges and trends with respect to network security and awareness, with no focus on MEC particularities. \\
\cite{yu_survey_2018} & Edge/IoT & No & Surveys how edge computing improves the performance of IoT networks, but also considers security issues in edge computing. & Analyzes the security attributes and proposes a framework for security evaluation for edge computing-based IoT.  \\ 
\cite{dependability-bagchi-2019} & Edge dependability & No & Explores dependability and deployment challenges in edge computing. Considers dependability in a wider meaning that includes security. & Presents new challenges in physical security and scalable authentication, considering both centralized and decentralized security mechanisms. \\
\cite{khan2019survey} & 5G security & No & Gives a security and privacy perspective on 5G, looking into several enabling technologies (e.g., cloud, fog, MEC, SDN, NFV, slicing). & Discusses MEC security and presents some MEC threats and recommendations, but without a special focus on MEC and mostly together with cloud-related security issues. \\
\cite{RJL2019} & MEC security & Yes & Discusses MEC from a security perspective. & Identifies seven security threat vectors and discusses possible solutions and approaches to secure MEC by design.\\
\cite{khan2019edge} & Edge general & No & Surveys edge computing, identifies requirements, and discusses open challenges. & Gives low importance to MEC security, which is very briefly discussed. \\
\cite{ni2019toward} & Edge/IoT security & No & Discusses security and privacy together with efficiency in data communication and processing computation for IoT at the edge. & Discusses general security threats, secure data aggregation, and secure data deduplication at the edge, with no focus on MEC, but basing the findings on fog nodes. \\
\cite{xiao2019edge} & Edge security & No & Reviews attacks and the corresponding defense mechanisms in edge computing systems. & Focuses on four types of attack (DDoS, side-channel attacks, malware injection attacks, and authentication and authorization attacks) and presents the related root causes, status quo, and challenges.\\
\cite{caprolu2019edge} & Edge security & No & Introduces the main technologies supporting the edge paradigm and discusses issues and solutions. & Focuses on virtualization technologies (e.g., containers and unikernels) and related security issues.\\
\cite{liu2019survey} & MEC/IoT security & Yes & Discusses data analytics in edge computing in the context of IoT. & Reviews the existing works on data analytics in edge computing highlighting pros and cons, proposes some requirements for secure IoT data analytics, and highlights open issues and future research directions.\\
\cite{pham_survey_2020} & MEC general & Yes & Surveys MEC fundamentals, discussing its integration within the 5G network and relation with similar UAV communication, IoT, machine learning, and others. & Discusses MEC security very briefly but it points to many related papers (e.g., in the topic of MEC for IoT, security and privacy in the context of MEC-enabled IoT in V2X, smart cities, and healthcare).\\
\cite{yahuza2020systematic} & Edge security & No & Aims to provide a systematic review of security and privacy requirements in edge computing, including a taxonomy of attacks and performance evaluation. & Surveys a significant number of papers, but suffers from some clear shortcomings with respect to content and presentation. \\
\cite{sha2020survey} & MEC/IoT security & Yes & Surveys existing IoT security solutions at the edge. &  Presents an edge-centric IoT architecture and reviews edge-based IoT research in terms of security and privacy.\\
\cite{alwarafy2020survey} & Edge/IoT security & No & Surveys the security and privacy issues in the context of
edge-computing-assisted IoT. & Defines security and privacy in the context of the edge-computing-assisted IoT, gives some classifications of attacks, and discusses countermeasures.\\
\cite{ranaweera-MEC-security-2021} & MEC security & Yes & Analyses the security and privacy of MEC. & Discusses threat vectors, vulnerabilities that lead to the identified threat vectors, and proposes solutions to overcome these. The paper can be perceived as an extension of \cite{RJL2019}. \\
\cite{alwakeel2021overview} & Edge/Fog security & No & Surveys security challenges, issues, and countermeasures in edge and fog computing. & Discusses security and privacy aspects in fog and edge computing. \\
\cite{nowak2021verticals} & MEC security & Yes & Surveys security aspects in relation to twelve representative vertical industries of 5G MEC. & Presents characteristics, threats and vulnerabilities, attacks and countermeasures for each identified vertical, and further correlates the impact to the required performance of the vertical. \\ 
\cite{ali2021multi} & MEC security & Yes & Reviews the MEC architecture in relation to security and privacy aspects. & Reviews the conceptual guidelines for MEC security architecture as well as security and privacy techniques, examines and categorizes significant threats, and considers possible safeguards. \\
\bottomrule
\end{tabular}
\label{tab:sec_surveys}
\end{scriptsize}
\end{center}
\end{table*}

\subsubsection{Academic publications}

Table \ref{tab:sec_surveys} lists recent surveys on security and privacy for MEC. Papers that refer to general MEC security aspects are also considered. The number of publications considering MEC security and privacy is large, so the table does not intend to be exhaustive. Not many papers are fully dedicated to MEC but consider it together with other technologies such as cloud or fog (sometimes only marginally such as in e.g., \cite{yousefpour2019all}). As mentioned in some of these (e.g.,~\cite{khan2019survey,roman_mobile_2018}), some specialized work on the security for MEC has been performed. However, the number of papers referring to particular aspects regarding general edge technologies (and, to some extent, applicable to MEC too) or general security issues that are not MEC specific is large and thus out of the goal of this paper. For example, a large number of papers are dedicated to defenses against (D)DoS in MEC (e.g., \cite{TLWX19,LW18,mamolar2018towards,krishnan2019sdnfv}) or usage of ML for MEC (e.g., \cite{xiao2018security,abeshu2018deep,singh2021machine}).
Security in 5G network slicing has been analyzed in \cite{5g-netslice-sec}, and some aspects are relevant for isolation in MEC too.
Nevertheless, papers that survey aspects of MEC privacy and security do exist, e.g., \cite{ali2021multi}, a comprehensive study performed in parallel and independent of our work, or \cite{nowak2021verticals} a MEC security analysis for each of the twelve considered vertical industries: (1) manufacturing industry, (2) financial sector, (3) healthcare, (4) education, (5) telecommunication, (6) authorities, (7) media and entertainment, (8) smart city, (9) agriculture and food industry, (10) logistics, (11) education, culture, and science, and critical infrastructure sectors.


\subsection{Challenges} \label{sec:sec-challenges}

In the following, we present security challenges for MEC by categorizing them at the MEC host level, MEC system level, and general challenges.

\subsubsection{MEC host level}

\paragraph{Physical security} 
The MEHs are located at the edge of the network, closer to the user and in open environments. The physical location of the MEHs becomes thus insecure~\cite{khan2019survey,dependability-bagchi-2019}, host-level devices being even more vulnerable to physical attacks than system-level equipment that is normally placed in a more physically secured area. Moreover, the tendency to deploy many MEHs to cover an area raises problems with respect to a good physical security level\cite{RJL2019}. This increases the risk of unauthorized physical access and hence physical deterioration or corruption of the devices, with direct consequences against the availability (e.g., DoS attacks) and the confidentiality (e.g., data leakage by both passive and active attacks) \cite{zte:sec}. 
The equipment might lack the hardware protection of commodity servers \cite{roman_mobile_2018}, but as a form of protection, the MEC devices should implement anti-theft and anti-damage measures \cite{zte:sec}. Surely, tamper resistance is a generally good security strategy to prevent the reading of confidential data (e.g., cryptographic keys) and thus should be adopted in the case of MEC equipment too \cite{porambage_survey_2018,dependability-bagchi-2019}. In this scenario, the well-known principle of the weakest link holds: the security of the overall system is given by the security of the weakest spot (the adversaries tend to attack weak spots). The MEHs with poor security can easily become points of attacks \cite{du2018big}.

\paragraph{Location privacy} 
Location tracking enabled by MEC can be seen as both a feature and a risk. 
Unauthorized access to the Location Application Programming Interface (API) for the MEC Location Service can leak sensitive information about the localization and tracking of users in time \cite{ETSI:MEC013}, similar to unauthorized access to the Radio Network Information in mobile networks (e.g., access to identifications, which might damage the privacy of the UE)~\cite{ETSI:MEC012}. MEHs (and, in consequence, end-users too) are thus directly exposed to location privacy risks \cite{ni2019toward}. To mitigate such risks, an important role is in the security of the APIs and the generated location reports or processed data.
On the other hand, the MEC localization service can be beneficial when GPS coverage is unavailable or for emergencies (e.g., for healthcare applications where monitoring devices can send signals requiring assistance to the closest MEC platform in case of emergency) \cite{thembelihle2017softwarization}.

\paragraph{Local defences} 
Due to their local character, attacks at the host level influence a geographically limited area, in the proximity of end users \cite{roman_mobile_2018}. This gives MEC capabilities to enforce security mechanisms and limit attacks in the local network segment~\cite{porambage_survey_2018}. MEC is suitable to deploy a defense perimeter, one example being against (D)DoS attacks when the adversary only targets a smaller volume of traffic, and the edge can alert the core network about the source of danger, resulting in overall better availability \cite{krishnan2019sdnfv}.
The local character of MEC can also enhance privacy protection by preventing data to arrive at centralized servers, thus avoiding a centralized point of trust. An example from \cite{porambage_survey_2018} consists of the processing of images with car plates at the edge and identification of the plate number only to forward to central processing (this prevents, for example, location leakage). At the same time, it is believed that local data exchange (as contrary to e.g., sending the data over the internet) reduces the exposure of data \cite{ni2019toward}, but of course can raise security risks when the number of nodes is high, because of high traffic and positioning at the edge of the network \cite{pham_survey_2020}. 
\cite{he2018security} discusses some security improvements that MEC can bring in the IoT scenario.

\paragraph{Virtualization security} 
Malicious virtual machines can try to exploit their hosts~\cite{roman_mobile_2018}. Attacks such as VMs manipulation might include a malicious insider with enough privileges to access and damage a VM or a malicious VM with escalated privileges \cite{OLAY18, roman_mobile_2018,porambage_survey_2018}. 
If a VM is running on multiple hosts, then a simple DoS attack can damage all hosts simultaneously~\cite{OLAY18}. As a protection against the DoS attacks, the VMs should be limited in resource consumption, and the resource consumption should be balanced among hosts.
With respect to the privacy of data, the user data is stored at MEC host level, so it might get leaked \cite{RJL2019}. Moreover, possible alteration of data requires adequate backup and recovery possibilities, in strong relation with dependability prevention.
Virtualization attacks can damage the orchestration at the host level, and a compromised VIM can lead to the disruption of MEC services~\cite{RJL2019}. 
Service manipulation is another example of corruption, with important consequences such as DoS or data leakage attacks~\cite{roman_mobile_2018}. If a host is corrupted (not necessarily by virtualization attacks but in general), the adversary might interfere at several levels (e.g., apps, services, resource consumption) and run a wide set of possible attacks.

\paragraph{Constrained resources}
 Computational expensive security mechanisms, including the usage of heavy cryptography, can be a problem. For example, the edge devices might have limited connectivity and resources, which impose restrictions on the security protocols that can be deployed and facilitate attacks against availability \cite{ngmn_sec}. This might conclude in restrictions in the deployment of highly secure protocols, for example for authentication \cite{roman_mobile_2018}. Usage of public-key cryptography and Public-Key Infrastructure (PKI) in particular might be an issue because of high computational costs and management \cite{dependability-bagchi-2019}. Lightweight cryptography should be considered. Data deduplication mechanisms at the edge (in the sense of detecting and discarding copies of data, or even preventing re-computations) would increase the performance on limited devices. But realizing this while maintaining security is normally possible via (Fully) Homomorphic Encryption ((F)HE), which by itself requires very high computational costs \cite{ni2019toward}. The European Agency for Cybersecurity (ENISA) itself identifies the complexity of the implementation of security solutions in 5G (caused by mixing technologies such as cloud, fog, and edge), as well as the efficient cryptography solutions (because of nodes constrained in resources) to be key topics in research and innovation of 5G security \cite{ENISA:research_topics}.

\newpage
\subsubsection{MEC system level}

\paragraph{Global defences}
The management and orchestration of the diverse security mechanisms is a complex issue, and enabling security mechanisms independently on multiple entities does not necessarily mean that the complete system is secured~\cite{roman_mobile_2018, pham_survey_2020}. A balance between local (decentralized) and global (centralized) defense mechanisms, between responsibility and autonomy has to be considered \cite{roman_mobile_2018,dependability-bagchi-2019}. A global monitoring that allows an overview of the MEC system should be set in place, and everything should be auditable \cite{roman_mobile_2018}. To provide privacy, end-to-end encryption becomes a necessity whenever applicable \cite{khan2019survey}.

\paragraph{MEO security}
MEO is exposed to virtualization attacks \cite{RJL2019}. A compromised MEO could have a critical impact on the functionality of the overall MEC system. Examples include the termination of MEC critical applications, on-boarding of malicious application packages, an unbalanced usage of the MEHs in terms of resources, and others.
Hypervisor introspection methods need to be applied, for Linux-based platforms, Security Enhanced Linux (SELinux) might be of use \cite{RJL2019}. More on virtualization defenses will be discussed in \ref{gen_challenges}, the subsection dedicated to general challenges.
However, in the rapidity of change in technology and attacks nowadays, it is considered suitable to approach by software programmable solutions than rigid hardware (to allow updates, moving targets, dynamic attacks, etc.) \cite{krishnan2019sdnfv}.  Security and privacy challenges in softwarization and virtualization are not specific to MEC, as flexible solutions are required to secure 5G in general \cite{khan2019survey}.
Solutions to prevent virtualization problems and security frameworks within the MEC-in-NFV architecture have been considered \cite{RJL2019,farris2017towards}. These include Trusted Platform Manager (TPM) to attest and validate MEC Apps and VNFs, as well as requests from the CFS portal~\cite{RJL2019}. Auto-configurable security mechanisms are proposed, as well as methods to secure VNF in NFV environments \cite{RJL2019,zarca2019security}. 
Ideally, the idea of a security orchestrator based on softwarization (SDN / VNF) would replace the need for manual configuration that is no longer feasible under the current circumstances~\cite{khan2019survey}. But how such a security orchestrator should be built and integrated with the architecture of MEC is still an open question.

\paragraph{Interconnection security}
At the MEC system level, OSS is exposed to threats outside the MEC system through the communication with the CSF Portal and device applications via the LCM proxy. This opens up for security risks, for example, the CSF Portal is prone to (D)DoS attacks \cite{RJL2019}. OSS can be subject to masquerading for adversaries that pretend to have legitimate access \cite{RJL2019}. A significant number of requests from the OSS to the MEO might (in the absence of proper security mechanisms) damage the functionality of the MEO.

\subsubsection{General challenges}
\label{gen_challenges}

\paragraph{Trust models and relations} 
Contrary to other technologies at the edge (e.g., edge cloud, fog), in MEC there is a lower number of owners that need to cooperate \cite{roman_mobile_2018}. However, it is of critical importance to clarify the trust models and relations between the entities involved (users, platforms, slices, apps, etc.) \cite{zte:sec}. In particular, trust needs to be considered in relation to mobility and network functions performed in the MEC and integration to the 5G standard too \cite{zte:sec}. Trust models have been considered for different edge technologies \cite{roman_mobile_2018,petri2012trust}. A flexible trust manager has been proposed as a solution to incorporate within MEC~\cite{porambage_survey_2018}. Other trust schemes have been proposed for MEC (e.g.,~\cite{he2018secure}).
Nevertheless, general protection mechanisms such as mutual authentication and access control mechanisms at all levels should be set in place. A good prevention is to minimize the data transmitted and stored on low reputation entities \cite{roman_mobile_2018}. The ownership of personal data must be assigned and clearly decided between different roles (e.g., stakeholders, MNO, third parties) \cite{khan2019survey}. To protect end users' data, techniques such as watermarking, visual cryptography, and biometrics were considered in the literature \cite{khan2019survey}. 

Nevertheless, security models (not only trust models) in accordance with the MEC requirements need to be developed and applied for security analysis. 
    
\paragraph{Network security} 
The heterogeneous nature of edge and its dynamic character introduces risks \cite{khan2019edge,pham_survey_2020,khan2019survey,ngmn_sec}. Security is prone to risks at the interconnection with other technologies, mostly within the 5G context. In particular, MEC should access the internet and establish connectivity to other MEC domains via the internet \cite{RJL2019}. 
Naturally, security should be considered for data in all forms (storage, computation, and transmission) and approached by appropriate cryptographical primitives and security technologies (including general approaches such as VPN communication, access control functions and policies, firewalls \cite{RJL2019}). Lightweight encryption can be used for performance enhancements whenever convenient~\cite{RJL2019}. However, there is still a place for improvement with respect to lightweight cryptography needed for solutions such as MEC.
MEC is exposed to attacks on the communication channels, mostly on the wireless channels close to the end-user~\cite{RJL2019}. A specific risk lies in the communication between the edge and the core \cite{khan2019survey,ngmn_sec}. Data has to be encrypted in transit (e.g., IPSec, TLS) \cite{zte:sec}, and end-to-end encryption is a way to prevent data leakage \cite{ni2019toward}. Adaptive security protocols could be used to enhance the security of communication channels \cite{RJL2019}. Software-Defined Virtual Private Local Area Networks (Soft-VPLS) can help in securing the communication between MEC components \cite{RJL2019}. Soft-VPLS allows different traffic categories (e.g., MEC service requests, user data, control statistics) to be routed via distinct tunnels, aiming to enhance both end-to-end security and overall communication performance \cite{RJL2019}. 
Proper isolation of network traffic, data, services, slices, etc. is required \cite{zte:sec, roman_mobile_2018}. 
The use of gateways at strategic points in the network is considered a good practice \cite{AKLOYG18}, and firewalls are now implemented as NFs \cite{thembelihle2017softwarization}. Techniques to provide physical layer security might turn out beneficial because of performance aspects \cite{pham_survey_2020,bai2019energy,xu2018exploiting}. More on communication security is discussed in \cite{khan2019survey}.

\paragraph{Monitoring and detection} 
 Mitigation techniques in network security include monitoring and logging, abnormal traffic analysis, malware and intrusion detection. In MEC, this should be performed at all levels. The collaboration between the edge nodes can be useful in this respect \cite{zte:sec}. Some work on access control and Intrusion Detection Systems (IDS) for MEC has been pointed out before \cite{roman_mobile_2018,mtibaa2015friend}. This can prevent or mitigate attacks such as DoS, malicious actions, and rogue entities \cite{roman_mobile_2018}. AI, in particular ML, could be successfully applied for intrusion and anomaly detection \cite{thembelihle2017softwarization,RJL2019}. Deep Learning (DL) was specifically considered for the detection of attacks, and Reinforcement Learning (RL) was proposed for edge caching security \cite{abeshu2018deep,pham_survey_2020,xiao2018security}. Federated Learning (FL) suits MEC because it allows the training data to be kept locally and privately among collaborating nodes. Thus, advances in using FL for constraint devices might be useful \cite{pham_survey_2020,lim2020federated,li2020federated}. More references about ML in IDS and against DDoS attacks can be found in \cite{khan2019survey}, also in correlation with SDN solutions.
The study of ML (and AI in general) for MEC security is a valid research direction~\cite{porambage_survey_2018}.

\paragraph{Virtualization security} 
General virtualization and softwarization issues need to be considered in MEC too. This includes security issues of both SDN and NVF \cite{roman_mobile_2018}, and in particular security aspects related to network slicing \cite{khan2019survey, 5g-netslice-sec}. SDN/NFV-based frameworks or approaches to provide security in relation to MEC and IoT have been considered \cite{porambage_survey_2018,farris2017towards}. Software-Defined Privacy (SDP), a solution currently in place for enforcing the security of Internet as a Service(IaaS) cloud customers, might be extended to provide privacy protection in MEC too \cite{porambage_survey_2018}.
Virtual Machine Introspection (VMI) and hypervisor machine introspection should monitor the activities in terms of resource utilization to prevent deprivation and DoS attacks \cite{RJL2019}. These should be run at both host and system level \cite{RJL2019}.  Examples of VMI include LibVMI \cite{RJL2019,tool_libvmi}. Artificial intelligence can be used to achieve better VMI solutions \cite{RJL2019,kumara2017leveraging}. 

\paragraph{Standardization and awareness} 
The necessity for standardized or universal security measures to ensure the security of the overall MEC system is considered to be an open problem~\cite{porambage_survey_2018}. At the same time, the human factor remains a risk, so it is important to raise awareness of the users that need to understand and apply security policies \cite{roman_mobile_2018}. A significant category is given by the application developers \cite{porambage_survey_2018}, and special focus should be put on the integration mechanisms.

\paragraph{Other security challenges} 
Many other security challenges exist in MEC (e.g., see \cite{khan2019survey} for some specific Backhaul threats). We next refer briefly to some of these.

Privacy of identity is known to be a challenge in mobile networks (including 5G), and it remains a challenge in MEC too \cite{khan2019survey,mjolsnes2019private}. More on Personally Identifier Information (PII) protection in the general context of \textit{General Data Protection Regulation (GDPR)} can be found in \cite{ETSI:CYBER_ABE}.
 
Usage of the appropriate technologies and primitives for security is always a challenge. For example, solutions nowadays tend to adopt blockchain-based technology, including in MEC~\cite{liao2021blockchain,rivera2020blockchain}. However, many times the blockchain technology is in fact not needed~\cite{wust2018you} and more suitable solutions (e.g., which introduce less complexity) are available. While ETSI has a dedicated group on Permissioned Distributed Ledgers (PDL) \cite{ETSI:PDL_group}, it remains open if blockchain is indeed a useful technology for MEC. Quantum security mechanisms have to be considered \cite{khan2019survey}. Cryptographical primitives currently used in 5G and MEC (mainly public-key primitives) are known to be vulnerable to quantum attacks. Although nowadays quantum attacks are still in their infancy, quantum-resistant cryptography is an active research field that aims to provide security against quantum adversaries. 

Mobility-related security challenges are of importance in MEC too. More about these will be discussed together with other aspects, in Section~\ref{sec:discussion}.

\section{Dependability} \label{sec:dependability}

If security is a well-known term, dependability is a less common term, whose meaning is sometimes ignored or confused. In this paper, we define dependability as in \cite{2004:dep-taxonomy}. Dependability is the \emph{ability to deliver a service that can justifiably be trusted}.

\subsection{Taxonomy}

\begin{figure}[!t]
  \centering
  \includegraphics[width=0.48\textwidth]{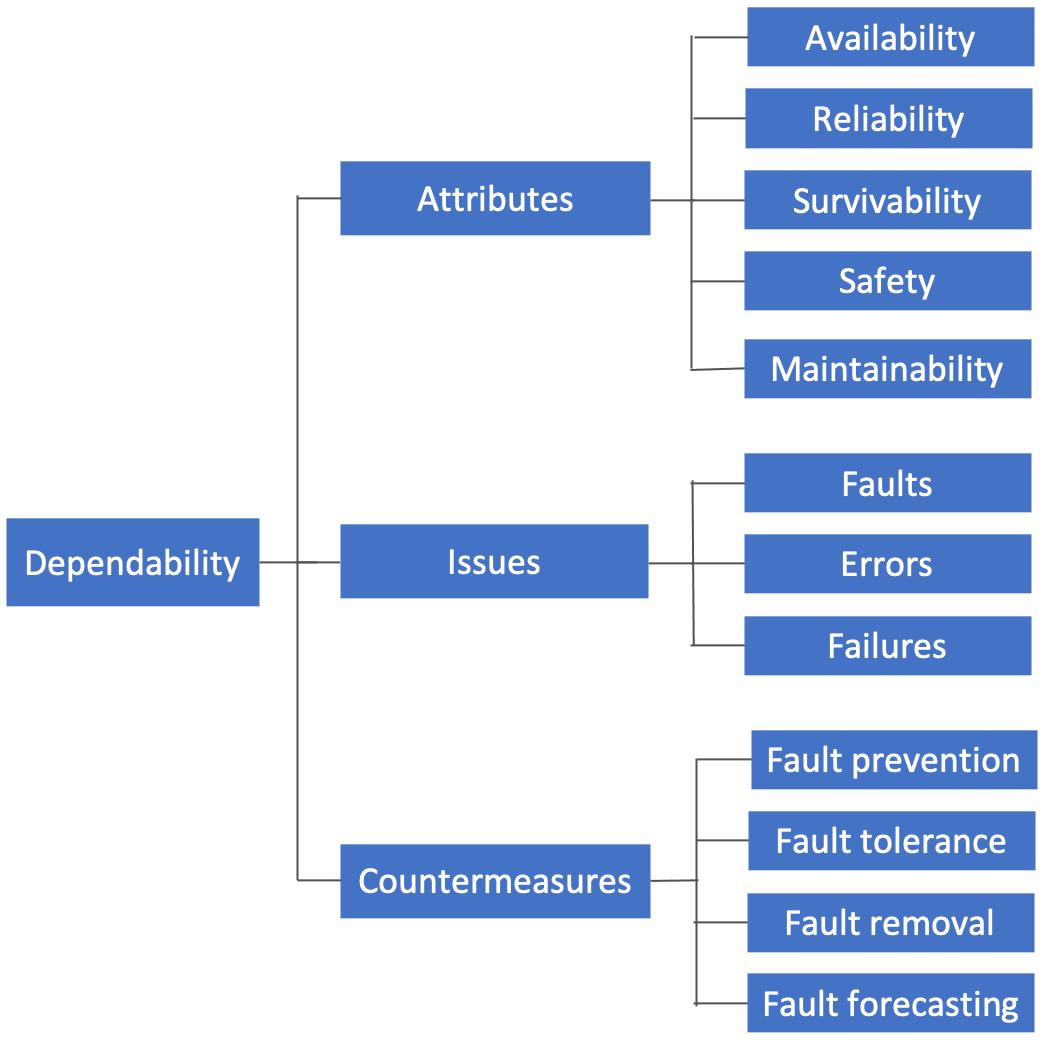}
  \caption{Dependability Taxonomy}
  \label{fig:dep-taxonomy}
\end{figure}

Figure \ref{fig:dep-taxonomy} represents the taxonomy related to dependability. 
The \emph{attributes} are the various ways to evaluate the dependability of a system.
The \emph{issues} are the causes that may lead to a lack of dependability.
The \emph{countermeasures} are the methods to enhance the dependability of a system.

Alternative definitions and taxonomies can be found in literature  \cite{resilinets,2009-trivedi}, where the term \emph{threat} is used instead of \emph{issue} and the term \emph{mean} is used instead of \emph{countermeasure}. Moreover, some works define a joint dependability and security taxonomy \cite{2004:dep-taxonomy, 2009-trivedi}. Instead, performance, or performability, is sometimes seen as one of the attributes of dependability\cite{2009-trivedi}. For the sake of clarity, we keep separate security, dependability, and performance, and we will jointly discuss them in Section \ref{sec:discussion}.

\subsubsection{Attributes}

The attributes consist in metrics that are able to characterize and measure specific properties of a system. The attributes can be applied to the MEC as a system, but also to MEC-based services.
The main attributes are listed in Table~\ref{tab:dep-attr}. The description of the attributes is based on previous definitions~\cite{2004:dep-taxonomy,2009-trivedi}. The most known attributes are \emph{availability} and \emph{reliability}.
A system is available when it is ready to deliver a service that complies with the service specifications. The simplest way to compute the availability of a system is the ratio between the expected uptime of the system and the aggregate time (sum of expected values of up and down times).
A system is reliable when it is able to continuously deliver a service that complies with the service specifications. The reliability can be computed based on the mean time to failure or the mean time between failures.
The \emph{survivability} and \emph{safety} are sometimes not listed among the attributes.
The \emph{maintainability} is associated with recovery mechanisms and can be measured as the mean time to repair.

\begin{table}[!t]
\caption{Dependability Attributes}
\begin{center}
\begin{tabular}{p{2.3cm}p{5.5cm}}
\toprule
\textbf{Attribute} & \textbf{Description}\\
\midrule
Availability (A) & Readiness for a correct service. \\
Reliability (R) & Continuity of correct service. \\
Survivability (S)&  Capability to continue to deliver a correct service in the presence of failures or accidents.\\
Safety (Sf)& Absence of catastrophic consequences on the user(s) and the environment.\\
Maintainability (M) & Ability to undergo modifications and repairs.\\
\bottomrule
\end{tabular}
\label{tab:dep-attr}
\end{center}
\end{table}

\subsubsection{Issues}

The dependability issues, fault and failure as listed in Figure \ref{fig:dep-taxonomy}, in the daily language are used interchangeably and mean that something that is not working.
In dependability taxonomy, they are not only independent causes of lack of dependability, but they represent a cause-effect sequence and they can be defined as follows \cite{2004:dep-taxonomy}: 
\begin{itemize}
 \item \emph{Fault} is the adjudged or hypothesized cause of an error;
 \item \emph{Error} is part of the system state that is liable to lead to a failure;
 \item \emph{Failure} is the deviation of the delivered service from the correct service (i.e., the service is no longer compliant with the specification).
\end{itemize}

An example of the cause-effect sequence is the following: an external fault flips a bit in the memory causing an error that manifests as a failure when that partition of the memory is accessed.
Table \ref{tab:faults-failures} lists the main types of faults and failures~\cite{2004:dep-taxonomy}.

\begin{table}[!t]
\caption{Main Types of Faults and Failures}
\begin{center}
\begin{tabular}{p{2.3cm}p{5.5cm}}
\toprule
\textbf{Type} & \textbf{Description}\\
\midrule
\textbf{Faults}&\\
Development fault & Fault occurring during development.\\
Physical fault & Fault that affects hardware.\\
Interaction fault & External fault.\\
\midrule
\textbf{Failures}&\\
Content failure & Deviation of the content of the information delivered.\\
Timing failure & Deviation of the arrival time or duration of the information delivered.\\
Erratic failure & The service is delivered (not halted) but is erratic.\\
Inconsistent failure & Some or all system users perceive differently incorrect service.\\
Catastrophic failure & The cost of harmful consequences is orders of magnitude, or even incommensurately, higher than the benefit provided by correct service delivery.\\
\bottomrule
\end{tabular}
\label{tab:faults-failures}
\end{center}
\end{table}

\subsubsection{Countermeasures}

In order to improve the dependability of a system, various categories of methods are used \cite{2004:dep-taxonomy}: 
\begin{itemize}
    \item \emph{Fault prevention} uses development methodologies, for both software and hardware, in order to reduce the number of faults;
    \item \emph{Fault tolerance} is carried out via error detection and system recovery in order to avoid failures;
    \item \emph{Fault removal} can be performed during the system development (via verification, validation, and testing) and during the system use (via corrective or preventive maintenance);
    \item \emph{Fault forecasting} is conducted by evaluating the system behavior (it can be both qualitative and quantitative).
\end{itemize}

\newpage
\subsection{State of the Art}

\subsubsection{Standards, regulations, and white papers}

The ETSI specifications that refer to MEC dependability aspects (listed in Table \ref{tab:etsi_standards}) are used as references for presenting the  state-of-the-art dependability requirements and regulations for MEC, listed in Table \ref{tab:MEC-dep-req}.
Note that N/A here (as in Table~\ref{tab:MEC-sec-req} for the security) does not mean that dependability requirements are out of scope or importance for the given element; N/A means that no explicit requirements are listed in the specifications.

Availability and reliability are mentioned as non-functional metrics in \cite{ETSI:MEC006}. Resiliency and high availability is mentioned in the white paper \cite{ETSI:5G-MEC_Software}.

In many ETSI specifications, \emph{service availability} is mentioned but not always with the same meaning in the dependability:
availability tracking API in \cite{ETSI:MEC010_2}; 
together with application availability in \cite{ETSI:MEC011}; 
testing the related query in~\cite{ETSI:MEC025};
testing API query in \cite{ETSI:MEC032_1, ETSI:MEC032_2};
together with 5G~\cite{ETSI:MEC031};
in connection with the coordination of inter-MEC systems and MEC-cloud systems \cite{ETSI:MEC035}.

The \emph{service continuity} in mobility is extensively addressed in \cite{ETSI:MEC018,ETSI:MEC022}.
It is also mentioned in \cite{ETSI:MEC016,ETSI:MEC032_2,ETSI:MEC035}
and in several white papers \cite{ETSI:5G-MEC_Evol,ETSI:5G-MEC, ETSI:5G-MEC_Enterprise, ETSI:5G-MEC_harmonizing}.

The \emph{fault management} is briefly addressed in \cite{ETSI:MEC003} and more in detail in \cite{ETSI:MEC010_1}.
Fault management in NFV implementation is addressed in~\cite{ETSI:MEC017}.
Testing of MEH fault management is presented in~\cite{ETSI:MEC025}.
API test for MEH fault management is discussed in~\cite{ETSI:MEC032_1}.
Fault management is also addressed in the white papers \cite{ETSI:5G-MEC_NetTransf} and \cite{ETSI:5G-MEC_AI}.

The \emph{network dependability} is addressed in \cite{ETSI:MEC015} and also in the white papers \cite{OpenStack,ETSI:5G-MEC_Enterprise}.

Several specifications refer to \emph{URLLC and mission-critical application}, for example in \cite{ETSI:MEC018,ETSI:MEC024} and the white papers \cite{ETSI:5G-MEC_Evol, ETSI:5G-MEC}. 
The specifications highlight the importance of MEC in mission-critical low-latency applications, such as Industrial IoT and Self-Driving Cars. These applications require communication with very high reliability and availability, as well as very low end-to-end latency going down to a millisecond level.

\emph{V2X communication} is maybe the most relevant use case in MEC. It is initially mentioned in\cite{ETSI:MEC002}, then more in detail in \cite{ETSI:MEC022}.
V2X communication is important because it has strict requirements in all three aspects. In particular, the ETSI specifications mention the reliability and availability (from both security and dependability perspectives) together with latency and throughput. Given the high mobility of the V2X users, one of the main topics to investigate is the handover between the MEHs, which has an impact on both service continuity and service availability~\cite{ETSI:MEC002}. Specifically, the predictive handover is investigated to meet the dependability and performance requirements \cite{ETSI:MEC022}. In~\cite{ETSI:MEC030}, service continuity is mentioned in the context of having multiple operators. 
The white paper \cite{5gaa:edge} highlights the high reliability and security requirements in the V2X communication in 5G-MEC.

\subsubsection{Academic publications}
Table \ref{tab:dep_surveys} lists recent surveys that address dependability aspects in MEC.
As for security, not many papers are fully dedicated to MEC but consider it together with other technologies such as cloud or fog.
A survey on dependability on MEC does not exist. A few papers are mainly focused on dependability aspects \cite{reliability-madsen-2013,dependability-bagchi-2019,roadmap-bakshi-2019} and others have parts dedicated to dependability aspects \cite{yousefpour2019all,li_survey_2019,khan2019edge} or shared with other aspects \cite{mahmud_fog_2018,mao_survey_2017,bilal2018potentials,elbamby-wireless-2019}. The other papers just mention it or use it as a requirement \cite{mao_survey_2017,li_survey_2019}, property \cite{shahzadi2017multi}, benefit \cite{habibi_fog_2020}, or challenge \cite{khan2019edge,pham_survey_2020}.

\renewcommand{\labelitemi}{-}
\begin{table*}[!t]
\caption{Dependability Requirements in the MEC Architecture According to Standardization Documentation}
\begin{center}
\begin{tabular}{p{1.8cm}p{13cm}}
\toprule
\textbf{Element} & \textbf{Dependability Requirements}\\
\midrule
\multicolumn{2}{l}{\textbf{General}}\\
 &
\vspace{-0.3cm}
\begin{itemize}[leftmargin=*]
	\item Additional tools are needed to generate workload and challenge the service in terms of service scalability, availability, and reliability \cite{ETSI:MEC006}.
	\item For alarm management, the following 3GPP-defined Integration Reference Points are used: ETSI TS 132 111-2 \cite{ETSI:132-111-2} and ETSI TS 132 332 \cite{ETSI:132-332} \cite{ETSI:MEC010_1}.
	\item The Multi-access Traffic Steering can be used by MEC apps and MEP for seamlessly steering/splitting/duplicating application data traffic across multiple access network connections \cite{ETSI:MEC015}.
\end{itemize} 
\\
\midrule
\multicolumn{2}{l}{\textbf{MEH}}\\
MEP &
\vspace{-0.3cm}
\begin{itemize}[leftmargin=*]
	\item The MEP shall allow MEC services to announce their availability \cite{ETSI:MEC002}.
	\item The MEP interacts with the MEC apps via the reference point Mp1, which provides functionalities, such as service availability and session state relocation support procedures \cite{ETSI:MEC003,ETSI:MEC011}.
	\item The MEP may use available radio and core network information to optimize the mobility procedures required to support service continuity \cite{ETSI:MEC021}.
\end{itemize} 
\\
Virtualization Infrastructure &
\vspace{-0.3cm}
\begin{itemize}[leftmargin=*]
	\item N/A
\end{itemize} 
\\
MEC App &
\vspace{-0.3cm}
\begin{itemize}[leftmargin=*]
	\item Some MEC applications expect to continue serving the UE after a location change of the UE in the mobile network. In order to provide continuity of service, the connectivity between the device application and the MEC application needs to be maintained \cite{ETSI:MEC003}.
	\item Each MEC service instance that has previously registered in MEP and is configured for heartbeat sends a heartbeat message to the MEP periodically in order to show that the MEC service instance is still operational \cite{ETSI:MEC011}.
\end{itemize} 
\\
\midrule
\multicolumn{2}{l}{\textbf{MEC Host-level Management}}\\
MEPM &
\vspace{-0.3cm}
\begin{itemize}[leftmargin=*]
	\item The MEPM also receives Virtualised resources fault reports and performance measurements from the VIM for further processing \cite{ETSI:MEC003}.
	\item In the NFV variant, the MEPM-V does not receive Virtualised resources fault reports and performance measurements directly from the VIM, but these are routed via the VNFM \cite{ETSI:MEC003, ETSI:MEC017}.
\end{itemize} 
\\
VIM & 
\vspace{-0.3cm}
\begin{itemize}[leftmargin=*]
	\item The VIM collects and reports the performance and fault information about the virtualised resources \cite{ETSI:MEC003}.
\end{itemize} 
\\
\midrule
\multicolumn{2}{l}{\textbf{MEC System-level Management}}\\
MEO &
\vspace{-0.3cm}
\begin{itemize}[leftmargin=*]
	\item The MEO maintains an overall view of the MEC system based on deployed MEHs, available resources, available MEC services, and topology \cite{ETSI:MEC003}.
	\item The MEO interacts with the MEPM via the reference point Mm3 to keep track of available MEC services \cite{ETSI:MEC003}.
\end{itemize} 
\\
OSS &
\vspace{-0.3cm}
\begin{itemize}[leftmargin=*]
	\item The OSS interacts with the MEPM via the reference point Mm2 for fault management \cite{ETSI:MEC003, ETSI:MEC017}.
	\item In the NFV variant, the OSS interacts with the NFVO for fault management \cite{ETSI:MEC017}.
\end{itemize} 
\\
User app LCM proxy &
\vspace{-0.3cm}
\begin{itemize}[leftmargin=*]
	\item N/A
\end{itemize} 
\\
\midrule
\multicolumn{2}{l}{\textbf{MEC Federation}}\\
MEF &
\vspace{-0.3cm}
\begin{itemize}[leftmargin=*]
	\item N/A
\end{itemize} 
\\
\bottomrule
\end{tabular}
\label{tab:MEC-dep-req}
\end{center}
\end{table*}


\begin{table*}[!t]
\caption{Academic Surveys Related to MEC Dependability}
\begin{center}
\begin{tabular}{p{0.6cm}p{2cm}p{0.5cm}p{6.3cm}p{6.3cm}}
\toprule
\textbf{Ref.} & \textbf{Aspect} & \textbf{MEC only} & \textbf{Main contribution} & \textbf{Relevance to MEC dependability} \\
\midrule
\cite{reliability-madsen-2013} &Fog reliability &No &Addresses the reliability in fog computing. &Presents the reliability challenges by combining the reliability requirements of cloud computing and networks of sensors and actuators. \\
\cite{taleb_multi-access_2017} &MEC general &Yes &Surveys the use cases and the enabling technologies. Focuses on orchestration and related challenges. & Mentions the five-nine availability. Briefly discusses the reliability aspect in deploying MEC services and suggests dubbed checkpoints or, for improving the scalability, the replication of MEC service instances. Discusses also resiliency as a service enhancement.\\
\cite{mahmud_fog_2018} &Fog general &No & Surveys taxonomy, existing works, challenges, and future directions of fog computing. & Identifies the reliability of fog computing as a poorly discussed topic and suggests better investigation of the consistency of fog nodes and availability of high-performance services.\\
\cite{mao_survey_2017} &MEC communication &No &Focuses on joint radio-and-computational resource management & Diffusely mentions reliability in various contexts (link, transmission, server). Focuses more in detail on mobility-aware fault-tolerant MEC. Mentions resiliency and high availability of MEC system as requirement.\\
\cite{shahzadi2017multi} &Edge general &No &Provides an overview of the state of the art and the future research directions for edge computing. &Inserts fault tolerance as a property in comparing the existing frameworks in edge computing. Mentions the availability of resources by stating that it is mostly dependent upon server capacity and wireless access medium for ensuring constant service delivery. \\
\cite{Shirazi_2017} &Edge security and resilience &No &Reviews some security and resilience aspects in MEC and fog in relation to cloud technologies. &Generally discusses resilience requirements and challenges in MEC together with fog and in comparison with the cloud.\\ 
\cite{yu_survey_2018} &Edge/IoT &No &Surveys how edge computing improves the performance of IoT networks &Addresses the dependability with respect to the storage by focusing on the recovery policy.\\
\cite{bilal2018potentials} &Edge general &No &Presents an overview of potential, trends, and challenges of edge computing. &Mentions fault tolerance (together with Quality of Service) as one of the challenges of edge computing. Discusses the need for proactive fault tolerance and automatic recovery.\\
\cite{porambage_survey_2018} &MEC/IoT security &Yes &Overviews the MEC technology for the realization of IoT applications. &Diffusely  mentions  reliability  in  various contexts.\\
\cite{khan2019edge} &Edge general &No &Surveys edge computing, identifies requirements, and discusses open challenges. &Presents the provision of low-cost fault-tolerant deployment models as an open challenge.\\
\cite{yousefpour2019all} &Edge general &No &Provides a tutorial on edge computing and related taxonomy and surveys the state-of-the-art efforts. &Considers the RAS (Reliability Availability Survivability) as objective and the resilient fog system design as a challenge.\\
\cite{li_survey_2019} & MEC Industrial Internet &Yes &Surveys the existing works on MEc for Industrial Internet. &Discusses the high reliability as a typical industrial requirement.\\
\cite{pham_survey_2020} &MEC general &Yes &Surveys MEC fundamentals, discussing its integration within the 5G network and relation to similar UAV communication, IoT, machine learning, and others. &Considers reliability as a challenge in MEC and diffusely discusses it.\\
\cite{dependability-bagchi-2019} &Edge dependability &No &Explores dependability and deployment challenges in edge computing. Considers dependability in a wider meaning that includes security. &Presents resiliency challenges, which include new failure modes, network impact, limited fail-over options, multi-tenancy support, and interoperability. \\ 
\cite{roadmap-bakshi-2019} &Fog dependability &No &Surveys the research efforts on fault tolerance and dependability in fog computing. & Presents redundancy models and fault management solutions. Discusses the dependability challenges and open problems. \\
\cite{habibi_fog_2020} &Edge general &Yes &Surveys different aspects of edge computing from an architectural point of view. &Mentions the improvement of reliability as a benefit of edge computing.\\
\cite{elbamby-wireless-2019} &Wireless edge &No &Discusses the feasibility and potential of providing edge computing services with latency and reliability guarantees. &Overviews the challenges and the enablers for realizing hugh reliability in wireless edge computing.\\
\cite{filali-MEC-surv-2020} &MEC general &Yes &Surveys the MEC and focuses on the optimization of the MEC resources &Diffusely mentions reliability, availability, and resilience.\\
\cite{liu-vehicular-edge-2020} &Vehicular Edge Computing & No &Surveys the state of the art of vehicular edge computing. &Diffusely mentions reliability.\\
\bottomrule
\end{tabular}
\label{tab:dep_surveys}
\end{center}
\end{table*}


\subsection{Challenges} \label{sec:dep-challenges}
In the following, we present dependability challenges for MEC by categorizing them on the MEC host level, MEC system level, and general challenges.

\subsubsection{MEC host level}

\paragraph{Physical dependability}
The MEH is practically consisting of a computer or a small server. For this reason, traditional techniques for making dependable a computer or server can be considered.
These techniques would act on both hardware and software.

Given the distributed nature of the MEC system, which is composed of multiple MEHs, the alternative is to consider a MEH expendable and provide fault tolerance by migrating to a new MEH in case of failure \cite{reliability-madsen-2013}.
In this case, careful deployment of MEHs and efficient failover mechanisms are needed. Deployment and failover mechanisms will be presented more in detail later.

\paragraph{Virtualization dependability}
An important characteristic of a MEH is virtualization. A MEH uses virtualization technologies, such as containers or VMs, which are managed by a VIM, such as Kubernetes or Openstack \cite{ETSI:MEC027}. Moreover, as already mentioned, the MEC architecture can be integrated with a virtualization architecture such as NFV \cite{ETSI:MEC017}.

The virtualization in the MEH needs to be considered in order to have a dependable MEC.
For example, the live VM migration in Openstack while injecting network failures and increasing the system pressure can be investigated \cite{hao-openstack-2019}. Note that the VM can be migrated to a different host or within the same host.
Regarding NFV, a problem that can be addressed is the VNF placement in a MEC-NFV environment in order to maximize the availability \cite{yala-vnf-2018}.
The most critical part of this work is to identify and include in the evaluation the necessary kinds of failures, e.g., the failures of the network, VMs, or physical machines.
Moreover, considering 5G network slicing, the protection of the network slices can be considered in the VNF placement~\cite{Chantre_2020}.

\subsubsection{MEC system level}

\paragraph{Deployment and failover mechanisms}
As already mentioned, a manner to make a MEC system resilient to the failure of MEH is to use failover mechanisms.
Failover mechanisms that need a proper deployment of the MEHs, i.e., the MEHs need to be close enough in order for a user to be able to reach multiple MEHs \cite{reliability-madsen-2013}.

For this reason, the deployment of MEHs can be performed by maximizing the failover capability \cite{li-deployment-2020} or by considering a 1+1 protection, where the users are able to connect to two MEHs (one is active and the other one is for backup) \cite{tonini-deployment-2019}.

Proactive failover mechanisms can be considered in order to reduce the impact of the failure \cite{huang-recovery-2019}.
In case a user is not able to reach another MEHs, other users can be used as a relay in order to reach active MEHs \cite{satria-recovery-2017}.

\paragraph{Resource allocation}
Another manner to improve the dependability of MEC is by a proper allocation of resources, usually computing and storage, in the different MEHs.
There are already several works that address this aspect \cite{chen-offloading-2015,chen-offloading-2016,liu-offloading-2017,liule-offloading-2020,merluzzi-offloading-2019,merluzzi-offloading-2020,liubennis-offloading-2019,liu-offloading-2018}.
Many of these works refer to the term \emph{task} and they are talking about \emph{offloading}.
The term task is used to refer to an application or procedure, instead offloading refers to the migration of the execution of a task from the mobile device to a MEH.

Most of the current works have as target energy efficiency and consider the dependability metrics as requirements.
The main difference between current works is the dependability metrics they are considering.
Some works consider the failure probability of MEHs to develop k-out-of-n allocation schemes, where the tasks are distributed among several MEH and the task is correctly executed if at least k out of n MEHs are not failed \cite{chen-offloading-2015,chen-offloading-2016}.
One work considers as reliability the probability of the delay bound violation, which is actually more similar to the performability \cite{liu-offloading-2017}.
Another work considers an offloading failure probability, which is connected to the error probability in the transmission between the user and the access where the MEH is located \cite{liu-offloading-2018}.
Another work considers also the execution reliability \cite{liule-offloading-2020}.
Finally, some works model the MEC system with queues and define the reliability as outage probability. There is an outage when a queue length exceeds a predefined threshold~\cite{merluzzi-offloading-2019,merluzzi-offloading-2020,liubennis-offloading-2019}.

Furthermore, there are works that address the task allocation together with the user-host association \cite{liubennis-offloading-2019} or consider that different users may have different dependability requirements~\cite{aposto-offloading-2020}.

\paragraph{MEO dependability}
The MEO is a critical element because it is a single point of failure of the MEC system.
A dependable MEO is important because if it fails the whole MEC system became unavailable.
Moreover, the MEO is also important to manage failure of the other MEC elements, in order to have a fault-tolerant MEC system.
For this reason, the design of the MEO and its functionality must be addressed carefully by considering the state-of-the-art technology \cite{ETSI:5G-MEC_NetTransf} and techniques of Artificial Intelligence \cite{ETSI:5G-MEC_AI}.
The challenges for the MEO are similar to the challenges for the NFVO, the same best practice should be followed \cite{NFVO-Dependability-Survey}.
For eliminating the single point of failure, the MEO can be designed as logically centralized but physically distributed, as for the SDN controllers \cite{sdn-consistency}. In this case, the coordination among the different MEOs becomes critical.

\paragraph{Consistency}
Consistency is defined as the property of multiple elements to have the same information and vision of the system.
The lack of consistency is a problem for a distributed implementation of the MEO, but also for naturally distributed elements as the MEH.
For example, the consistency regarding the agreement in the event of a failure needs to be addressed \cite{wang-consistency-2019}.

\subsubsection{General challenges}
\paragraph{Dependability modelling}
A first way of evaluating the dependability of a new system is to realize dependability models. For example, this approach has been used to evaluate the availability of SDN~\cite{sdn-model} and NFV \cite{nfv-model}.
A dependability model can allow to evaluate the impact of the elements composing the MEC system and identify the critical issues. Advanced models can also consider the dependability correlation bet. The more common models techniques are derived by the Petri Nets, such as the Stochastic Activity Networks and Stochastic Rewards Nets (SRN), and can be implemented by using tools such as M\"{o}bius\footnote{https://www.mobius.illinois.edu/} or SHARPE\footnote{https://sharpe.pratt.duke.edu/}. Some works have addressed the modeling of reliability and availability in edge computing \cite{2021-model-opt, 2022-model-survey}.
For example, SRN has been used to model the availability of an edge system~\cite{raei-model-2017},  a semi-Markov model has been used to evaluate the impact of VNF aging in MEC~\cite{2022-aging}, and a two-level model has been used to evaluate the availability of a MEC system \cite{thilina}.

\paragraph{Network dependability}
The network connectivity is an important element in the MEC system \cite{ETSI:MEC015}. It includes the access network to which the users are connected, the connection between the access and the MEHs, the connections between the MEHs, and the connection between the MEHs and the MEO.
Unreliable network connectivity can have a huge impact on the dependability of the MEC system \cite{ETSI:5G-MEC_Enterprise,OpenStack}.
Reliable network connectivity can be provided via physical redundancy and dedicated protocols.

\paragraph{Service continuity and user mobility}
One property that is carefully addressed in the ETSI specification is the service continuity \cite{ETSI:MEC018,ETSI:MEC022}.
The importance is also due to the nature of MEC, or more precisely MEC with mobile access networks. 

ETSI defines the service continuity as \emph{"the perception of the out-of-service time and can be measured by the latency between the terminated instance and the resumed instance of the same service, maintaining the instance state"} \cite{ETSI:MEC018}.
The causes of a lack of continuity are application software failure, malfunction of MEC system, loss of connectivity due to network failure, and user's voluntary or involuntary action. 

Depending on the scenario and the application, ETSI defines different levels of service continuity: no continuity, low continuity, soft continuity, and hard continuity \cite{ETSI:MEC018}.

\paragraph{Failure management}
MEC might introduce more complex failure modes \cite{dependability-bagchi-2019,roadmap-bakshi-2019}.
This situation is a common problem in modern complex ICT systems \cite{complex}.
Moreover, faults and failures in MEC (as generally in computing platforms) are hard to detect \cite{reliability-madsen-2013}.

The fault management is diffusely addressed in the ETSI specifications \cite{ETSI:MEC010_1}.
Anyway, there are aspects that need to be properly addressed: uncontrolled error propagation by defining error-containment regions; recovery of faulty components\cite{roadmap-bakshi-2019, 2022-recovery}; extreme event control \cite{elbamby-wireless-2019}.

\paragraph{Dependable architecture}
Beyond the ETSI architecture, alternative edge computing architectures have been proposed in order to improve dependability.
One architecture aims to deliver failure resistant and efficient applications~\cite{le-arch-2017}. Another work proposes a dependable edge computing architecture customized for smart construction \cite{kocho-arch-2018}.

\section{Performance} \label{sec:performance}


The performance is usually the main target of a new technology, where security and dependability are aspects that need to be guaranteed. For this reason, the research community has focused on the performance of MEC and many works are on the topic although the number of surveys on MEC performance is limited.

\subsection{Taxonomy}

\begin{figure}[!t]
  \centering
  \includegraphics[width=0.48\textwidth]{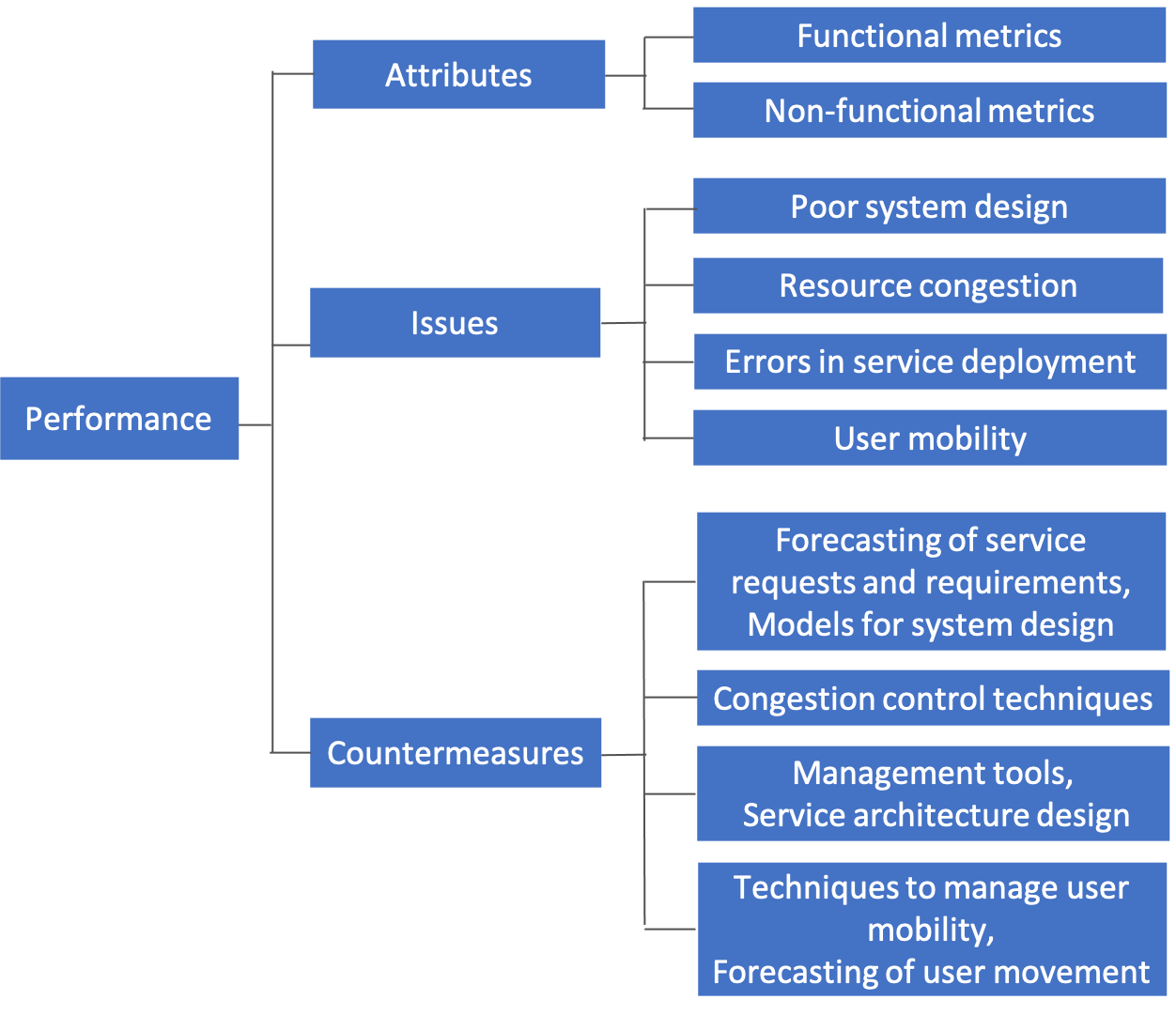}
  \caption{Performance Taxonomy}
  \label{fig:perf-taxonomy}
\end{figure}

Given the wide nature of the performance, this subsection presents the performance taxonomy specific to MEC.

Figure \ref{fig:perf-taxonomy} represents the taxonomy related to the performance. 
The \emph{attributes} are the various metrics to evaluate the performance of a system.
The \emph{issues} are the causes that may lead to a lack of performance.
The \emph{countermeasures} are the methods to address the performance issues.

The attributes are presented first from a general perspective, then according to what has been defined by ETSI~\cite{ETSI:MEC006}.
The issues and the countermeasures are related to a networking and computing context (which MEC belongs to) because having a general perspective would have resulted in a too broad introduction.

\subsubsection{Attributes}

The performance of a system can be evaluated by means of metrics. There are different well-known metrics that can be divided into two classes. A first class contains general metrics focused on only one aspect, such as data transport service. The following well-known metrics belong to this class:
\begin{itemize}
\item \emph{Throughput} is the number of bits or messages successfully delivered per unit of time.
\item \emph{Latency} is the delay introduced for completing a service, e.g., delivering a block of data between two points of the system. 
\item \emph{Jitter} allows quantifying the latency variation.
\item \emph{Loss rate} refers to the number of bits or messages per unit of time that are lost during the service.
\end{itemize}
All these metrics can be measured at different layers of the protocol stack, depending on the particular service considered in the performance evaluation. For example, measuring the latency at the network layer allows us to quantify the delay introduced only by the network. This kind of metric is not sufficient to quantify the performance of a Voice-over-IP (VoIP) service because it does not take into account the latency added by the application during the elaboration of the received bits~\cite{Malas_2011}.
The second class is composed of metrics that aim to summarize in a value the performance of a complex service that requires the interaction of a set of components. Most of the metrics belonging to this class are related to the concept of Quality of Service (QoS) or Quality of Experience (QoE). These metrics can be classified as follows~\cite{Malas_2011}: 
\begin{itemize}
\item \emph{Subjective metrics}, which require the involvement of humans for quantifying the experimented performance of the service. The quantification is performed using some reference scale, such as that defined in the Mean-Opinion-Score (MOS) procedure and its evolution~\cite{Tobias_2016}.
\item \emph{Objective metrics}, which allow quantifying the performance by using machine-executable algorithms, such as PSNR (Peak Signal-to-Noise Ratio used for quantifying the video signal quality). 
\end{itemize}

Starting from this general definition, the ETSI document~\cite{ETSI:MEC006} has defined a set of metrics specifically designed for the MEC systems. Given the high flexibility of MEC architecture, the MEC metrics have been defined with the following two goals:

\begin{itemize}
  \item Evaluate the performance increase given by a MEC solution with respect to a non-MEC one;
  \item Compare the performance of different MEH locations within the network in order to select the most suitable MEH for the considered use case.
\end{itemize}

\begin{figure}[!t]
  \centering
  \includegraphics[width=0.5\textwidth]{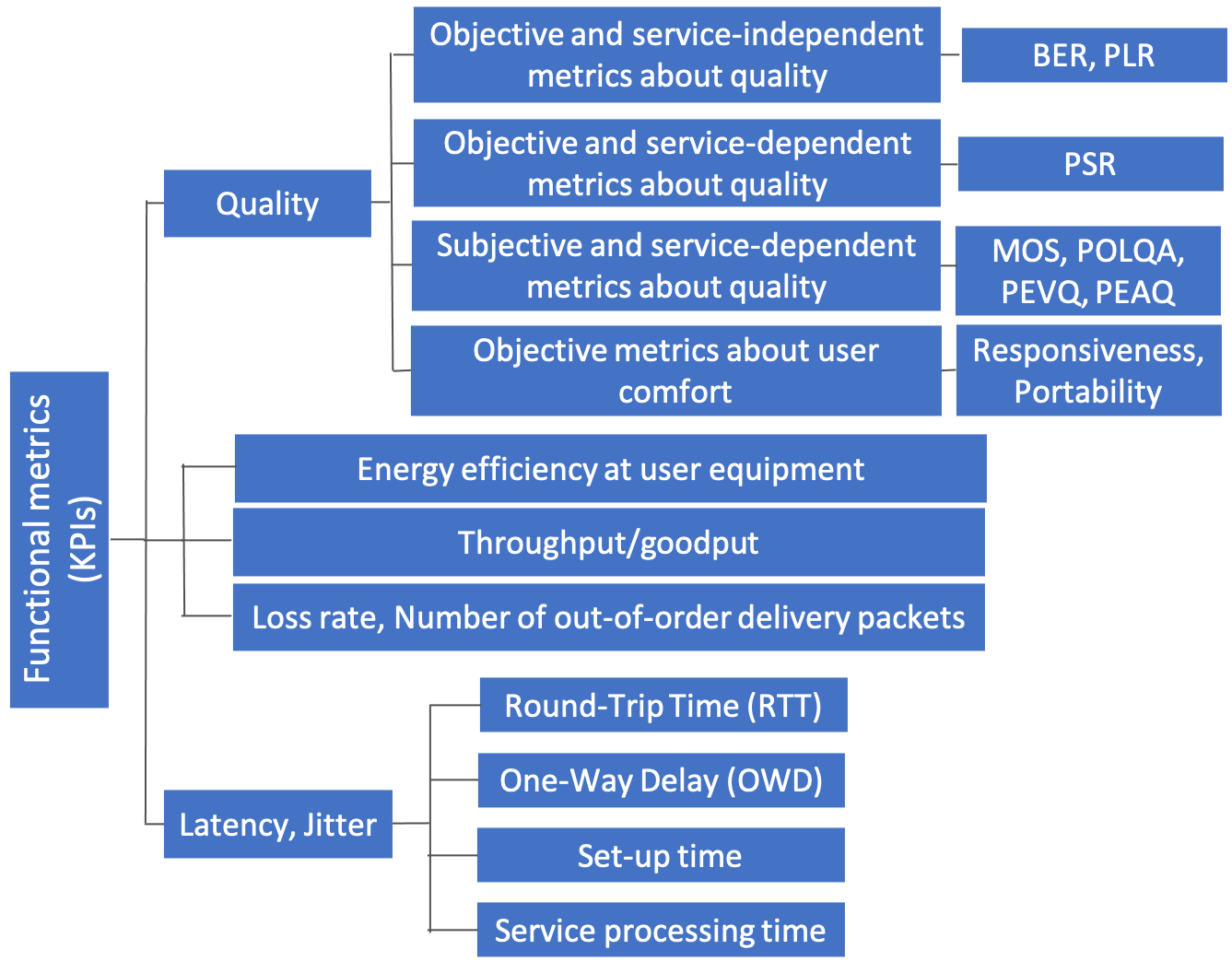}
  \caption{Functional Metrics} 
  \label{fig:Functional}
\end{figure}

\begin{figure}[!t]
  \centering
  \includegraphics[width=0.47\textwidth]{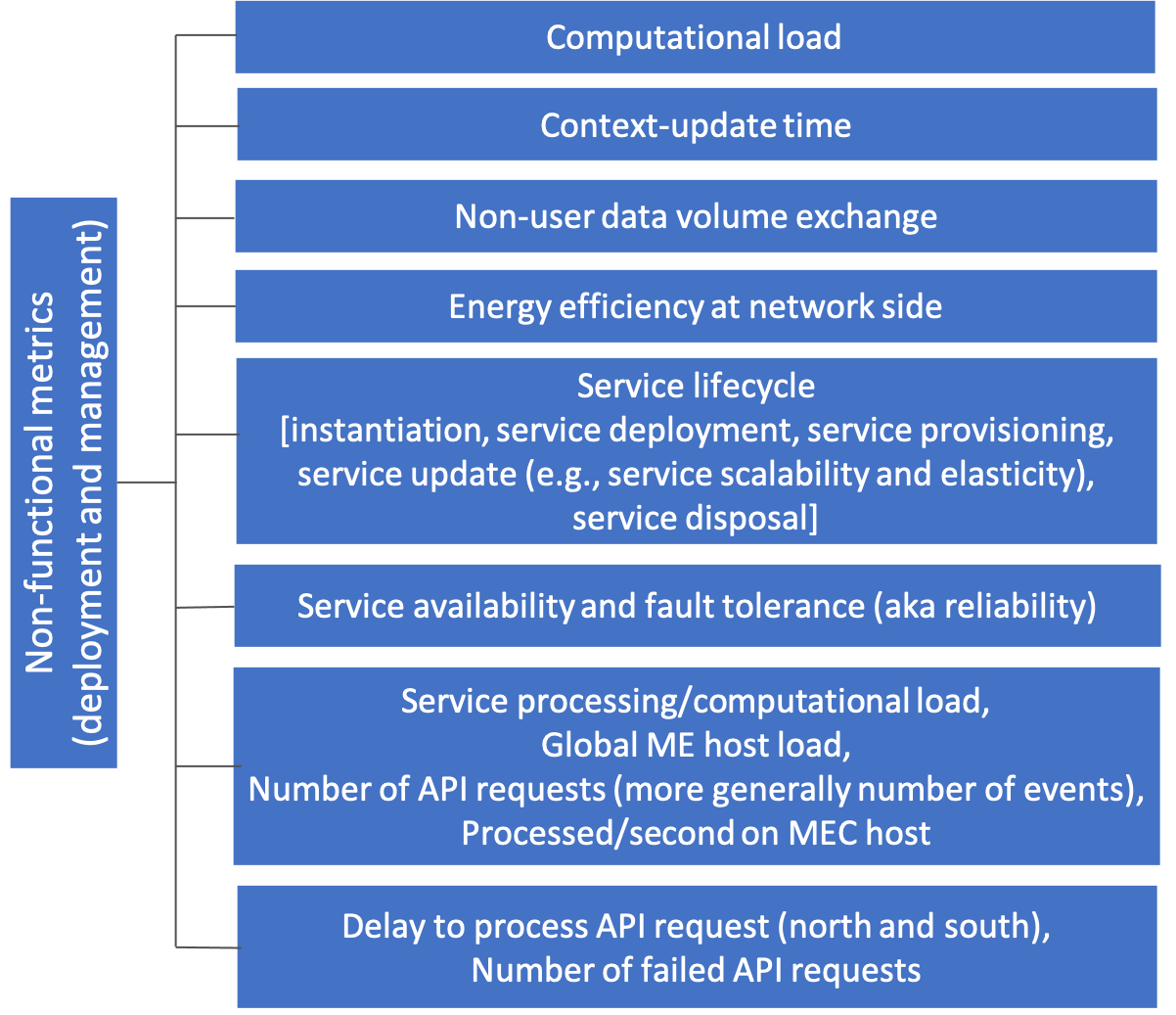}
  \caption{Non-Functional Metrics} 
  \label{fig:NonFunctional}
\end{figure}

Taking into account these goals, ETSI defined two classes~\cite{ETSI:MEC006}, which we will further present by summering the content of the ETSI document: 
\begin{itemize}
  \item \emph{Functional metrics}: they quantify MEC performance impacting on user perception (often called Key Performance Indicators, KPIs). The set of KPIs is shown in Figure~\ref{fig:Functional}.
  \item \emph{Non-functional metrics}: they are related to the performance of the service in terms of deployment and management. Figure~\ref{fig:NonFunctional} describes the set of non-functional metrics.
\end{itemize}

As shown in Figure~\ref{fig:Functional}, the functional metrics consider the performance by both taking into account single aspects of the service (i.e., latency, jitter, loss rate, throughput, and energy efficiency) and referring to metrics that summarize the interaction of multiple components forming a complex service (i.e., all the metrics under the umbrella indicated as \emph{quality}). The non-functional metrics in Figure~\ref{fig:NonFunctional} are strictly related to the aspects that are important for the network operator, i.e., deployment and management aspects. In some cases, these aspects are not related to the performance perceived by the users, such as \emph{non-user data volume exchange} or \emph{energy efficiency at network side}. Indeed, these performance parameters impact the costs necessary for the network operator to manage the service. On the other hand, some metrics, such as \emph{delay to process API request} and \emph{number of failed API requests}, negatively impact the user experience. Indeed, a high delay in processing API requests can increase the \emph{set-up time} of a new service request and/or the \emph{service processing time}. 
For both classes, the metrics need to be adapted to the particular MEC use case. In particular, the actual assessment of these metrics can depend on the particular service and/or application. For example, the latency in localization (time to fix the position) is different from the latency in content delivery.
In both cases, one could measure all the statistics over the above metrics. In fact, all metrics are in principle time-variable and could be measured in a defined time interval and described by a profile over time or summarized through the following values: maximum value, mean and minimum value, standard deviation, and value of a given percentile.

\begin{table*}[!t]
\caption{Summary of Performance Issues and Countermeasures}
\begin{center}
\begin{tabular}{p{2.9cm}p{6.9cm}p{6.8cm}}
\toprule
\textbf{Issue} & \textbf{Description} & \textbf{Countermeasure} \\
\midrule
\textbf{System design} & \vspace{-0.3 cm}
\begin{itemize}[leftmargin=*]
    \item \textbf{Poor networking, memory and computation resources} negatively impact latency, loss rate, throughput and quality.
    \item \textbf{Bad resources location} can negatively effects the performance also in the case of resource overprovisioning (but badly located).
\end{itemize} 
& \vspace{-0.3 cm} \begin{itemize}[leftmargin=*]
\item \textbf{Forecasting of the necessary resources} for each class of services, defined by taking into account the performance requirements and the offered load. 
\item Usage of \textbf{models for system design} starting from the forecasted service requests and requirements. 
\end{itemize}\\
\midrule
\textbf{Resources congestion} & \vspace{-0.3 cm}
\begin{itemize} [leftmargin=*]
\item The \textbf{resources congestion} can deeply impact the latency, throughput, loss rate, and quality at different layers. 
\end{itemize} 
& \vspace{-0.3 cm}
\begin{itemize}[leftmargin=*]
\item \textbf{Reactive congestion control} techniques aim to \textbf{detect the congestion status} of the system and then react by using strategies aimed at reducing the time for which the system is in this state, in order to reduce the negative effects of a congestion event on the user service. 
\item \textbf{Proactive congestion control} techniques aim to prevent the congestion state of the system resources.
\end{itemize}\\
\midrule
\textbf{Service deployment} & \vspace{-0.3 cm}
\begin{itemize}[leftmargin=*]
\item \textbf{Service deployment errors and problems}: at the communication layer (e.g., IP address plan or routing protocols configuration errors, and radio channel selections) and at the service layer (e.g., software and hardware architecture selected for the service deployment). 
\item The network softwarization and, in general, the service virtualization add further issues related to the used \textbf{virtualization technology}: these technologies differ in terms of required physical resource (e.g., memory required by the image and CPU utilization), and offered performance (e.g., latency and throughput). 
\end{itemize}
& \vspace{-0.3 cm}
\begin{itemize}[leftmargin=*]
\item \textbf{Usage of management and monitoring tools} at different layers helps to detect and solve issues related to deployment errors. 
\item These performance issues can be avoided by \textbf{proactively evaluating the performance of alternative hardware and software architectures}, e.g. service implementation in dedicated hardware, used CPU and OS, connections of different components of complex services, used virtualization technique (e.g., hypervisor-based, containers, and Unikernel) in the case of service implemented in general-purpose hardware.
\end{itemize}\\
\midrule
\textbf{User mobility} & \vspace{-0.3 cm}
\begin{itemize}[leftmargin=*]
\item \textbf{High variability of the edge resource utilization} can lead to temporary congestion, degradation of latency and, in some cases, loss of data. 
\item The \textbf{user may move away} from the location of the resources supporting the requested services, degrading the performance and, in some cases, interrupting the service when there is no (or insufficient) resource in the new position. 
\end{itemize}
& \vspace{-0.3 cm}
\begin{itemize}[leftmargin=*]
\item \textbf{Definition of techniques for managing user mobility} at network layer, such as the handover management defined in mobile networks, at service layer, such as live service migration permitted by agile virtualization technology. 
\item In both cases, \textbf{forecasting the user's movements} can allow proactive reservation of resources and the initiation of all operations related to connection handover and/or service migration, thereby reducing the issues related to user mobility significantly.
\end{itemize}\\
\bottomrule
\end{tabular}
\label{tab:performance_issues}
\end{center}
\end{table*}

\subsubsection{Issues}

\emph{Poor system design} is one of the most important performance issues. The system design deals with the definition of the resources (and their location) necessary for supporting the user services. For example, if there are no (or insufficient) resources in the area where the user requires URLLC services, the service may need to be supported on a remote server, where the propagation delay is higher than the maximum accepted delay. The remoteness of resource location for video-on-demand services implies the involvement of many network resources with consequently a consumption increase of communication resources and energy.

\emph{Resources congestion} can impact the different performance parameters, depending on the kind of congested resources. For example, the congestion of communication resources increases the latency for data transferring, the lack of storage resources can add data loss, whereas insufficient computation resource adds delay in processing data.

\emph{Service deployment} can manifest problems at the network layer and service layer due to configuration errors or bad architecture (hardware and software) choices, which have a deep impact on performance.

\emph{User mobility} adds variability to the features of the connection with the resources providing the service. The user movement can degrade the data rate available at the access network and increase the latency for achieving the service location. In some cases, in the new user location, there are no resources to support the service, leading to service interruption.

\subsubsection{Countermeasures}

To solve the presented issues, there are some general approaches. \emph{Poor system design} issues can be coped with a preventive analysis of the amount of services and related features, in terms of offered load and service requirements. This information is then used as input in models developed for system design. The models differ from the main target of the design. The most common target is to reduce the capital expenditures (CAPEX) and the operating expenses (OPEX) by maximizing the number of services with satisfied requirements. Recently, some models aim to reduce energy consumption. 

To cope with the \emph{resources congestion} issues, there are two big classes of congestion control approaches: reactive and proactive. The TCP congestion control is an example of a reactive approach, whereas admission control of new service requests is an example of proactive \emph{congestion control}. This control consists in rejecting new service requests when their admission degrades the performance of the already accepted services. The congestion is prevented by analyzing the available resources and estimating the required ones by the newly requested service. 

The usage of management and monitoring tools at different layers represents an important countermeasure against service deployment problems. They help to detect and solve issues related to device, network, and service configuration. The analysis of alternative hardware and software architectures is necessary to prevent performance issues related to bad choices on deployment aspects, such as service implementation in dedicated hardware, the connection among the different components of a complex service, and virtualization technology used for service (or of one component) implementation in a central or distributed cloud.

The issues related to \emph{user mobility} are well-known in mobile networks, where different techniques have been designed to maintain the network performance during user movements. Similar issues need to be considered at the service layer in the case of critical services, where the time necessary to achieve the resources offering the services is higher than the minimum latency requirements. In this case, the service migration is necessary. In all cases, \emph{forecasting the user movements} can allow for proactively reserving resources and starting all operations related to the connection handover and/or service migration, reducing the issues related to user mobility. 

A summary of performance issues and countermeasures is shown in Table~\ref{tab:performance_issues}.

\subsection{State of the Art}

\subsubsection{Standards, regulations, and white papers}

Table \ref{tab:MEC-per-req} shows the ETSI specifications that represent the state-of-the-art of performance requirements and regulations for MEC. This list has been obtained by analyzing the documents shown in Table~\ref{tab:etsi_standards}). 
Note that N/A here (as in Table~\ref{tab:MEC-sec-req} for the security and Table~\ref{tab:MEC-dep-req} for dependability) does not mean that performance requirements are out of scope or importance for the given element; N/A means that no explicit requirements are listed in the specifications.

Several API-focused standards \cite{ETSI:MEC015,ETSI:MEC021} describe some performance requirements to implement in some MEC elements.
The standards that define the performance metrics is \cite{ETSI:MEC006}.
A lot of use cases with related performance requirements are described in \cite{ETSI:MEC002}. Whereas studies for the MEC support of some important services and technical aspects, such as network slicing \cite{ETSI:MEC024}, NFV \cite{ETSI:MEC017}, mobility~\cite{ETSI:MEC018}, and alternative virtualization technologies \cite{ETSI:MEC027}, indicate the performance required to the MEC elements. An important document describes the MEC 5G integration \cite{ETSI:MEC031}. 

The white papers listed in Table \ref{tab:etsi_standards} mainly discuss the problems and solutions related to the MEC deployment with 4G, 5G, and cloud RAN (CRAN) \cite{ETSI:5G-MEC_Evol}\cite{ETSI:5G-MEC}\cite{ETSI:5G-MEC_Cloud}. In these documents, some performance issues are presented, but no explicit performance requirements are clearly defined for the MEC elements. The enhanced DNS can have a key role in the performance of MEC services because it enriches the list of deployment options suitable to support the distributed MEC environment, in terms of providing the connectivity between devices and application instances \cite{ETSI:5G-MEC_dns}.

\renewcommand{\labelitemi}{-}
\begin{table*}[!t]
\caption{Performance Requirements in the MEC Architecture According to Standardization Documentation}
\begin{center}
\begin{tabular}{p{1.7cm}p{15.1cm}}
\toprule
\textbf{Element} & \textbf{Performance Requirements}\\
\midrule
\multicolumn{2}{l}{\textbf{General}}\\
 &
 \vspace{-0.3cm}
\begin{itemize}[leftmargin=*]
	\item A large set of use cases is described with the goal of deriving useful requirements. However, some requirements are defined by design constraints and do not originate from use cases \cite{ETSI:MEC002}.
	\item A number of performance metrics that can be used to demonstrate the benefits of deploying services and applications on a MEH are presented in \cite{ETSI:MEC006}.
	\item The MEC 5G integration implies several key issues. The issues and their related solution proposals are discussed in \cite{ETSI:MEC031,ETSI:5G-MEC}.
	\item The current QoS-related information in RNI API is not sufficient in order to allow necessary prediction regarding the QoS performance (e.g. latency, throughput, reliability). Therefore, potential enhancements on RNI API for the prediction should be considered including both relevant measurements in RAN or processed results for the prediction \cite{ETSI:MEC022}.
\end{itemize} 
\\
\midrule
\multicolumn{2}{l}{\textbf{MEH}}\\
\vspace{-0.3cm}
MEP &
\vspace{-0.3cm}
\begin{itemize}[leftmargin=*]
\item The MEP on a MEH provides a framework for delivering MEC services and platform essential functionality to MEC applications running on the MEH \cite{ETSI:MEC002}.
\end{itemize} 
\\
Virtualization Infrastructure &
\vspace{-0.3cm}
\begin{itemize}[leftmargin=*]
	\item Multiple Alternative Virtualization Technologies (AVTs) need to be supported in the same MEH to satisfy the performance requirements of heterogeneous MEC services \cite{ETSI:MEC027}. The different AVT solutions have an impact on MEC framework and on MEC management APIs \cite{ETSI:MEC027}.
\end{itemize} 
\\
MEC App &
\vspace{-0.3cm}
\begin{itemize}[leftmargin=*]
	\item When MEC is deployed in an NFV environment, the most simple possibility to realize the Data Plane needs the support of a physical network function or VNF or a combination thereof, and its connection to the network service that contains the MEC app VNFs \cite{ETSI:MEC017}.
\end{itemize} 
\\
\midrule
\multicolumn{2}{l}{\textbf{MEC Host-level Management}}\\
MEPM &
\vspace{-0.3cm}
\begin{itemize}[leftmargin=*]
	\item The MEPM receives Virtualized resources fault reports and performance measurements from the VIM for further processing \cite{ETSI:MEC003}.
	\item In the NFV variant, the MEPM-V does not receive virtualized resources fault reports and performance measurements directly from the VIM, but these are routed via the VNFM \cite{ETSI:MEC003, ETSI:MEC017}. 
	\item The MEPM shall be able to collect and expose performance data regarding the virtualization environment of the MEH related to a specific MEC app and a specific MEC app instance \cite{ETSI:MEC002}. These data can be used to verify how well the Service-Level Agreements (SLAs) are met.
	\item Different MEC applications running in parallel on the same MEH may require specific static/dynamic up/down bandwidth resources, including bandwidth size and bandwidth priority. To this aim, optional traffic management services are described in \cite{ETSI:MEC015}.
	\item The MEPM should be able to provide different sets of features/services in distinct Network Slice Instances (NSIs) \cite{ETSI:MEC024}.
	\item The MEPM should be able to provide the same feature/service differently in distinct NSIs \cite{ETSI:MEC024}.
	\item The MEPM collects usage and performance data per NSI \cite{ETSI:MEC024}.
    \item The MEPM-V can access the performance management and fault information of virtualized resources related to a particular ME app VNF instance that is lifecycle-managed by the VNFM \cite{ETSI:MEC017}.
\end{itemize} 
\\
VIM & 
\vspace{-0.3cm}
\begin{itemize}[leftmargin=*]
	\item The VIM collects and reports the performance and fault information about the virtualized resources \cite{ETSI:MEC003}.
\end{itemize} 
\\
\midrule
\multicolumn{2}{l}{\textbf{MEC System-level Management}}\\
MEO &
\vspace{-0.3cm}
\begin{itemize}[leftmargin=*]
	\item The MEC system management should support the management of MEHs, including suspend, resume,
configure, add and remove, by an authorized third-party \cite{ETSI:MEC002}.
    \item The functional requirements related to application mobility are reported in \cite{ETSI:MEC021}. Examples of functional requirements are the maintaining of the connectivity between a UE and an application instance when the UE performs a handover to another cell associated/not associated with the same MEH, the support of two instances of a MEC application running on different MEHs to communicate with each other, or the use of radio/core network information for optimizing the mobility procedures required to support service continuity.
    \item The MEO is able to trigger the application relocation owing to external relocation request, unsatisfied performance requirements, load balancing and disaster recovery \cite{ETSI:MEC018}.
    
\end{itemize} 
\\
MEAO & 
\vspace{-0.3cm}
\begin{itemize}[leftmargin=*]
	\item The MEC system management shall be able to expose up to date performance data of the application to the authorized third-parties such as application developers and application providers \cite{ETSI:MEC002}.
	\item When MEC is deployed in an NFV environment, for performance enhancements, it can make sense to re-use the SFC functionality provided by the underlying NFVI for traffic routing. Differently from the simple solution based on the MEC App VNFs, in such a deployment, the Data Plane is obtained using the MEAO without a dedicated MEC App. The MEAO will need to translate the traffic rules into a Network Forwarding Path (NFP) and send it to the NFVO. The MEP will not control the traffic redirection directly but will pass requests to activate/deactivate/update traffic rules to the MEPM-V which will forward them to the MEAO. When receiving such a request, the MEAO will request the NFVO to update the NFP accordingly \cite{ETSI:MEC017}.
\end{itemize} 
\\
OSS &
\vspace{-0.3cm}
\begin{itemize}[leftmargin=*]
	\item N/A 
\end{itemize} 
\\
User app LCM proxy &
\vspace{-0.3cm}
\begin{itemize}[leftmargin=*]
	\item N/A
\end{itemize} 
\\
\midrule
\multicolumn{2}{l}{\textbf{MEC Federation}}\\
MEF &
\vspace{-0.3cm}
\begin{itemize}[leftmargin=*]
	\item N/A
\end{itemize} 
\\
\bottomrule
\end{tabular}
\label{tab:MEC-per-req}
\end{center}
\end{table*}

\subsubsection{Academic publications}

We can find the discussion of performance issues (and, in some cases, related solutions) of MEC deployment in many of the technical surveys dedicated to MEC. Among these, in Table~\ref{tab:perf_surveys}, we summarize the most updated and important from the performance perspective.

\begin{table*}[!t]
\caption{Academic Surveys Related to MEC Performance}
\begin{center}
\begin{tabular}{p{0.6cm}p{2cm}p{0.5cm}p{6.3cm}p{6.3cm}}
\toprule
\textbf{Ref.} & \textbf{Aspect} & \textbf{MEC only} & \textbf{Main contribution} & \textbf{Relevance to MEC performance} \\
\midrule
\cite{taleb_multi-access_2017} &MEC general &Yes &Surveys the use cases and the enabling technologies. Focuses on orchestration and related challenges. & Discusses the role of MEO in orchestrating resource allocation and service placement for assuring efficient network utilization and QoE. Provides a qualitative comparison of different orchestrator deployment options.\\
\cite{mao_survey_2017} &MEC communication &No &Focuses on joint radio-and-computational resource management. & Provides a summary of future research directions for the joint radio-and-computational resource management referring to the performance issues related to the deployment of MEC systems, the Mobility management for MEC, and the Green MEC.\\
\cite{shahzadi2017multi} &Edge general &No &Provides an overview of the state of the art and the future research directions for edge computing. & Provides a comprehensive review and comparison of the prevalent MEC frameworks. The comparison is based on different parameters, such as system performance, network performance, overhead of deployment, and system migration overhead. \\
\cite{yu_survey_2018} &Edge general &No &Presents an overview of potential, trends, and challenges of edge computing referring to the IoT Applications. &Discusses performance challenges related to IoT applications. \\
\cite{porambage_survey_2018} & MEC for IoT & Yes & Holistic overview on the exploitation of the MEC technology for the realization of IoT applications. & Presents research topics related to MEC service level congestion control, latency-aware routing, and dynamic application routing.\\
\cite{khan2019edge} &Edge general &No &Surveys edge computing, identifies requirements, and discusses open challenges. &Presents a comparison of MEC solutions aimed to optimize the execution cost and deployment, reduce network latency, minimize energy consumption, and maximize throughput.\\
\cite{pham_survey_2020} &MEC general & Yes &Surveys MEC fundamentals, discussing its integration within the 5G network and relation with similar UAV communication, IoT, machine learning, and others. & Discusses the lessons learned, open challenges, and future directions on the performance of MEC integrated with the forthcoming 5G technologies.\\
\cite{filali-MEC-surv-2020} &MEC general &Yes &Surveys the MEC and focuses on the optimization of the MEC resources. & Provides a state-of-the-art study on the different approaches that optimize the MEC resources and its QoS parameters (i.e., processing, storage, memory, bandwidth, energy, and latency).\\
\cite{elbamby-wireless-2019} &Wireless edge &No &Discusses the feasibility and potential of providing edge computing services with latency and reliability guarantees. &Overviews the challenges and the enablers for achieving low-latency and high-reliability networking; discusses network resources optimization techniques for a selection of use cases.\\
\cite{liu-vehicular-edge-2020} &Vehicular Edge Computing & No &Surveys the state of the art of vehicular edge computing. & Presents a summary of techniques for caching, task offloading, and data management highlighting the considered performance parameter.\\
\cite{Moura2019} & MEC and Game Theory & Yes & Discusses how Game Theory can address the emerging
requirements of MEC use cases. & Overviews of Game Theory models for achieving a performance-cost balance in realistic edge network scenarios, and prospected future trends and research directions in the application of game theory in future MEC services. \\
\cite{Qiu2020} & Edge and IIoT & No & Discusses the opportunities and challenges of edge computing in IIoT. & Overviews of schemes for routing and task scheduling for performance improvements. Describes challenges and potential solutions of applying some new technologies to edge-based IIoT.\\
\cite{mancuso2020} & MEC for Industrial Verticals & Yes & Explores how the MEC is used and how it will enable industrial verticals. & Discusses MEC deployment bottlenecks and related solutions for performance improvements in a smart metropolitan area, where disparate verticals can coexist.\\
\bottomrule
\end{tabular}
\label{tab:perf_surveys}
\end{center}
\end{table*}


\subsection{Challenges}
In the following, we present performance challenges for MEC by categorizing them on MEC host level, MEC system level, and general challenges.

\subsubsection{MEC host level}

\paragraph{Virtualization performance}
The MEC architecture allows not only to improve the performance of network functions implemented in the form of VNF, close to the users, but also to be prepared to host third-party services, creating a new market for the operator~\cite{Mijumbi_2016}. The virtualization technologies can deeply impact the performance of the MEC system, as shown by some works (such as the most recent \cite{Aggarwal_2020,Perez_2021, Valsamas2022} and reference therein), which presented an experimental comparison of virtualization technologies for implementing VNF and edge services. This well-known result spurs the study of new virtualization technologies and deployment paradigms in order to improve performance and resource efficiency. From container-based virtualization to micro virtual machines, new virtualization solutions claim to offer performance close to bare metal, with quick deployment and startup times. 

Recent works analyzed the performance of multiple virtualization technologies, including VMs, containers, unikernels, and Kata Containers \cite{Aggarwal_2020}.
VMs have traditionally been the primary technology for VNF deployment, creating an isolated and secure environment with high associated overhead. To reduce the memory clutter of VMs, containers represent an interesting alternative solution, as they pack the application and its dependencies into a light and agile entity that can be run on any platform \cite{Bentaleb2022}. However, due to the underlying shared kernel, containers face security problems in a multi-tenant environment \cite{Manco_2017}. To address the security issues related to the shared container kernel, Kata Containers have recently been proposed. Kata Containers act and perform like classic containers but provide stronger workload isolation by using hardware virtualization technology as a second layer of defense~\cite{katacont_2019}.
Unikernels are lightweight machine images that enclose the application and require only OS libraries and dependencies. It works in a single address space and has a much smaller attack surface due to its minimal nature.
Each of these technologies faces a trade-off between isolation and agility. The studies presented in~\cite{Aggarwal_2020,Behravesh_2019,Perez_2021} show the performance in terms of the following parameters:
\begin{itemize}
\item Image size of the service, which impacts the amount of storage required to host the application; 
\item CPU and memory utilization of the service, which has a great impact on the number of services that a physical server can run simultaneously;
\item Throughput of service requests;
\item Delay in responding to user requests, which can drastically degrade the user's satisfaction with the service (a website in the experimental analysis).
\end{itemize}

For example, the results of \cite{Aggarwal_2020} show a boot time of the container (0.623 s) lower than that of  VM (32 s), whereas the service throughput of the implementation of a HTTP server with VM (130 req./s) is higher than that of Container (143.4 req./s)

Kata Containers and unikernels are in their development phase, but they provide serious competition to the ecosystem of containers and VMs. Indeed, they offer a lightweight solution for the deployment and migration of virtualized services, improving the performance in terms of latency, throughput, and quality. These improvements are more evident in the presence of user mobility, where the agility and the lightweight of the virtualized services are key factors. However, Kata Containers are not mature yet for using them in production environments managing data-intensive workloads, as shown by the experimental results presented in \cite{Perez_2021}. Indeed, this paper confirms the results shown in~\cite{Aggarwal_2020}: Kata Containers are not really efficient in terms of memory consumption and speed, while still being a good deployment choice in security-sensitive multi-tenant environments.

\paragraph{Performance isolation}
As shown in \cite{Gupta_2006} performance isolation can be a very critical issue to take into account when VM shares the physical resources. With some initial experiments to understand the co-existential or neighbor-dependent behavior of VMs, the authors inferred that the problem of performance isolation can be improved by reducing the resource contention amongst the VMs on the same physical host. Performance isolation can be drawn to the lowest level abstraction of shared resources, like CPU, memory, network, and disk. For example, the disk is continuously being used by multiple processes waiting in the I/O queue. Thus, the I/O scheduler will play a vital role in resource contention, as studied in \cite{Cherkasova_2007}\cite{Li_J_2019}. In~\cite{KOTULSKI_2020}, the authors present a systematic overview of existing isolation techniques in nodes and networks, especially in RAN and CN of 5G systems.

\subsubsection{MEC system level}
\paragraph{MEC deployment}

The network operator selects the number of MEHs and their location analyzing different technical and business parameters, such as available site facilities, supported applications and their requirements, measured or estimated user load, etc. As shown in Figure~\ref{fig:5G-MEC-depl}, the network operator has four different deployment options.

MEHs can be deployed in different network locations: from near the gNB to a remote data network. Although running MEHs far from the edge can be useful in scenarios in which compute power requirements are stricter than latency ones, the most interesting scenario is represented by locating MEHs close to the user (e.g., at the gNBs of a 5G system). This scenario adds two very important features for enabling new services: 

\begin{itemize}
 \item the reduction to low values of the delay between the end-user device and the MEH hosting the application, which enables low-latency services; 
 \item the access to user context, such as the user channel quality conditions or user location, which enables context-aware services, e.g. services adaptive to network conditions.
\end{itemize}

The design of MEHs location and the instantiation of MEC services require the analysis of multiple trade-offs for efficient usage of physical and virtualized resources. Depending on the use case, the processing power demands and the latency requirements can be very heterogeneous. Other than the QoE of the users, the location of MEHs must consider the Total Cost of Ownership (TCO). On one hand, the centralized cloud decreases the TCO. On the other hand, the centralized cloud fails to address the low latency requirements. 
A good trade-off can be achieved by utilizing existing infrastructures such as Telco towers, central offices, and other Telco real estates. In the case of a mobile network, MEHs can be located in various parts of the network architecture, ranging from uCPEs (mainly at customer premises) to RAN-edge (co-located with base stations), Smart Central Offices (where the MEH could be hosted co-located with CRAN aggregation points), or edge data centers at the local/regional level.

The deployment problem has been analyzed by considering different optimization approaches and performance parameters. For example, using Shanghai Telecom’s base station dataset, some works consider the MEH placement problem with the goal of minimizing the energy consumption~\cite{Li_Y_2018_41_partenza} or balancing the workloads of MEHs while minimizing the access delay between the mobile user and MEH~\cite{WANG_2019_23_partenza}\cite{Kasi_2020}. Other recent works aim to minimize the cost of service providers while guaranteeing the completion time of services~\cite{Li_bo_2021} or to minimize the number of MEHs while ensuring some QoS requirements~\cite{Zeng_20_partenza}. In general, the deployment of MEHs requires defining multi-objective problems, such as i) finding a Pareto front optimizing the time cost of IoT applications, load balance, and energy consumption of MEHs~\cite{WANG_2019_23_partenza}, ii) optimizing the response time taking into account heterogeneity of MEC/cloud systems and the response time fairness of base stations, which may significantly degrade the system quality of services to mobile users~\cite{Cao_K_2021}, iii) finding an optimized trade-off between response delay and energy consumption~\cite{Li_bo_2020}. An interesting issue is the update of the MEC infrastructure. As an example, in ~\cite{Loven_2020} the authors consider the problem of scaling up an edge computing deployment by selecting the optimal number of new MEHs and their placement and re-allocating access points optimally to the old and new MEHs. In this case, the considered performance is the Quality of Experience of users and the QoS of the network operator. As concerning the integration MEC-5G, an interesting problem is the determination of the MEH and UPFs optimal number and locations to minimize overall costs while satisfying the service requirements~\cite{Leyva_2019_29_partenza}.

\paragraph{Resource allocation}
Taking into account the available resources, the network operator can solve the service placement problem. Placing MEC services over a number of MEHs can prove to be critical for the user QoE and should take into account gravity points, e.g., shopping malls, which attract a plethora of users. Indeed, a bad design of the service placement can lead to the saturation of the available resources with the consequent rejection of service requests or a worsening of the user experience. 
The flexible availability of resources plays a crucial role in the performance of a service. For example, wireless bandwidth and computing resources can be dynamically assigned to a service for achieving optimal benefits from MEC system. Orchestrating a MEC system in terms of resource allocation and service placement is critical for assuring efficient network resource utilization, QoE, and reliability.

A set of works considers the data caching problem in the edge computing environment, proposing schemes that maximize the data caching revenue of the operator~\cite{Liu_2020}, or that improve content delivery speeds, network traffic congestion, cache resource utilization efficiency, and users' quality of experience in highly populated cities~\cite{Sinky_21_partenza}. Referring to the VNF architecture, the SFC and VNF placement can be performed by reducing the execution time and the resource utilization~\cite{Wang_Meng_2020}, taking into account both service requirements and the resource capacity in the edge~\cite{Gu_2020}, maximizing revenue at the network level while matching demand~\cite{siew_2021}, minimizing both energy consumption and resource utilization~\cite{Chen_2021}, or maximizing the number of user request admissions while minimizing their admission cost (i.e., computing cost on instantiations of requested VNF instances and the data packet traffic processing of requests in their VNF instances, and the communication cost of routing data packet traffic of requests between users and the MEH hosting their requested VNF instances)~\cite{Ma_Y_2020}.

\paragraph{User association}
The most simple strategy for the user-MEH association is to allocate the nearest MEH that offers the requested service. Indeed, this approach 
agrees on the proximity strategy that is desired from the performance perspective, e.g. latency and energy consumption. However, this simple strategy can lead to performance degradation when other aspects, such as the load of the MEHs, the available transmission capacity, and the users' mobility are not considered. For example, an increase in requests for MEC services in a given area can lead to an overload of MEC resources, generating performance bottlenecks. To avoid this problem, the user-MEH association strategy needs to take into account the load status of the MEC infrastructure. Different works propose optimization solutions for some metrics, such as latency and QoE, that simultaneously distribute the load between different servers, e.g.~\cite{Liu_Qi_2020}. Furthermore, in the case of mobile users, the selection of the edge cloud becomes crucial due to the uncertainty of movement and wireless conditions. In this scenario, other than the MEC resources, the user mobility impacts also the available network capacity, given the uncertainty of the number of users sharing the resources and, in the case of the wireless access network, of the wireless channel that impacts the radio resource efficiency.

The user-MEH association problem can consider different aspects, such as i) the enhancement of the user allocation rate, server hiring cost, and energy consumption by means of an association scheme able to consider that edge users’ resource demands arrive and depart dynamically~\cite{Chunrong_Wu_2020}, ii) the dynamic balancing of the computation load for neighboring MEHs~\cite{Wang_L2020}. Other approaches jointly consider the MEH placement and association problem with the goal of minimizing the deployment cost~\cite{Liu_Zhen_2020} or number of MEH~\cite{Lee_2019} while guaranteeing certain end-to-end service latency.

\subsubsection{General challenges}

\paragraph{Service continuity and user mobility}
Service mobility is a key aspect that can impact MAC performance in a mobile network scenario. Indeed, to maintain a high QoE it is of primary importance not only to establish a connection between the end user and MEC resources but also to maintain it throughout all the necessary stages. Optimal end-to-end session connectivity needs to be maintained for the entire course of service usage. To achieve this goal, the MEC services should be able to migrate quickly depending on user movements. The user movements can frequently change the anchor points of MEC services (e.g., from one MEH to another). In this scenario, ensuring optimum QoE for a delivered MEC service becomes challenging, especially for delay-sensitive applications. 
At the IP level, the Distributed Mobility Management (DMM)~\cite{Giust_2015} represents a notable solution towards managing user mobility, overcoming also the scalability and reliability drawbacks of centralized mobility schemes. At the service level, the management of IP mobility is not sufficient to avoid quality degradation due to the redirection of MEC service requests of the mobile user to a distant edge hosting the service. The MEC architecture aims for a single-hop connectivity to the service. When the user is moving away from the MEH serving its request, the MEC service should migrate from the old to a new MEH closer to the new user location for maintaining the same performance.
Procedures for service migration take time, especially when these require moving large amounts of data. The user can experience a degradation of application performance, and in some cases, the service continuity cannot be guaranteed. 
Numerous studies have addressed the problem of VM migration, but new technical challenges have emerged from the analysis of the problem from a service point perspective~\cite{Bittencourt_2015}. In particular, the time it takes to prepare a VM for the target MEH, transfer it over the network, and finally deal with the problem of changing the IP address after relocating the VM, makes it difficult to achieve service continuity.
The analysis of this problem starts with the study of the features of different virtualization technologies. Other than presenting remarkable overhead in terms of both processing and storage capabilities, hypervisor-based virtualization techniques show high latency for start-up activation and migration procedures. Container-based virtualization enables high-density deployment of services, and has limited features to support stateful service migration between different host MEHs. In both classes, the main drawbacks derive from the stateful nature of the application, which implies that the state of the application and the network stack must be preserved in the case of migration or failure. To alleviate these drawbacks, the stateful "Follow me Cloud" paradigm has been proposed \cite{Taleb_2013}. Based on this new paradigm, some recent works propose mechanisms for fast container-based live migration, such as \cite{Addad_2020,Ma_Li_2019}. In \cite{Abdah_2019}, the authors present a survey on the techniques for service migration at the edge of the network.  
The development of stateless applications relies on user inputs or distributed shared storage, avoiding the storage of internal states. This feature allows a stateless application to be replicated on different MEHs. Based on the specific offloading request and current connectivity quality, the most appropriate instance could be selected. Examples of studies on the performance of stateless migration are~\cite{Avino_2017}\cite{Doan_2019}. 

\section{Multi-aspect Challenges} \label{sec:discussion}

Dependability, security, and performance are all three important aspects in 5G MEC. These aspects are usually addressed individually but they are not independent and they can be actually conflicting (i.e., a solution for improving one aspect may impact the others).

\begin{figure}[!t]
  \centering
  \includegraphics[width=0.4\textwidth]{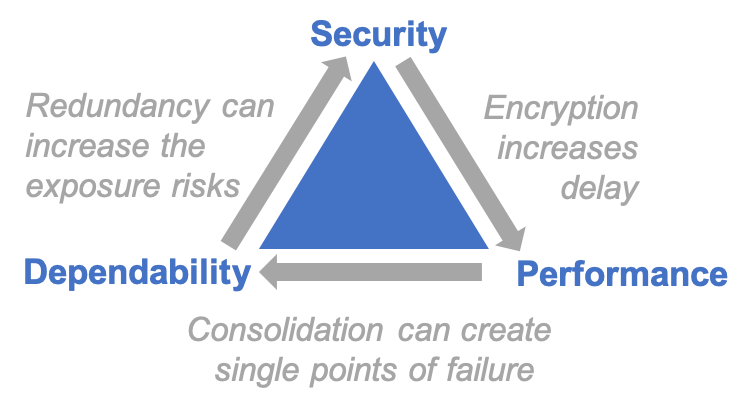}
  \caption{Examples of Conflicts}
  \label{fig:conflicts}
\end{figure}

Three examples of conflicts are shown in Figure~\ref{fig:conflicts}. Reading the figure clockwise, the examples are the following: the usage of encryption to gain security (more precise, confidentiality) causes a delay and therefore a reduction of the performance; the reduction of the energy consumption can be achieved by consolidation (i.e., reduction of the active elements), but the consolidation can create a single point of failure and therefore a reduction of the dependability; the dependability can be achieved by redundancy (i.e., deployment of multiple elements that perform the same functionality), the redundancy can increase the exposure risk since more elements can be attacked.


All potential conflicts will need to be studied in the design and operations of future 5G-MEC systems.
In this section, we present the challenges when these three aspects are considered together. The presented challenges are summarized in Table~\ref{tab:challenges}.

\begin{table*}[!t]
\caption{Summary of the Multi-aspect Challenges}
\begin{center}
\begin{tabular}{p{3cm}p{13.9cm}}
\toprule
\textbf{Topic} &\textbf{Challenges}\\
\midrule
\multicolumn{2}{l}{\textbf{MEC Host Level}}\\
Physical device & 
\vspace{-.3cm}
\begin{itemize}[leftmargin=*]
	\item Evaluate the impact on performance and security when \emph{redundant hardware and software} is used to increase the dependability of a MEH.
	\item Investigate the use of cryptographic primitives (e.g., secret sharing schemes) to achieve dependability with a decreased rate of duplicated data and lightweight cryptography to achieve better performance.
\end{itemize} 
\\
Virtualization technology &
\vspace{-.3cm}
\begin{itemize}[leftmargin=*]
	\item Develop \emph{VM migration} and \emph{VNF allocation} algorithms able to jointly take into the requirements in all three aspects.
	\item Explore new methods aimed to \emph{secure the container-based virtualization} without degrading its performance.
	\item Develop \emph{new lightweight and easy-to-instantiate virtualization. technologies} or \emph{new mixing and nesting models of virtualization technologies} able to support MEC URLLC services with security constraints. 
\end{itemize} 
\\
\midrule
\multicolumn{2}{l}{\textbf{MEC System Level}}\\
Deployment and design &
\vspace{-.3cm}
\begin{itemize}[leftmargin=*]
	\item Evaluate the \emph{deployment of MEHs} including dependability (failover), performance, and security requirements.
	\item Develop new \emph{failover mechanisms} able to jointly take into account security and performance constraints.
	\item Define robust design models aimed to optimize performance while satisfying dependability constraints.
\end{itemize} 
\\
Resource allocation &
\vspace{-.3cm}
\begin{itemize}[leftmargin=*]
	\item The allocation of computing and storage resources, also called \emph{task offloading}, should not only aim to performance targets (such as energy efficiency or latency reduction), but also consider dependability and security requirements.
\end{itemize}
\\	
User association &
\vspace{-.3cm}
\begin{itemize}[leftmargin=*]
	\item New multi-constrained path computation algorithms need to be developed for establishing user association methods able to jointly satisfy the constraints imposed by the three aspects.
	\item Other than performance-related metrics, new metrics need to be defined for security and dependability to find finding Pareto frontier when the three aspects are jointly considered.
\end{itemize} 
\\
MEO &
\vspace{-.3cm}
\begin{itemize}[leftmargin=*]
	\item The reliability and the performance of the MEO system can be increased by the design of \emph{distributed MEO architecture}, which must be deeply analyzed also from the security perspective.
	\item The \emph{MEO functionalities} should aim to orchestrate and manage the MEC system by jointly considering the three aspects.
\end{itemize} 
\\
Consistency &
\vspace{-.3cm}
\begin{itemize}[leftmargin=*]
	\item Consistency is an important requirement given the distributed and redundant nature of the system. These two features aim to improve the scalability and dependability of the system, but their impact on performance and dependability should be investigated.
	\item Deeply analyse new methods proposed for guaranteeing consistency (even in the case of active adversaries).
\end{itemize} 
\\
\midrule
\multicolumn{2}{l}{\textbf{MEC Federation Level}}\\
Heterogeneous scenario &
\vspace{-0.3cm}
\begin{itemize}[leftmargin=*]
	\item The interconnection between the MEC systems that are geographically distributed and focused on different use cases should guarantee proper levels of security, dependability, and performance.
\end{itemize} 
\\
Multiple operators &
\vspace{-0.3cm}
\begin{itemize}[leftmargin=*]
	\item Having MEC systems that belong to multiple operators has several challenges: from privacy concerns to end-to-end performance guarantees; from security exposure to end-to-end dependability; from compatibility issues to waste of resources.
\end{itemize} 
\\
Multi-domain orchestration&
\vspace{-0.3cm}
\begin{itemize}[leftmargin=*]
	\item The MEC federation can be over multiple domains (edge and cloud). Infrastructure-as-Code can be used to implement a multi-domain orchestration.
\end{itemize} 
\\
\midrule
\multicolumn{2}{l}{\textbf{General Challenges}}\\
Modelling &
\vspace{-.3cm}
\begin{itemize}[leftmargin=*]
	\item Develop new models for a joint evaluation of the 5G-MEC system from the three different perspectives.
\end{itemize} 
\\
Network connectivity &
\vspace{-.3cm}
\begin{itemize}[leftmargin=*]
	\item The network connectivity has an important impact on dependability and performance requirements and is the source of potential threats.
	\item New technologies are under development for increasing connectivity performance, such as Teraherz communications. This activity must be integrated with the development of new models for analyzing the impact of these technologies on the MEC system reliability and performance jointly.
	\item Develop new solutions for mMTC and URLLC use cases, such as those based on NOMA, able to improve system capacity while satisfying the constraints on dependability and security, in scenarios characterized by a large number of users exchanging small blocks of data.
\end{itemize} 
\\
Service continuity and user mobility& 
\vspace{-.3cm}
\begin{itemize}[leftmargin=*]
	\item The solutions for achieving service continuity (even in the case of user mobility) may lead to conflicts between the three aspects. New solutions must be developed considering the effects on these three aspects jointly, trying to achieve Pareto optimal.
\end{itemize}
\\
Monitoring and detection &
\vspace{-.3cm}
\begin{itemize}[leftmargin=*]
	\item The monitoring and detection for security, dependability, and performance should be jointly executed to exploit the cross information. For joint monitoring and detection, AI-based techniques can provide extra capabilities.
\end{itemize} 
\\
Ubiquitous Pervasive Intelligence &
\vspace{-.3cm}
\begin{itemize}[leftmargin=*]
	\item The MEC should be managed and orchestrated via secure, dependable, and performing intelligence.
        \item The security and dependability can be provided via edge intelligence solutions.
\end{itemize} 
\\
Joint KPIs &
\vspace{-.3cm}
\begin{itemize}[leftmargin=*]
	\item Availability should include both security and dependability causes, moreover a system should be considered up when only when the performance requirements are met.
\end{itemize} 
\\
\bottomrule
\end{tabular}
\label{tab:challenges}
\end{center}
\end{table*}


\subsubsection{MEC host level}

\paragraph{Physical device}
Section~\ref{sec:dep-challenges} has introduced that physical dependability can be achieved by using redundant hardware and software within a MEH.  Similarly, Section~\ref{sec:sec-challenges} has introduced the challenges related to physical security at the MEH level. The redundancy can have an impact on performance and security. For example, adding redundant physical devices increases the vulnerability related to direct physical access (e.g., the attacker can perform physical modifications on the device or on the communication link). 
The countermeasures against these threats increase costs and can impact performance. On the other hand, it was already mentioned that possible alteration or deletion of data (as a consequence of an attack) requires adequate backup and recovery techniques, which are feasible by means of dependability. Cryptographic primitives such as secret sharing schemes can be useful to allow recovery capabilities with lower duplication rates.
Moreover, physical security mechanisms are normally beneficial for performance, as they usually introduce less latency \cite{roman_mobile_2018,pham_survey_2020,bai2019energy}.

\paragraph{Virtualization technology}
In the previous sections, the importance of virtualization for each aspect has been introduced.

From a dependability perspective, the live VM migration in OpenStack \cite{hao-openstack-2019} and VNF placement in a MEC-NFV environment \cite{yala-vnf-2018,Chantre_2020} have been introduced in Section~\ref{sec:dep-challenges}. In both cases, dependability is investigated together with the performance.
In~\cite{hao-openstack-2019}, the authors analyze the impact of the system pressures and network failures on the performance of VM live migration by considering the migration time.
In \cite{yala-vnf-2018}, the authors address the problem of VNF placement by considering two conflicting objectives, namely minimizing the access latency and maximizing the service availability.

The MEC scenario can be characterized by lower computations and storage resources with respect to the cloud scenario. The hypervisor-based virtualization becomes unsuitable given that the image files of VMs are large and its overhead is non-negligible. Furthermore, when MEC is integrated with 5G, user mobility adds new requirements to the virtualization techniques. In particular, the service continuity and the fast migration of service between MEHs extend the popularity of container-based technology, which allows to easily build, run, manage, migrate, and remove containerized applications. The different container solutions (e.g. LXC, LXD, Singularity, Docker, Kata Containers, etc.) offer diverse performance in terms of CPU and memory load, security, networking bandwidth, and disk I/O, and configuration options that can further improve the performance in some specific scenarios \cite{Casalicchio2020}. The isolation mechanism of existing container-based technologies is weak, due to the sharing of one kernel among multiple isolated environments. This feature adds four types of problems to consider for container security: (i) protecting a container from applications inside it, (ii) inter-container protection, (iii) protecting the host from containers, and (iv) protecting containers from a malicious or semi-honest host. A big challenge is to explore new methods (e.g., using trusted images, managing container secret, securing the runtime environment, and vulnerability scanning), which can secure the container-based virtualization without degrading the performance in terms of agility, resource consumption, and computation delay. Alternatively, the unikernel technology guarantees security by the isolation provided by the underlying operating system or hypervisor. Unikernel is more secure than the container. However, from the performance perspective, unikernel presents low usability given that it does not have a shell and does not support online debugging, online upgrades and updates either. In the case of failure of a unikernel application, only the reboot solves the problem. Application and/or configuration updates require recompiling the source code to produce a new unikernel and deploy a new version. This action can be very costly and sometimes prohibitive. This weak point of unikernel technology represents a big challenge that needs to be considered in order to evaluate the most suitable virtualization technologies for MEC, able to achieve a good tradeoff between performance, dependability, and security. This evaluation can converge to a single new virtualization technology or a set of them, each one achieving the best tradeoff in some specific use cases. 


\subsubsection{MEC system level}

\paragraph{Deployment and design}
The problem of the deployment of MEHs in a particular area is commonly addressed with the aim of improving performance. Many existing studies focus on different goals, for example, to minimize the overall latency of mobile users, to maximize the overall throughput of the MEH network, etc.~\cite{Li_Y_2018_41_partenza}, \cite{Kasi_2020}, \cite{Cao_K_2021}.
Many of these studies assume that the MEHs are free of failures. However, in a dynamic and volatile network scenario, MEHs are subject to runtime errors caused by various events, such as software exceptions, hardware errors, cyber attacks, etc., similar to what occurs in cloud servers \cite{Benkacem_2018}. When these failure events happen, the design objectives and corresponding constraints can easily be violated. The quality of experience will decrease immediately and significantly, especially those of latency-sensitive applications. During a failure event, mobile users can be disconnected from the MEC infrastructure and the related services are unavailable.
This scenario can have a disruptive impact on the perceived quality of the offered services particularly in areas with a high density of users. Given the negative effects of the MEH failures, the design of the MEC infrastructure must consider the reliability aspects of the system. The robustness of the MEC infrastructure requires coverage of a specific area with a number of MEHs greater than strictly necessary to guarantee performance in ideal situations. Making a MEC design considering only the performance, without taking into account the robustness of the infrastructure, leads to localizing the MEHs, minimizing the overlap of the coverage of each of them. Simply maximizing the collective coverage of the MEC infrastructure could lead to situations where no MEH can take control of mobile users disconnected from failed MEHs. Such a MEC infrastructure is highly vulnerable to runtime errors. 

As already presented in Section~\ref{sec:dep-challenges}, to alleviate this problem, recent studies \cite{li-deployment-2020,tonini-deployment-2019} investigate the dependability in the deployment of the MEHs because of its impact on the effectiveness of the failover mechanisms.
Of course, considering only the dependability target is also not correct. The dependability should be jointly considered with the performance. For example, if both dependability and latency prefer a denser deployment of MEHs, energy consumption, and economic costs push for a less dense deployment. For this reason, the deployment strategies will aim to find the best trade-off in the given scenario. 

Moreover, the failover mechanisms themselves should consider performance but also security aspects. For example, the failover mechanisms should be fast (for example, it can be proactive \cite{huang-recovery-2019}) in order to reduce the delay. Moreover, using other users to rely on to reach distant MEHs \cite{satria-recovery-2017} might introduce severe security threats.

\paragraph{Resource allocation}
As we have already mentioned in previous sections, another challenge addressed by current studies is the allocation of resources, usually computing and storage, in the different MEHs.
As mentioned in Section~\ref{sec:dep-challenges}, the resource allocation in MEC is often called task allocation, since an application or procedure (task) is moved (offloaded) from the local execution in the mobile device to a remote execution on a MEH.

Most of the current works have as a target the energy efficiency. 
As mentioned in Section~\ref{sec:dep-challenges}, several works also consider the dependability metrics as requirements~\cite{chen-offloading-2015,chen-offloading-2016,reliab-qos-cache}.
Other works implicitly consider reliability and latency by focusing on the queue length~\cite{merluzzi-offloading-2019,merluzzi-offloading-2020,liubennis-offloading-2019}.
Some works consider both reliability and latency as constraints~\cite{liu-offloading-2017,liule-offloading-2020,2022-reliab-delay} or as targets~\cite{liu-offloading-2018}.

To the best of the authors' knowledge, no work considers security aspects in depth. The importance must be better investigated. For example, the task can have different security requirements, and the MEHs have different security guarantees. A general good practice is the adding of a threshold for resource consumption and guarantee of some resource availability for security functionalities~\cite{5g-netslice-sec}. Otherwise, for example, by compromising the user apps an adversary can use too many resources and thus affect performance. More specific approaches consider for example the usage of a low resource Intrusion Detection System (IDS) \cite{RJL2019} or lightweight cryptography, as already discussed in Section~\ref{sec:sec-challenges}.
A first work that tries to address all three aspects (security, dependability, and performance) together is ~\cite{edge-policy}.

\paragraph{User association}
As mentioned in Section~\ref{sec:dep-challenges}, some works address the task allocation together with the user-host association~\cite{liubennis-offloading-2019}. For some critical applications requiring low latency and high reliability, such as in some V2X and Industrial IoT scenarios, the design of user association algorithms jointly considering the three aspects is of paramount importance. Multi-constrained optimal path computation algorithms have been proposed for solving routing problems~\cite{Garroppo2010}. Some of these can be extended to solve the user association problem. Other than performance-related metrics, new metrics summarizing the security and the reliability level of a link and a node need to be studied and jointly considered in the path computation towards alternative MEHs. Through these tools, path computation solutions towards different MEHs can be then compared to select the MEH satisfying the user requirements on the three different aspects. To reduce the complexity added by the path computation problem, similar metrics can be defined only for the MEH to develop a user association algorithm aimed at finding the best trade-off among the three aspects.

Concerning security and performance, whenever possible, perform user security services locally, with no need for centralization (e.g., identification and authentication \cite{roman_mobile_2018}).

\paragraph{MEO}
The MEO is an important element of the MEC architecture. It is a logically centralized element that is the main responsible for the system-level management of MEC.
The ETSI MEC architecture \cite{ETSI:MEC003} specifies that the MEC applications have a certain number of resource requirements, and these requirements are validated by the MEO.

As introduced in Section~\ref{sec:dep-challenges}, the MEO is crucial for the availability and reliability of the MEC system because it orchestrates the creation of a MEC and the fault management of the other MEC elements.
The solutions aimed to have a dependable MEO, such as physical distributed MEO, can generate challenges from a security perspective given the increment of the exposure risk. As discussed in Section~\ref{sec:sec-challenges}, special attention must be given to the virtualization security.

\paragraph{Consistency}
In Section~\ref{sec:dependability}, consistency has been presented as a property needed when we use redundant elements for improving the dependability of a system. This is valid for the MEO but also for the MEHs (during the handover). Consistency is also important in the case of distributed systems, which aim, not only to increase dependability but also to increase scalability. 
For achieving consistency, the redundant and/or distributed elements need to exchange information about their states in order to have a consistent global view of the system. This information exchange can have an impact on the security and performance of the system and can be a source of risks that a potential attacker may exploit (i.e., an active attacker might aim to break the consistency of the system). Moreover, the information exchange can use resources and impact the performance of the system.


\subsubsection{MEC federation level}

\paragraph{Heterogeneous scenario}
The MEC federation will operate in a heterogeneous scenario where the MEC systems are geographically distributed and focused to different use cases.
This scenario brings new challenges, for example the interconnection between the MEC systems should guarantee proper levels of security, dependability and performance, which can be different from use case to use case. Therefore, the interconnectivity should provide different guarantees and isolation among the flow.

\paragraph{Multiple operators}
One of the possible business cases of the MEC federation considers MEC systems belonging to multiple operators~\cite{ETSI:MEC-federation}. This scenario implies many challenges in managing the MEC federation in all aspects including privacy concerns, end-to-end performance guarantees, security exposure, end-to-end dependability, compatibility issues, and waste of resources.

\paragraph{Multi-domain orchestration}
Moreover, the MEC federation can be over multiple domains (edge and cloud). This scenario requires a multi-domain orchestration~\cite{ETSI:MEC-federation}.
One of the possible orchestration methods can be the Infrastructure-as-Code, which works as a common tool that allows abstracting diverse provisioning methods
(API, CLI, etc.) used in the individual domains and activate infrastructure components by using a high-level language~\cite{ETSI:MEC-federation}. Infrastructure-as-Code can be used to implement a combined MEO/MEPM/VIM, as deployed in the Linux Foundation Edge (LFE) Akraino Public Cloud Edge Interface (PCEI) blueprint\footnote{https://www.lfedge.org/projects/akraino/release-4-2/public-cloud-edge-interface-pcei/}.


\subsubsection{General challenges}

\paragraph{Modelling}
A joint evaluation and analysis of the security, performance, and dependability of 5G-MEC is needed.
In the literature, there are several works that focus on joint modeling: performance and dependability \cite{1992-modelling}; security and dependability \cite{2009-trivedi,2004-depsecmod}.
There are also tools that aim to jointly evaluate these aspects, such as M\"{o}bius \cite{2009-mobius}.
To the best of the authors' knowledge, there is no current work that jointly evaluates the three aspects of MEC.
Work focuses on dependability and security in a medical IoT context~\cite{model-iot}.

\paragraph{Network connectivity}
The network is a key part of a MEC system. It includes the network access to reach the user, and the edge network to interconnect the MEHs and other MEC elements.
The network has an impact on performance (lack of requirements), dependability (lack or degraded connectivity), and security (the MEC system is exposed via the network). 

Recently, new studies are focused on the definition of new wireless communications systems, such as mmWave and TeraHertz. This trend is spurred by the need of having a very high data rate, necessary for reducing the delay related to task offloading of applications. However, these new communication systems are vulnerable to blocking events due to obstacles or beam collisions, which can interrupt the data exchange. As a concern, this trend, one set of challenges is related to the study of new beamforming techniques based on antenna arrays both at the transmitter and at the receiver side, and to the usage of multi-link communications. Another set of challenges is related to the simultaneous study of the performance and the reliability of the network connectivity with the goal of improving some performance parameters, such as latency, energy consumption, and deployment costs while achieving the reliability requirements. As an example, the reliability can be improved allowing the UE to send information to different MEHs, via different mmWave beams and over all the available UE-MEHs links simultaneously. More effort is necessary to define strategies, such as the Parallel Redundancy Protocol (PRP), in order to find an acceptable trade-off between network resource consumption and achieved reliability.

For mMTC and some URLLC use cases, the Non-Orthogonal Multiple Access (NOMA) represents a key technology of 5G for improving the system capacity and the connection density~\cite{NOMA2020}. The key idea of NOMA is to share a given resource slot (e.g., time/frequency) among multiple users, and apply Successive Interference Cancellation (SIC) to decode the transmitted information. NOMA allows grant-free transmission. This feature is particularly important for the uplink transmission of small data blocks, such as in the IoT scenarios. In this case, the control signaling overhead is reduced, as well as the transmission latency and the power consumption of the terminal device. However, NOMA adds some security issues that need to be carefully studied.  In the case the NOMA connection is used by two users for offloading tasks to a MEH at the same time when SIC is performed, one user can decode the other user’s message. During this period, an eavesdropper or an attacker may attempt to decode the mobile user’s message. To cope with this problem, the key challenges are to combine physical-layer security and NOMA-MEC in order to find solutions that can achieve a good trade-off between security and performance. Given the NOMA features, the performance is related to the latency, system capacity, and energy consumption on the user side.

Concerning both security and performance, Soft-VPLS (discussed in Section~\ref{sec:sec-challenges}) allows different traffic categories (e.g., MEC service requests, user data, control statistics) to be routed via distinct tunnels, aiming to enhance both end-to-end security and overall communication performance~\cite{RJL2019}.

\paragraph{Service continuity and user mobility}
The service continuity has been defined in Section~\ref{sec:dependability} and is critical even because the user(s) might be mobile.
A lack of service continuity has a huge impact on both performance and dependability.
However, it is often not easy to implement securely while satisfying performance needs.

Solutions to improve the service continuity can lead to a conflict between performance and dependability.
In MEC-host-assisted proactive state relocation for UE in connected mode~\cite{ETSI:MEC018}, the improved latency is accomplished at the price of relocation success, because failed handover operation may nullify the state transition preparation made by the MEHs. Therefore, a trade-off between latency and reliability is needed.

Similarly, solutions to achieve secure user mobility can lead to a conflict between security and performance. Special security protocols need to be set in place to allow secure mobility of users (e.g., from moving from one MEH to another) so that this process does not expose sensible data (e.g., identifiers, keys, sensible data). Forward and backward security play an important role in these scenarios (e.g., compromising a key shared between the user and a MEH should not expose previous or further keys shared by the user with past or future MEHs). Protocols such as key update or setting up a new security context (if needed) introduce some latency that, if not properly adopted, might affect performance.

\paragraph{Monitoring and detection} 
As presented in Section~\ref{sec:sec-challenges}, monitoring and detection are key elements for the security in 5G MEC, but they are also important for fault and performance management. The monitoring and detection for security, dependability, and performance should be addressed jointly in order to exploit the cross information and provide advanced features.
As already mentioned for security, AI-based techniques can be developed to provide advanced monitoring and detection functionalities \cite{ETSI:5G-MEC_AI}. Moreover, Virtual Machine Introspection (VMI), and hypervisor machine introspection should monitor the activities in terms of resource utilization to prevent performance issues and DoS attacks \cite{RJL2019}. As already discussed in Section~\ref{sec:sec-challenges}, a good tradeoff between local and global monitoring techniques would be beneficial for security in relation to performance.

\paragraph{Ubiquitous Pervasive Intelligence}

As introduced in Section~\ref{sec:background}, the main innovation of 6G is the ubiquitous and pervasive use of intelligence. MEC will help the achievement of such intelligence \cite{6g-edge-intelligence}. MEC will provide the computing resources close to the end users that will be used to provide the intelligence. MEC will use intelligence to manage and orchestrate its resources. In using and proving intelligence, there is the need to take care of security, dependability, and performance. The 6G requirements show extreme performances and 7-nine values of availability and reliability. These performance and dependability targets need to be jointly addressed in order to provide new specific use cases that are the "extreme" combination of the 5G use cases \cite{docomo}. Security will need to be also considered because trustworthiness is one of the main 6G features. The 6G security will be enabled by trust foundations, privacy-enhancing technologies, AI/ML assurance and defense, distributed ledger technologies, quantum security, and physical-layer security \cite{hexa-x}.

\paragraph{Joint KPIs}
New KPIs can be defined to jointly consider all the three aspects: security, dependability, and performance. 
Given the Tables~\ref{tab:sec_prop} and \ref{tab:dep-attr}, both security and dependability have availability as an attribute, even if its definition in the two contexts is different. Availability can be jointly defined as the readiness to access a correct service by legitimate parties. A correct service must also meet the performance requirements (which can be related to the metrics defined in Figure~\ref{fig:Functional}), and the causes of failing the performance requirements can be both failures and attacks, not only load dynamics.


\section{Discussion and Conclusions} \label{sec:conclusions}

In this survey, the state of the art and the challenges of the 5G MEC have been studied with respect to three aspects: security, performance, and dependability.

First, the ETSI MEC architecture has been introduced including the NFV-based version and the integration with 5G.

Second, for each aspect, the taxonomy, the state of the art, and the challenges have been presented.

The taxonomy has given a general introduction to the aspects to the readers that are only experts in the other aspects.
Comparing the taxonomy of the three aspects, we can notice differences - especially for the issues and the countermeasures, but also similarities - such as the presence of availability as an attribute for both security and dependability.

The state of the art has introduced the standards and the current surveys that address the investigated aspect of 5G MEC. It resulted that security is the most studied aspect of 5G-MEC systems, while dependability is the least studied.
The requirements for each element of the ETSI MEC architecture have also been highlighted, as indicated in the specifications. It was discovered that ETSI does not specify yet the requirements for important elements, such as the virtualization infrastructure, the user app LCM proxy, and the MEF.

The challenges of the investigated aspect of 5G MEC have been presented by categorizing them in MEC host level, MEC system level, and general challenges.
This study has shown that, although the main target of 5G-MEC systems is to improve the performance of network services, many works have addressed the security and strict security requirements have been specified. Fewer works have addressed the dependability of 5G MEC, even though URRLC services and mission-critical applications have strict requirements on dependability.
Moreover, several challenges, such as deployment, resource allocation, virtualization, service continuity, and MEO, are common to multiple aspects, but they are not jointly addressed yet.

Finally, the challenges of jointly addressing security, dependability, and performance have been investigated by using the same categorization (MEC host level, MEC system level, and general challenges) plus the MEC federation level. The investigation has shown the importance of jointly addressing the three aspects and how focusing on only one aspect can cause the failure of meeting the requirements on the other aspects. The new concept of MEC federation makes the integration of the three aspects even more important. The orchestration of heterogeneous resources and services at all levels becomes enormously complex, which may be efficiently managed by advanced AI techniques. In this context, the MEO becomes a critical element, especially from the security and dependability points of view. In the future perspective, the ubiquitous pervasive intelligence will help to manage the complexity of the 5G-MEC systems towards the 6G.




\bibliographystyle{IEEEtran}
\bibliography{./bib/general-survey.bib,./bib/standards-whitepapers.bib,./bib/sec-refs.bib,./bib/dep-refs.bib,./bib/perf-refs.bib,./bib/mix-refs.bib}

\vfill\null

\begin{IEEEbiography}[{\includegraphics[width=1in,height=1.25in,clip,keepaspectratio]{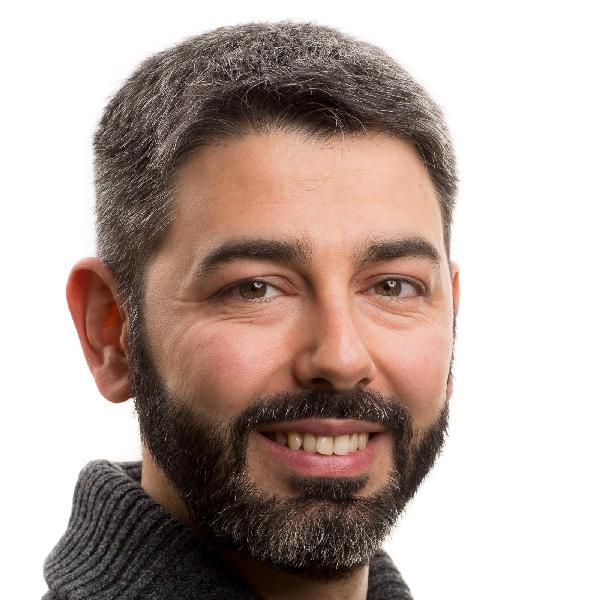}}]{Gianfranco Nencioni} is Associate Professor with the University of Stavanger, Norway, from 2018. He received the M.Sc. degree in telecommunication engineering and the Ph.D. degree in information engineering from the University of Pisa, Italy, in 2008 and 2012, respectively. In 2011, he was a visiting Ph.D. student with the Computer Laboratory, University of Cambridge, U.K. He was a Post-Doctoral Fellow with the University of Pisa from 2012 to 2015 and the Norwegian University of Science and Technology, Norway, from 2015 to 2018. He is currently the head of the Computer Networks (ComNet) research group and leader of the 5G-MODaNeI project funded by the Norwegian Research Council. His research activity regards modelling and optimization in emerging networking technologies (e.g., SDN, NFV, 5G, Network Slicing, Multi-access Edge Computing). His past research activity has been focused on energy-aware routing and design in both wired and wireless networks and on dependability of SDN and NFV.
\end{IEEEbiography}
\vfill\null

\begin{IEEEbiography}[{\includegraphics[width=1in,height=1.25in,clip,keepaspectratio]{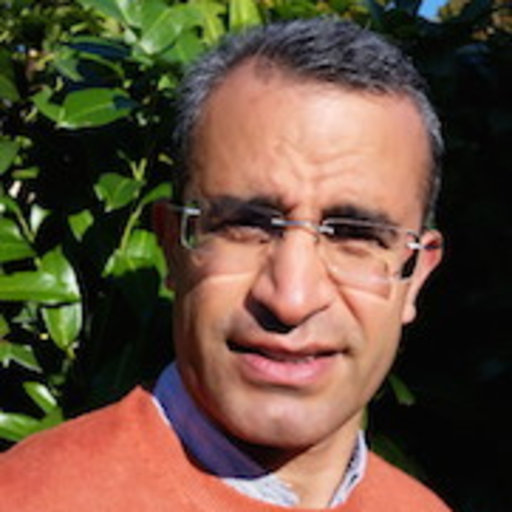}}]{Rosario G. Garroppo} is Associate Professor at the Dipartimento di Ingegneria dell’Informazione of the University of Pisa. His expertise is on networking and his main research activities are focused on experimental measurements and traffic modelling in broadband and wireless networks, MoIP systems, traffic control techniques for multimedia services in wireless networks, network optimization, and green networking. On these topics, he has published more than 100 peer-reviewed papers in international journal and conference proceedings, and won a Best Paper Award at the 4-th International Workshop on Green Communications (2011). He served as Technical Program Committee member of several international conferences on wireless networks, and as referee for several international journals and conferences. He was co-creator and co-organizer of the international IEEE Workshop on advanced EXPerimental activities ON WIRELESS networks and systems (EXPONWIRELESS), held in conjunction with IEEE WoWMoM since 2006 until 2009. 
\end{IEEEbiography}
\vfill\null

\begin{IEEEbiography}[{\includegraphics[width=1in,height=1.25in,clip,keepaspectratio]{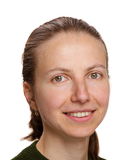}}]{Ruxandra F. Olimid} is Associate Professor at the Department of Computer Science, University of Bucharest.
She received her Ph.D. in Computer Science from the University of Bucharest in 2013. She has a background in both computer science (BSc and MSc from the University at Bucharest, 2008 and 2010) and telecommunications (BSc from the University Politehnica of Bucharest, 2009). She was a post-doctoral researcher and an Adjunct Associate Professor at the Department of Information Security and Communication Technology, Norwegian University of Science and Technology, Norway, from 2016 to 2018 and from 2018 to 2022, respectively. Her past experience includes Cisco certifications (CCNA, WLAN/FE) and almost 10 years in Orange Romania. Her research interests include cryptography and privacy, with a current focus on privacy and security in communication networks.
\end{IEEEbiography}
\vfill\null

\EOD

\end{document}